\documentclass[11pt]{article}
\usepackage{amsmath,amssymb,bm,epsf,epsfig,graphicx,hyperref,physics,cleveref,subfig,tensor}

\numberwithin{equation}{section}
\usepackage[backend=bibtex,natbib,style=numeric-comp,sorting=none, doi=false, isbn=false,url=false]{biblatex}
\input epsf.sty
\def \be {\begin{equation}}
\def \ee {\end{equation}}
\def \bea {\begin{eqnarray}}
\def \eea {\end{eqnarray}}
\def \bdm {\begin{displaymath}}
\def \edm {\end{displaymath}}

\def\0{\nonumber}

\begin{document}
\begingroup\allowdisplaybreaks
{}~ \hfill\hbox{CERN-TH-2025-196} \break
\begin{center}
 {\large \bf $\,$\\
\vskip2cm
Observables of boundary RG flows from string field theory
}

\vskip 1.1cm

{\large Jaroslav Scheinpflug\footnote{Email:
jscheinpflug at g.harvard.edu}$^{(a)}$
,  Martin Schnabl\footnote{Email:
schnabl.martin at gmail.com}$^{(b,c)}$, Jakub Vošmera\footnote{Email: jakub.vosmera at ipht.fr}$^{(d)}$
} \vskip 1 cm

$^{(a)}${\it {Jefferson Physical Laboratory, Harvard University} \\
{Cambridge, MA 02138, USA}}

\vskip .5 cm

$^{(b)}${\it {FZU - Institute of Physics of the Czech Academy of Sciences } \\
{Na Slovance 1999/2, 182 00 Prague 8, Czech Republic}}

\vskip .5 cm

$^{(c)}${\it {CERN, Theoretical Physics Department, CH-1211 Geneva 23, Switzerland} }

\vskip .5 cm

$^{(d)}${\it Institut de Physique Théorique\\
	CNRS, CEA, Université Paris-Saclay\\
 Orme des Merisiers, Gif-sur-Yvette, 91191 CEDEX, France}

\end{center}

\vspace*{6.0ex}

\centerline{\bf Abstract}

\vskip .2 cm

We present simple explicit formulae for the change of the $g$-function, boundary state, boundary spectrum and structure constants between the endpoints of short boundary RG flows at next-to-leading order. The formulae are derived using open string field theory and tested on integrable RG flows 
between conformal boundary states in unitary Virasoro minimal models.



\bigskip

\vfill \eject

\baselineskip=15pt

\tableofcontents

\section{Introduction and summary}
\label{sec:intro}

It was realized that string field theory (SFT) provides a consistent all-order Wilsonian scheme for two-dimensional conformal perturbation theory \cite{Mukherji:1991tb, GHOSHAL1991295, Sen:2019jpm,  Scheinpflug:2023osi, Scheinpflug:2023lfn, Maccaferri:2024cjb,Mazel:2024alu}. The key step is to represent the endpoint of an RG flow as a classical solution of the equations of motion of a SFT built on top of the UV CFT. Morally speaking, the reason for this is that the SFT equations of motion represent conditions for the vanishing of eta-functions \cite{Gaiotto:2024gii} i.e.\ the conditions for the endpoint to possess BRST symmetry. Since the BRST invariance we have in mind comes from gauging worldsheet diffeomorphisms by fixing conformal gauge, the equations of motion can also be thought of as representing the conditions for the endpoint to be a CFT (to the extent that the SFT around the classical solution comes from a worldsheet theory in conformal gauge \cite{Sen:1990hh, Sen:1990na, Sen:1992pw, Sen:1993mh}). The above is, of course, a bit heuristic because the worldsheet theory, on top of which the SFT is built, remains undeformed and one works in target space. 

The analogues of conformal perturbation theory data are then captured by off-shell string amplitudes, which are evaluated as UV CFT correlation functions integrated over regions of moduli space of punctured spheres with a suitable regulator and 
an integration measure. The regulator is imposed by recognizing that UV divergences of conformal perturbation theory are analogous to IR divergences in the target space SFT. In particular, they can be identified with degenerating Feynman diagrams, so they can be dealt with in a standard way \cite{Sen:2019jpm} i.e. by explicitly inverting the kinetic operator, rather than using a Schwinger parametrization. The Wilsonian cutoff corresponds to the size of regions of moduli space centered at the points where divergences develop, the so-called Feynman regions. A consistent partitioning of the off-shell amplitudes into contributions from Feynman diagrams and vertices (from a partitioning of the moduli space into Feynman and vertex regions) is determined by solving the so-called geometric BV master equation. Every such solution corresponds to a choice of string field frame \cite{Hata:1993gf}, and morally speaking a choice of conformal perturbation theory renormalization scheme. This leads to a geometrization of Wilsonian conformal perturbation theory, whose consistency is guaranteed by the geometric BV master equation. The Wilsonian cutoff can be easily varied by the procedure of adding stubs to the string vertices \cite{Erbin:2023hcs,Erler:2023emp, Schnabl:2023dbv, Schnabl:2024fdx, Maccaferri:2024puc, Stettinger:2024hkp}.  Given a classical solution and the SFT amplitudes, various IR CFT observables can be computed, as will be illustrated in what follows.\footnote{JS thanks Davide Gaiotto and Xi Yin for many useful discussions on the conceptual connection between SFT and conformal perturbation theory.}


Our primary focus in this paper will be on finding the endpoints of the RG flows resulting from the boundary CFT being perturbed by the integrated operator
\begin{align}
    \lambda \int_{-\infty}^\infty dx\, V(x)\,,
\end{align}
where $V$ is a boundary field with scaling dimension $h_V=1-y$. For $|y|\ll 1$, we will call such operators \textsl{nearly marginal}. As we will aim to use $y$ as an expansion parameter, we will assume that $y$ can be parametrically tuned to $0$. For the sake of keeping our presentation simple, we will also assume that the deformation by $V$ does not excite deformations by other quasi-primary operators of dimension close to 1 in the boundary spectrum. At the level of the $VV$ collision, this amounts to demanding that the only quasi-primary operator of dimension $1+\mathcal{O}(y)$ produced in the OPE is again $V$ (the RG flow is then often called \textsl{Fibonacci type}). As we will review in detail in Appendix \ref{app:opes}, this implies that the singular part of the OPE in the limit $y\to 0$ takes the form\footnote{We use the notation $\mathsf{T}\{\ldots\}$ to explicitly enforce correct ordering of operators as they collide along the boundary. In particular, we define
\begin{align}
    \mathsf{T}\big\{\psi_i(x)\psi_i(y)\big\} = \bigg\{\begin{array}{ll}
    \psi_i(x)\psi_j(y)     &  \text{if $x>y$} \\[2mm]
      \psi_j(y)\psi_i(x)    & \text{if $x<y$}
    \end{array}\,.
\end{align}
}
\begin{align}
    \mathsf{T}\big\{V(\xi) V(0)\big\} = \ldots + C_{VVV}\frac{V(0)}{|\xi|}+\text{less singular terms}\,,\label{eq:VVOPE}
\end{align}
where $\ldots$ stands for propagation of relevant states.\footnote{In particular, in Appendix \ref{app:opes} we will show that general algebraic structure of the OPE ensures that the derivatives of relevant quasi-primaries propagating in the $VV$ collision do not contribute to the simple pole in the limit $y\to 0$.}

Furthermore, in order to ensure tractability of perturbation theory, we wish to make additional assumptions which would guarantee the presence of a nearby-lying IR root $\lambda^\ast(y)$ of the beta function, or, equivalently, the existence of a solution $\Psi^\ast(y)$ of string field theory equations of motion which can be tuned to the perturbative string vacuum $\Psi=0$ as we take $y\to 0$. We will call such flows \textsl{short}. In particular, we will observe below (see also \cite{Zamolodchikov:1987ti,Ludwig:1987gs, Affleck:1991tk,Affleck:1992ng,Scheinpflug:2023osi, Scheinpflug:2023lfn, Mazel:2024alu}) that choosing $V$ such that 
\begin{align}
    C_{VVV}\neq 0
\end{align}
is sufficient to ensure existence of an IR fixed point already at cubic order in interactions and thereby a solution for $\lambda^\ast(y)$ which starts at linear order in $y$. This, in turn, will guarantee that the IR BCFT data can be expanded in integer power series in $y$ such that they can be seen to agree with the UV data as we take $y\to 0$.

As we would like to extract as complete an information about the IR BCFT as possible (including the OPE structure constants), we will invest substantial effort into identifying the spectrum of primary operators of the IR BCFT. For the sake of simplicity, we will assume that our BCFTs of interest are compact so that their spectrum is discrete.
As long as the RG flow at hand is short, it is then to be expected that the IR spectrum will be close, parametrically in $y$, to the UV spectrum. These provisions enable one to individually track the boundary primaries $\psi_i$ as the theory flows from the UV to the IR. 
To further streamline our analysis, we will restrict ourselves to the cases where there is no mixing between boundary operators with similar conformal dimension. To the first non-trivial order in $y$, this no-mixing property will turn out to follow from requiring that the $V\psi_i$ collision does not produce primary operators of dimension $h_i+\mathcal{O}(y)$ other than $\psi_i$. This, in turn, is reflected by the singular part of the OPE (in the limit $y\to 0$)
\begin{align}
\mathsf{T}\big\{V(\xi) \psi_i(0)\big\} = \ldots + C_{iiV}\frac{\psi_i(0)}{|\xi|}+\text{less singular terms}\label{eq:Vpsi_coll}
\end{align}
where $\ldots$ only contains contributions from states of weight $h<h_i+\mathcal{O}(y)$.\footnote{Again, the analysis of Appendix \ref{app:opes} reveals that derivatives of quasi-primaries $\psi_j$ appearing in the $V\psi_i$ collision are incapable of contaminating the simple pole of the $y\to 0$ collision \eqref{eq:Vpsi_coll}.}


The main outcome of this work is a set of practical formulae which enable one to compute (up to next-to-leading order in $y$) the BCFT data of the IR fixed point in terms of the UV BCFT data and integrated correlators in the UV BCFT involving the deforming operator $V$. The basic building blocks for our formulae are the following ``zero-momentum amplitudes'':
\begin{itemize}
    \item \textsl{3-point boundary amplitude}
\begin{align}
    \tilde{\mathcal{A}}_{ijk}= \big\langle \psi_i(\infty)\,\psi_j(1)\,\psi_k(0)\big\rangle = g C_{ijk}\,,  \label{eq:Amp3pt}
\end{align}
where $C_{ijk}$ is the boundary OPE structure constant of boundary primaries $\psi_i$, $\psi_j$, $\psi_k$ with generic dimensions $h_i$, $h_j$, $h_k$ and $g=\langle\mathbf{1}(0)\rangle$ is the $g$-function of the UV BCFT. 
\item \textsl{4-point boundary amplitude}
\begin{align}
    \tilde{\mathcal{A}}_{ijkV}&=\int\limits_{\substack{-\infty \\ \text{a.c.}}}^\infty d\xi\, \bigg[\big\langle\mathsf{T}\big\{\psi_i(\infty)\,\psi_j(1)\,\psi_k(0)\,V(\xi)\big\}\big\rangle-\Sigma_{ijk}(\xi)\bigg]\,,\label{eq:Amp4pt}
\end{align}
where $\mathsf{T}$ imposes the cyclic ordering of operator insertions along the boundary. Since the amplitude $\tilde{\mathcal{A}}_{ijkV}$ will only contribute at the highest order in $y$ which we will consider, we can afford to compute the integrand in the strict $y=0$ limit.
The logarithmic divergences, which would have generally appeared in the collisions of $V(\xi)$ with the boundary operator insertions at $0$, $1$ and $\infty$, are then avoided by introducing the subtraction
\begin{align}
   \Sigma_{ijk}(\xi)&= \frac{1}{2}\bigg(\frac{1}{|\xi|}+\frac{1}{|\xi-1|} - \frac{1}{|\xi||\xi-1|}\bigg)gC_{iiV}C_{ijk}+\nonumber\\
    &\hspace{4cm} +\frac{1}{2}\bigg(\frac{1}{|\xi-1|} + \frac{1}{|\xi||\xi-1|}-\frac{1}{|\xi|}\bigg)gC_{jjV}C_{ijk}+\nonumber\\
    &\hspace{4cm}+ \frac{1}{2}\bigg(  \frac{1}{|\xi||\xi-1|}+\frac{1}{|\xi|}-\frac{1}{|\xi-1|}\bigg)gC_{kkV}C_{ijk}\,.\label{eq:Sigmaijk}
\end{align}
On the other hand, power divergences are avoided by analytically continuing the integral around the collisions in the dimension $h$ of the propagating state. This is indicated by the subscript ``a.c.'' in \eqref{eq:Amp4pt}. In particular, one is meant to evaluate integrals using the following prescription: considering a function $f(\xi)$ which can be expressed as
\begin{align}
    f(\xi)= \sum_a\frac{C_a}{(\xi-\xi_0)^{\mu_a}}+\tilde{f}(\xi)\,,
\end{align}
where $\mu_a >1$ and $\tilde{f}(\xi)$ is integrable on some interval $[\alpha,\beta]$ around $\xi_0$, we define
\begin{align}
    \int\limits_{\substack{\alpha \\ \text{a.c.}}}^\beta d\xi\, f(\xi) \equiv \sum_a\frac{C_a}{1-\mu_a}(\beta-\xi_0)^{1-\mu_\alpha} -\sum_a\frac{C_a}{1-\mu_a}(\alpha-\xi_0)^{1-\mu_\alpha} +\int\limits_{\alpha}^\beta d\xi\, \tilde{f}(\xi)\,.\label{eq:ac_prescription}
\end{align}
This prescription is equivalent to instead computing the integral of $(\xi-\xi_0)^\varepsilon f(\xi)$ at a value of $\varepsilon$ such that $\varepsilon>\mu_a-1$ for all $a$ (so that the integral converges) and then analytically continuing the result to the value $\varepsilon=0$. In Appendix \ref{app:powers}, we will further discuss an explicit form of subtractions (see \eqref{eq:Piijk}) which can be introduced in \eqref{eq:Amp4pt} to implement the prescription \eqref{eq:ac_prescription} so as to avoid power divergences in practical computations. In particular, notice that when we send $\alpha\to -\infty$ and $\beta\to \infty$ then, assuming that $\tilde{f}(\xi)$ is integrable on $[-\infty,\infty]$, the above analytic-continuation prescription amounts to just dropping the power-divergent part of $f$ in the integral.

We also note that a special case of the log subtractions \eqref{eq:Sigmaijk} was already discussed in the context of conformal perturbation theory in general dimension in \cite{Komargodski:2016auf}, where it was derived by invoking one-loop exactness of the generalized free-field beta function.

\item \textsl{2-point bulk-boundary amplitude}
\begin{align}
    \tilde{\mathcal{B}}_{ai} = 2^{\Delta_a-h_i}\big\langle \phi_a(i,-i)\,\psi_i(0)\big\rangle = gB_{ai}\,,\label{eq:Amp2ptBB}
\end{align}
where $B_{ai}$ is the bulk-boundary OPE structure constant corresponding to the bulk primary field $\phi_a$ (with scaling dimension $\Delta_a$) and the boundary primary field $\psi_i$.

\item \textsl{3-point bulk-boundary amplitude}
\begin{align}
    \tilde{\mathcal{B}}_{aiV} = \frac{1}{2}\int\limits_{\substack{-\infty \\ \text{a.c.}}}^\infty d\xi\,\bigg[2^{\Delta_a-h_i}\, \big\langle \mathsf{T}\big\{\phi_a(i,-i)\,\psi_i(0)\,V(\xi)\big\}\big\rangle- \Sigma_{ai}(\xi)\bigg]\,,\label{eq:Amp3ptBB}
\end{align}
where, as in the case of the 4-point boundary amplitude \eqref{eq:Amp4pt}, we will only need to compute the integrand at $y=0$. The log divergences of the integrated correlator are removed by the subtraction
\begin{align}
   \Sigma_{ai}(\xi) =  gC_{iiV}B_{ai}\bigg(\frac{1}{|\xi|}-\frac{1}{|\xi+i|}\bigg) \,,\label{eq:Sigmaai}
\end{align}
while the power divergences are again handled by analytically continuing in the dimensions of the states propagating in the collision of $V$ with $\psi_i$.
\end{itemize}

All correlators are understood as having been computed on the upper-half plane in the UV BCFT. We normalize all UHP correlators so as to have $\langle\mathbf{1}(0)\rangle=g$, where $\mathbf{1}$ is the boundary identity and $g$ is the UV BCFT $g$-function. We also normalize all bulk and boundary primaries to have their 2-point functions on the sphere and disk, respectively, normalized to unity (upto a factor of $g$ in the case of the disk 2-point function).

In terms of the atomic ``amplitudes'' \eqref{eq:Amp3pt}, \eqref{eq:Amp4pt}, \eqref{eq:Amp2ptBB} and \eqref{eq:Amp3ptBB}, we then derive the following results for the IR BCFT data (denoted by an asterisk $^\ast$ in the superscript):
\begin{itemize}
    \item \textsl{$g$-function}
    \begin{align}
        g^\ast-g = -\frac{\pi^2}{3}\bigg[\bigg(\frac{y}{C_{VVV}}\bigg)^3 \tilde{\mathcal{A}}_{VVV}+\frac{1}{2}\bigg(\frac{y}{C_{VVV}}\bigg)^4 \tilde{\mathcal{A}}_{VVVV}+\mathcal{O}(y^5)\bigg]\,,
        \label{gFinal}
    \end{align}
    \item \textsl{anomalous dimensions}
    \begin{align}
        (h_i^\ast -h_i)\,g= 2\,\bigg(\frac{y}{C_{VVV}}\bigg)\tilde{\mathcal{A}}_{iiV}-\bigg(\frac{y}{C_{VVV}}\bigg)^2\bigg(\tilde{\mathcal{A}}_{iViV}-\frac{2}{3}\frac{C_{iiV}}{C_{VVV}}\tilde{\mathcal{A}}_{VVVV}\bigg)+\mathcal{O}(y^3)\,,
        \label{anomalDim}
    \end{align}
    \item \textsl{boundary-state coefficients}
    \begin{align}
        \langle \phi_a \| B^\ast\rangle\!\rangle-\langle \phi_a \| B\rangle\!\rangle  &= -2\pi\bigg[\bigg(\frac{y}{C_{VVV}}\bigg) \tilde{\mathcal{B}}_{aV}+\nonumber\\
        &\hspace{2cm}-\bigg(\frac{y}{C_{VVV}}\bigg)^2 \bigg(\tilde{\mathcal{B}}_{aVV}-\frac{1}{3}\frac{B_{aV}}{C_{VVV}}\tilde{\mathcal{A}}_{VVVV}\bigg)+\mathcal{O}(y^3)\bigg]\,,
        \label{bStateShift}
    \end{align}
     \item \textsl{boundary OPE structure constants}
    \begin{align}
        (C_{ijk}^\ast-C_{ijk})\,g = -\bigg(\frac{y}{C_{VVV}}\bigg)\tilde{\mathcal{A}}_{ijkV}+\mathcal{O}(y^2)\,,
        \label{boundaryOPE}
    \end{align}
    \item \textsl{bulk-boundary OPE structure constants}
    \begin{align}
        (B_{ai}^\ast-B_{ai})\,g = -2\frac{y}{C_{VVV}}\tilde{\mathcal{B}}_{aiV}+\mathcal{O}(y^2)\,.
        \label{bulkBoundaryOPE}
    \end{align}
\end{itemize}
The leading $\mathcal{O}(y^3)$ term for the change in the $g$-function was originally found by Affleck and Ludwig \cite{Affleck:1991tk, Affleck:1992ng} while a formula for the subleading correction was first derived by two of the authors in \cite{Scheinpflug:2023osi}, and recast in the form (\ref{gFinal}) in \cite{Scheinpflug:2023lfn} (modulo a careful treatment of boundary ordering). Also note that in contrast to the rest of the BCFT data, only the leading-order corrections to the structure constants are found, as they already involve the computation of an integrated correlator (see also \cite{Burrington:2023vei} for a recent investigation of the change of structure constants under bulk deformations).

As we will discuss at length below, we use Witten's open string field theory to derive the above formulae for the IR BCFT. To streamline this derivation and to avoid complications connected with the propagation of BRST-exact descendant fields, we will make one additional simplifying assumption: consider that the collisions of the deforming operator $V$ with other boundary primaries $\psi_i$ (including $V$ itself) produce quasi-primaries $\psi_j$ with $h_j<h_i$ which, in the limit $y\to 0$, satisfy $h_j-h_i\in\mathbb{Z}$. We will then demand that the corresponding structure constants satisfy
\begin{align}
  C_{Vij} =(-1)^{h_i  - h_{j}+1} C_{iVj}\,.\label{eq:res}
\end{align}
From the discussion in Appendix \ref{app:collV} it will become apparent that if the condition \eqref{eq:res} was violated, the open-string amplitude involving $V$ and $\psi_i$ on the external legs would propagate BRST-exact states which cannot be dealt with geometrically by connecting disks (representing elementary vertices) with open-string propagator strips. Instead, one would need to algebraically invert the SFT kinetic operator on such states to correctly obtain their contribution to the full amplitude (see \cite{Erbin:2019spp,Erbin:2020eyc,Sen:2020cef,Sen:2020eck,Sen:2021qdk} for other examples of situations where the need for such algebraic propagators arises). While doable in principle, this procedure would bring in additional technical difficulties which we choose not to address in this paper. Hence, while our general results for the IR BCFT data seem to make sense more universally (indeed, the condition \eqref{eq:res} does not seem to play any special role from the pure 2d CFT point of view), we should stress the caveat that they were only derived assuming \eqref{eq:res}. Note that in the particular case when $\psi_i=V$, the condition \eqref{eq:res} reduces to demanding that the $VV$ OPE does not propagate states $\psi_j$ with odd negative dimension (in the $y\to 0$) limit, while no constraint is put on propagating states $\psi_j$ with even dimension.
Also note that the condition \eqref{eq:res} holds in all of the explicit examples in Virasoro Minimal Models which we discuss in Section \ref{sec:mm}.

The rest of this paper is organized as follows. First, in Section \ref{sec:rg}, we review the construction of classical solutions representing the endpoints of short RG flows within Witten's open SFT (OSFT). We then follow by Section \ref{sec:g_function}, which discusses the computation of the $g$-functions in terms of OSFT action evaluated on such a classical solution. Section \ref{sec:anomalous_dimensions} then outlines the procedure of computing the IR BCFT spectrum by solving a deformed BRST cohomology problem. In Section \ref{sec:ellwood}, we compute the shifted boundary state via computing the so-called Ellwood invariant. Section \ref{sec:structure} then computes boundary and bulk-boundary structure constants of the IR fixed point via the computation of on-shell amplitudes of linearized fluctuations around the open-string background background given by the classical OSFT solution descriving the RG flow. Finally, we illustrate the applicability of the above general results on integrable flows between boundary states in unitary minimal models in Section \ref{sec:mm}. Section \ref{sec:disc} then concludes with a discussion of future directions. For many of the technical details, the reader is to refer to the appendices.

\section{RG flow as tachyon condensation}
\label{sec:rg}

Let us begin by reviewing the construction of \cite{Scheinpflug:2023lfn} whose idea is to describe the RG flows between two boundary conformal field theories in terms of condensing an open-string tachyonic excitation of a D-brane on a critical string background. This is achieved by embedding the UV BCFT with central charge $c$ into a critical open-string worldsheet theory by supplementing it with a suitable auxiliary theory $\text{BCFT}_\text{aux}$ with central charge $c_\text{aux} =26-c$ and the $(b,c)$-ghost sector $\text{BCFT}_{(b,c)}$, namely
\begin{align}
    \text{BCFT}\longrightarrow \text{BCFT}\oplus \text{BCFT}_\text{aux}\oplus \text{BCFT}_{bc}\,.\label{eq:embedding}
\end{align}
One then hopes to realize the IR fixed point of a short boundary RG flow triggered by a nearly-marginal operator $V$ by constructing a classical solution 
\begin{align}
    \Psi^\ast(y)=\lambda^\ast(y)\, cV+\mathcal{O}\big((\lambda^\ast)^2\big)
\end{align}
of the equations of motion of the SFT for the background \eqref{eq:embedding}, such that it only excites the identity module in the auxilliary sector $\text{BCFT}_\text{aux}$. It will be the subject of this section to describe the procedure of constructing $\Psi^\ast(y)$ in some detail. 

Here one needs to bear in mind that by itself, neither the critical coupling $\lambda^\ast(y)$, nor the classical solution $\Psi^\ast(y)$ are physical observables. In fact, they will depend on the off-shell data of the particular string field theory one works with. This can be understood in the broader context of conformal perturbation theory by realizing that quantities such as the beta function and the value of the coupling at the fixed point are not universal and may depend on which renormalization scheme we pick. The freedom of choice of string field theory vertices therefore corresponds to the choice of the particular renormalization scheme in which we conduct intermediate steps of our calculation.
To eventually extract physical information about the corresponding IR BCFT, one then needs to compute various string field theory observables such as partition functions and on-shell scattering amplitudes for the new critical string background given by the solution $\Psi^\ast(y)$. A number of such observables will be analyzed below in detail in Sections \ref{sec:g_function}, \ref{sec:anomalous_dimensions}, \ref{sec:ellwood} and \ref{sec:structure}.

\subsection{Witten's cubic open string field theory}
\label{sec:sft_intro}

In order to address numerous questions in string theory, ranging from non-perturbative corrections to string amplitudes \cite{Balthazar:2019rnh, Balthazar:2019ypi, Sen:2020cef, Sen:2021qdk, Alexandrov:2021shf, Alexandrov:2021dyl, Eniceicu:2022nay, Eniceicu:2022dru, Alexandrov:2022mmy, Alexandrov:2023fvb, Kaushik:2025neu, Alexandrov:2025pzs} and the construction of otherwise difficult to access curved string backgrounds \cite{Cho:2018nfn, Cho:2023mhw, Kim:2024dnw,Cho:2025coy, Frenkel:2025wko} to dealing with time-dependent backgrounds \cite{Sen:2002nu, Sen:2023qya, Cho:2023khj} and, including, tachyon condensation \cite{Sen:1999nx, Gaiotto:2002wy, Yang:2005rx, Schnabl:2005gv, Hellerman:2008wp, Erler:2009uj,Kudrna:2018mxa, Erler:2019xof, Scheinpflug:2023lfn, Mazel:2024alu}, it is beneficial to have at our disposal an off-shell field theory for strings, namely string field theory (see \cite{Erler:2019vhl,Erbin:2021smf,Maccaferri:2023vns,Sen:2024nfd, Yin_stringbook_2025} for recent reviews). An example of such a framework for open-string degrees of freedom is provided by the Witten's cubic open string field theory \cite{Witten:1985cc} whose action reads
\begin{align}
    S[\Psi]=-\frac{1}{2} \big\langle \Psi, Q \Psi \big\rangle -\frac{1}{3}\mathcal{V}_{0,3}(\Psi,\Psi,\Psi)\,.\label{eq:SWitten}
\end{align}
Here $\Psi$ is the open-string field which takes its values in the Hilbert space $\mathcal{H}_\mathrm{o}$ of the combined matter+ghost worldsheet BCFT corresponding to some consistent open-string vacuum (D-brane) of our choice. An inner product on this Hilbert space is provided by the BPZ inner product $\langle \cdot,\cdot\rangle$ which can be defined in terms of a BCFT correlator on the upper-half plane and the $SL(2,\mathbb{R})$ map $\mathcal{I}(z) = -1/z$ as
\begin{align}
    \big\langle \Psi_1,\Psi_2\big\rangle  \equiv \big\langle \mathcal{I}\circ \Psi_1 (0)\,\Psi_2(0)\big\rangle_\mathrm{UHP}\,.
\end{align}
The kinetic term of the action involves the worldsheet BRST charge
\begin{align}
    Q=c_0 L_0 -b_0\sum_{n\neq 0} nc_{-n}c_{n}+\sum_{n\neq 0} c_{-n} L_n^{\mathrm{(m)}}+\frac{1}{2}\sum_{\substack{m,n\neq 0\\ m+n\neq 0}} (m-n) \,c_{m} c_{n} b_{-m-n}\,,
\end{align}
where $L_n$ are the modes of the total worldsheet stress-energy tensor while $L_n^{\mathrm{(m)}}$ are the modes of just the matter BCFT stress-energy tensor. Finally, the cubic interaction term in \eqref{eq:SWitten} can be expressed in terms of the Witten $\ast$-product as
\begin{align}
    \mathcal{V}_{0,3}(\Psi_1,\Psi_2,\Psi_3) \equiv \big\langle \Psi_1,\Psi_2\ast\Psi_3 \big\rangle \equiv \big\langle F_1\circ \Psi_1(0)\, F_2\circ \Psi_2(0)\, F_3\circ \Psi_3(0)\big\rangle_\mathrm{disk}\,,\label{eq:star_def}
\end{align}
where the functions $F_k(w)$ map a unit upper-half disk (local coordinate patch $w$) into the unit disk and are given by
    \begin{align}
        F_1(w) = \bigg(\frac{1+iw}{1-iw}\bigg)^\frac{2}{3}\,,\qquad
        F_2(w) =e^{\frac{2\pi i }{3}}F_1(w)\,,\qquad
        F_3(w) =e^{\frac{-2\pi i }{3}}F_1(w)\,.\label{eq:Fs}
    \end{align}
Given these definitions, it is possible to establish the following properties
\begin{enumerate}
    \item $Q$ is nilpotent
    \begin{align}
        Q^2=0\,,
    \end{align}
    \item $Q$ is a derivation of the $\ast$-product
    \begin{align}
        Q(\Psi_1\ast \Psi_2) = (Q\Psi_1)\ast\Psi_2+(-1)^{|\Psi_1|}\Psi_1\ast Q\Psi_2\,,
    \end{align}
    where $|\Psi|$ denotes the ghost number (grassmanality) of $\Psi$.
    \item $\ast$-product is associative
    \begin{align}
        \Psi_1\ast (\Psi_2\ast\Psi_3) = (\Psi_1\ast\Psi_2)\ast\Psi_3\,.
    \end{align}
\end{enumerate}
The triplet $(\mathcal{H}_\mathrm{o},Q,\ast)$ therefore constitutes a differential graded algebra. Noting the cyclic definition \eqref{eq:Fs} of the local-coordinate maps $F_k$, it also holds that
\begin{align}
    \big\langle \Psi_1,\Psi_2\ast\Psi_3\big\rangle = \big\langle \Psi_2,\Psi_3\ast\Psi_1\big\rangle=\big\langle \Psi_3,\Psi_1\ast\Psi_2\big\rangle\,.
\end{align}
The BPZ inner product $\langle\cdot,\cdot\rangle$ therefore equips $(\mathcal{H}_\mathrm{o},Q,\ast)$ with a cyclic structure. 
In addition, we note that the local coordinate maps \eqref{eq:Fs} defining the Witten vertex enjoy the so-called \emph{twist symmetry}. In particular, the twist transformation implements the reflection of the open string around its midpoint. In terms of the local coordinate $w$ on the unit upper-half disk, it amounts to the involution
\begin{align}
    t:\quad w \mapsto -w^\ast\,.\label{eq:t}
\end{align}
We say that the 3-vertex is twist symmetric if, for $k=1,2,3$, there exist involutions $T_k$ such that
\begin{align}
    T_k \circ F_k = F_k \circ t\,,\label{eq:Tcov}
\end{align}
which leave the 3-punctured disk invariant (of course, if \eqref{eq:Tcov} holds for, say, $k=1$, then cyclicity guarantees that it holds also for $k=2,3$).  
For the particular case of the Witten 3-vertex, we observe we can simply set
\begin{align}
    T_1(Z)= Z^\ast\,,\qquad  T_2(Z) = e^{\frac{2\pi i}{3}} T_1(e^{-\frac{2\pi i}{3}} Z)\,,\qquad  T_3(Z) = e^{-\frac{2\pi i}{3}} T_1(e^{\frac{2\pi i}{3}} Z)
\end{align}
to see that it is indeed twist-symmetric. Expressing the $\ast$-product of two states $\varphi_1(0)|0\rangle$ and $\varphi_2(0)|0\rangle$ in terms of an OPE, one can see that twist symmetry of the vertex implies that the two states are inserted in the OPE symmetrically around the origin: in particular, one can use the definition \eqref{eq:star_def} to show that for the Witten vertex, we can write
\begin{align}
    \varphi_1(0)|0\rangle\ast \varphi_2(0)|0\rangle = U_3^\ast\,g_1\circ \varphi_1(0)\,g_2\circ \varphi_2(0)|0\rangle\,,
\end{align}
where we have introduced the functions
\begin{subequations}
    \begin{align}
        g_1(x)&=\cot\big(\tfrac{2}{3}(+\tfrac{\pi}{2}-\arctan x)\big)\,,\\
        g_2(x)&=\cot\big(\tfrac{2}{3}(-\tfrac{\pi}{2}-\arctan x)\big)\,,
    \end{align}
\end{subequations}
which map 0 to $\pm 1/\sqrt{3}$. Here we define $U_r$ to be operators on the worldsheet boundary CFT Hilbert space implementing the conformal mapping 
\begin{align}
    f_r(z) = \tan\big(\tfrac{2}{r}\arctan z\big)\,,
\end{align}
that is 
\begin{align}
    U_r\, \phi(z)\,U_r^{-1} \equiv  f_r \circ\phi(z)\,.
\end{align}
(See e.g.\ \cite{Schnabl:2002gg} for more details.) $U_r^\ast$ then denotes the BPZ conjugate of $U_r$. 

Varying the action \eqref{eq:SWitten} w.r.t.\ the string field $\Psi$, one obtains the equation of motion
\begin{align}
    Q\Psi +\Psi\ast\Psi=0\,.\label{eq:EOMWitten}
\end{align}
The classical solutions $\Psi^\ast$ of \eqref{eq:EOMWitten} are conjectured to be in one-to-one correspondence with consistent open-string vacua (D-branes) for a given closed-string background which stays fixed. See \cite{Erler:2014eqa,Erler:2019fye} for a construction of classical solutions for any known BCFT. To identify the vacuum which the solution corresponds to, we have to evaluate  observables on $\Psi^\ast$, examples of which we will discuss in detail below in Sections \ref{sec:g_function}, \ref{sec:anomalous_dimensions}, \ref{sec:ellwood} and \ref{sec:structure}.

Furthermore, using the Feynman rules dictated by the action \eqref{eq:SWitten} it is possible to reproduce all on-shell tree-level $n$-point open-string amplitudes. 
Indeed it is possible to show that the algebraic properties of the $\ast$-product enable one to cover the moduli spaces of these amplitudes using only the cubic vertices and open-string propagators. 
Moreover, it was also shown \cite{Zwiebach:1992bw} that adding to the Feynman rules of \eqref{eq:SWitten} the Shapiro-Thorn vertex \cite{Shapiro:1987ac,Shapiro:1987gq}
\begin{align}
    \mathcal{V}_{1,1}(\Phi,\Psi) \equiv  \big\langle \Phi(i,-i) I, \Psi\big\rangle\,,
\end{align}
the theory correctly accounts for all disk amplitudes involving an arbitrary number of on-shell open-string states as well as on-shell closed-string states $\Phi$.
Here $I$ stands for the string field which acts as the identity of the $\ast$-algebra. As a state, it can be expanded in terms of the total worldsheet Virasoro descendants of the vacuum as
\begin{align}
    |I\rangle = U_1^\ast |0\rangle  = e^{L_{-2}-\frac{1}{2}L_{-4}+\frac{1}{2}L_{-6}-\frac{7}{12}L_{-8}+\ldots} |0\rangle\,.\label{eq:ShapiroThorn}
\end{align}
Also note, that in order for the open-string midpoint insertion of $\Phi$ on $I$ to be well-defined, we need to restrict $\Phi$ to be a weight $(0,0)$ primary state.
In general, expanding the action \eqref{eq:SWitten} in small fluctuations around the classical solution $\Psi^\ast$, the on-shell amplitudes provide one with a large spectrum of observables which one may use to characterize the vacuum represented by $\Psi^\ast$.



\subsection{Perturbative solution for short RG flows}

To find the classical solution $\Psi^\ast(y)$ representing the short RG flow, we find it convenient to decompose the string field $\Psi$ as
\begin{align}
    \Psi = R+X\,,
\end{align}
where $R\equiv P_0 \Psi$ is the component of $\Psi$ in the image of $P_0$ which is defined to project onto states with total worldsheet conformal dimension $\mathcal{O}(y)$. In principle, this may include not only the deforming state $cV$, but also the Nakanishi-Lautrup field $\partial c$, other almost marginal fields in the matter BCFT, as well as states constructed from descendants of matter primaries with negative conformal dimension. Nevertheless, recalling the assumed form \eqref{eq:VVOPE} of the $VV$ collision, which is free of mixing with other almost marginal fields in the theory, as well as the fact that the Witten vertex is twist-symmetric, we quickly realize that the most general form of $R$ reads
\begin{align}
    R=\lambda(y) cV+ R^\prime\,,\label{eq:ansatzPsi}
\end{align}
where the string field $R^\prime$ is defined to contain other possible states with weight $\mathcal{O}(y)$ which can be generated at high order in interactions. 
We will assume that the solution is \textsl{triggered} by the slightly relevant state $cV$, so that both the $1-P_0\equiv\bar{P}_0$ component $X$ of the solution, as well as the field $R^\prime$, only enter at higher order in interactions.

We will now substitute the ansatz \eqref{eq:ansatzPsi} into the OSFT equation of motion \eqref{eq:EOMWitten} to find expressions for the string fields $R$ and $X$ perturbative in $y$. In particular, we will work in a scheme where we will compute successively better approximations in $y$ to the true solution $\Psi^\ast(y)$. To this end, we will denote by $\Psi^{(k)}$
a string field (or, more precisely, an equivalence class of string fields) which agrees with the solution $\Psi^\ast(y)$ up to and including the order $y^k$. Also, for the sake of concreteness, from now on we will focus on the case of such RG flows for which $\lambda(y)$ starts at order $y$, that is
\begin{align}
    \lambda(y)=\lambda_1 y +\lambda_2y^2 +\ldots\,.
\end{align}
We will see in a moment that this corresponds to the case where the perturbing operator $V$ experiences a non-vanishing cubic potential. Probing the SFT equation of motion \eqref{eq:EOMWitten} up to and including order $k+1$ then amounts to verifying the relation\footnote{Alternatively, we could have chosen to solve the equation of motion \eqref{eq:EOMWitten} as in \cite{Scheinpflug:2023osi} by expanding the string field as 
\begin{align}
    \Psi^\ast(y) = \sum_{k=1}^\infty \Psi_k
\end{align}
and find conditions for $\Psi_k=\mathcal{O}(\lambda^k)$ order by order in $\lambda$.
}
\begin{align}
    Q\Psi^{(k+1)} +\Psi^{(k)}\ast \Psi^{(k)}=0\,,\label{eq:EOMapprox}
\end{align}
where the equality should be enforced up to and including terms of order $y^{k+1}$.
Furthermore, we note that hitting the equation \eqref{eq:EOMapprox} with the projectors $\bar{P}_0$ and $P_0$ and realizing that they commute with $Q$, we can split \eqref{eq:EOMapprox} into the $P_0$ and $\bar{P}_0$ components as
\begin{subequations}
\begin{align}
  QX^{(k+1)} +\bar{P}_0(\Psi^{(k)}\ast \Psi^{(k)})&=0  \,,\label{eq:EOMPbar}\\
  QR^{(k+1)} +{P}_0(\Psi^{(k)}\ast \Psi^{(k)})&=0  \,.\label{eq:EOMP}
\end{align}
\end{subequations}
In particular, once the approximation $\Psi^{(k)}$ of the solution up to the order $y^k$ has been found, the $\bar{P}_0$-part $X^{(k+1)}$ of the refined string field $\Psi^{(k+1)}$ can be straightforwardly constructed by inverting $Q$. For instance, in Siegel gauge, we write
\begin{align}
    X^{(k+1)} = -\frac{b_0}{L_0}\bar{P}_0 (\Psi^{(k)}\ast \Psi^{(k)})\,.
\end{align}
Since we assume that the solution only starts at linear order in $y$ inside the image of $P_0$, its $\bar{P}_0$ projection only starts receiving contributions at the order $y^2$. That is, we can put 
\begin{align}
    X^{(1)}=0 \qquad \implies \qquad \Psi^{(1)}=R^{(1)}\,.\label{eq:X10}
\end{align}
On the other hand, finding the string fields $R^{(k)}$ requires more care. For $k=0$, the relation \eqref{eq:EOMP} is solved by putting
\begin{subequations}
\begin{align}
    \lambda^{(1)}(y) &= \lambda_1 y\,,\\
    {R^\prime}^{(1)} &= 0\,,
\end{align}
\end{subequations}
where, for the time being, $\lambda_1$ remains undetermined. A more non-trivial constraint then comes in for $k=1$. In this case, using \eqref{eq:X10}, we can rewrite \eqref{eq:EOMP} as
\begin{align}
    Q R^{(2)}+P_0\big(R^{(1)}\ast R^{(1)}\big)=0\,.\label{eq:eom2}
\end{align}
Here one has to realize that for a generic $VV$ OPE of the form \eqref{eq:VVOPE}, the projected star product $P_0(R^{(1)}\ast R^{(1)})$ may not produce just a state proportional to $c\partial c V$. Indeed, referring to discussion in the summary part of Section \ref{sec:intro} and the detailed analysis of Appendix \ref{app:opes}, if the $VV$ collision produces a quasi-primary $\psi_i$ with nearly-odd negative dimension, that is $h_i+\mathcal{O}(y)\in 2\mathbb{Z}+1$ such that $h_i<0$, the result will be contaminated by the descendants of $c\partial c\, \psi_i$. Even though these states would be typically trivial in the BRST cohomology,
one would still have to go through the trouble of algebraically inverting $Q$ on them to find the corresponding state ${R^\prime}^{(2)}$ which would restore balance in the equation of motion \eqref{eq:eom2}. While one should not find any in-principle obstruction while performing this procedure, we expect it to be technically demanding to handle in full generality. Hence, to streamline our derivation, we will choose to assume that the $VV$ collision does not produce quasi-primaries with nearly-odd negative dimensions. One then finds that \eqref{eq:eom2} can be solved by putting
\begin{subequations}
\begin{align}
    \lambda^{(2)}(y) &= \frac{y}{C_{VVV}}+\lambda_2 y^2+\mathcal{O}(y^3)\,,\label{eq:lambda(2)}\\
    {R^\prime}^{(2)} &= 0\,,
\end{align}
\end{subequations}
for some $\lambda_2$, at least as long as we choose to restrict ourselves to flows for which the deforming operator $V$ further satisfies
\begin{align}
    C_{VVV}\neq 0\,,\label{eq:CVVVneq0}
\end{align}
namely that the deforming operator $V$ experiences cubic effective potential.
While this is not a necessary condition for the flow to be short, it is sufficient to guarantee that the solution for the critical coupling $\lambda^\ast(y)$ exists at the leading order in the expansion in interactions and, ultimately, that it can be expanded as an integer-power series in $y$. We will focus on the flows satisfying \eqref{eq:CVVVneq0} from now on. If we had $C_{VVV}=0$, then one would have to search for the solution at higher order in interactions. Then it typically happens that $\lambda^\ast(y)$ is expanded in fractional powers of $y$. 

Finally, in order to completely fix the form of the solution $\Psi^\ast(y)$ at the order $y^2$, let us work out the correction to the critical coupling $\lambda^\ast(y)$ at next-to-leading order, that is, the value of the undetermined factor $\lambda_2$ from \eqref{eq:lambda(2)}.  For this, we only need to solve the projection of the $k=2$ equation of motion \eqref{eq:EOMP} along the direction spanned by the states proportional to $c\partial c V$. Overlapping \eqref{eq:EOMP} for $k=2$ with the state $cV$ and carefully evaluating the Witten 3-vertex (bearing in mind that the state $cV$ has dimension $-y\neq 0$) one finds that it translates into the equation
\begin{align}
    -yg\lambda^{(3)}(y)+ (\lambda^{(2)}(y))^2\, gC_{VVV}K^{3y}-\frac{1}{3}(\lambda^{(1)}(y))^3\mathcal{A}_{VVVV}=0\,,\label{eq:lambda_eq}
\end{align}
for $\lambda^{(3)}(y)$. Here $K = \frac{3 \sqrt{3}}{4}$ is the mapping radius of the Witten vertex, which is effectively setting the value of the Wilsonian cutoff in the corresponding renormalization scheme, and we have also introduced the notation 
\begin{align}
   \mathcal{A}_{VVVV}= 6\,\Big\langle cV\ast cV,\frac{b_0}{L_0}\bar{P}_0\, (cV\ast cV)\Big\rangle\label{eq:AVVVV}
\end{align}
for the Witten OSFT Feynman diagram with one propagator and four identical external legs $cV$. Using the detailed computations of Appendix \ref{app:feyn}, \eqref{eq:AVVVV} can be evaluated as
\begin{align}
    \mathcal{A}_{VVVV}&= \tilde{\mathcal{A}}_{VVVV}+12gC_{VVV}^2\log K\,,\label{eq:AVVVVeval}
\end{align}
where $\tilde{\mathcal{A}}_{VVVV}$ is the zero-momentum 4-point boundary amplitude \eqref{eq:Amp4pt} evaluated in the case when $\psi_i=\psi_j=\psi_k=V$. More explicitly, we can write $\tilde{\mathcal{A}}_{VVVV}$ as an integrated correlator
\begin{align}
    \tilde{\mathcal{A}}_{VVVV}&=\int\limits_{\substack{-\infty \\ \text{a.c.}}}^\infty d\xi\, \bigg[\big\langle\mathsf{T}\big\{V(\infty)\,V(1)\,V(0)\,V(\xi)\big\}\big\rangle-\frac{1}{2}\bigg(\frac{1}{|\xi|}+\frac{1}{|\xi-1|} + \frac{1}{|\xi||\xi-1|}\bigg)gC_{VVV}^2\bigg]\,,\label{eq:AVVVVt}
\end{align}
where all logarithmic divergences due to propagation of the (nearly) marginal state $V$ in the $VV$ collisions are explicitly subtracted. Moreover, as explained in detail in Appendix \ref{app:plumbing}, power divergences can be removed by analytically continuing in the weights of the propagating states, as indicated by the subscript ``a.c.''. In Appendix \ref{app:powers}, we derive an explicit form \eqref{eq:Piijk} of subtractions which can be performed to implement this analytic continuation in practice. Also note that even though \eqref{eq:AVVVVt} was derived assuming the restriction \eqref{eq:res} on states propagating in the $VV$ OPE, ensuring that $P_0(cV\ast cV)$ is proportional to $c\partial c V$, the result \eqref{eq:AVVVVt} for $\tilde{\mathcal{A}}_{VVVV}$ seems to make sense more generally: indeed, whatever negative-dimension quasi-primary states are generated in the $V(\xi)V(0)$ collision, their descendants never contaminate the simple pole $\frac{1}{\xi}$, as substantiated by \eqref{eq:poch0} (similarly for the collisions at $1$ and $\infty$). Hence, the explicit subtractions on the r.h.s.\ of \eqref{eq:AVVVVt} correctly cancel all log divergences.

Solving \eqref{eq:lambda_eq} for $\lambda^{(3)}(y)$, we can finally write the full solution $\Psi^\ast(y)$ (up to the order $y^2$) for a short RG flow induced by $V$ as
\begin{align}
    \Psi^\ast(y) = \lambda^\ast (y)\, cV - (\lambda^\ast (y))^2\frac{b_0}{L_0}\bar{P}_0  (cV\ast cV)+\mathcal{O}(y^3)\,,\label{eq:PsiStar}
\end{align}
where the critical coupling $\lambda^\ast(y)$ takes the form
\begin{align}
    \lambda^\ast(y) = \frac{y}{C_{VVV}}+\frac{1}{3}\bigg(\frac{y}{C_{VVV}}\bigg)^2\bigg(\frac{\tilde{\mathcal{A}}_{VVVV}}{g C_{VVV}}+3C_{VVV}\log K\bigg)+\mathcal{O}(y^3)\,.
\end{align}
In accordance with general expectation, we observe that the solution $\Psi^\ast(y)$ bears explicit dependence on the local coordinates of the Witten vertex. To see that this dependence drops our from various observables computed on this solution will provide a non-trivial check on the validity of our results in the following sections.

\section{Boundary degeneracy from on-shell action}
\label{sec:g_function}

The boundary degeneracy $g$, often called the $g$-function, is one of the most elementary characteristics of a conformal boundary condition. Suppose we put the BCFT on a Euclidean cylinder and that we take the bulk-channel thermodynamic limit where the cylinder is taken to be spatially long (compared to the scale set by the temperature). The $g$-function then corresponds to an intensive contribution to the free energy, correcting the extensive Casimir energy contribution of the bulk. This description makes it clear that it is also the coefficient of the identity Ishibashi state in the conformal boundary state. In string theory, where conformal boundary conditions of the worldsheet BCFT correspond to D-branes, it is proportional to the disc partition function, thus giving a leading contribution to the tension of the D-brane.

Thus, when studying boundary RG flows, determining the change of the $g$-function is one of the first computations one would like to perform. Despite this, direct computations in boundary conformal perturbation theory have proven to be cumbersome, done only at the leading loop order as in the original work of Affleck and Ludwig \cite{Affleck:1991tk, Affleck:1992ng}. 

Using the SFT approach to two-dimensional conformal perturbation theory that was broadly described in the introduction, \cite{Scheinpflug:2023osi} was able to perform next-to-leading order computations of the shift of the $g$-function. A simplification immediately presents itself in that, by the Sen's conjecture, the shift of the $g$-function corresponds to the value of the OSFT action evaluated on the classical solution representing a given RG fixed point. With the choice of a Wilsonian scheme that is specified by working with Witten's cubic vertex, we can compute the on-shell value of the action \eqref{eq:SWitten} as
\begin{align}
    g^\ast-g = -2\pi^2 S[\Psi^*]  = -\frac{\pi^2}{3} \langle \Psi^*, \Psi^*\ast \Psi^* \rangle\,,\label{eq:OnShellS}
\end{align}
where we have made use of the equation of motion $Q \Psi^* + \Psi^*\ast \Psi^* = 0$. 
This fact makes it manifest that the leading contribution is suppressed at least by a third power of the perturbation, which was not manifest in the original treatment of Affleck and Ludwig \cite{Affleck:1991tk, Affleck:1992ng}, nor is manifest when one directly computes the $g$-function by deforming the conformal boundary state. One thus effectively skips a few orders in the conformal perturbation theory, reducing the leading order computation to just a few lines of algebra.


In particular, substituting the classical solution \eqref{eq:PsiStar} into the cubic OSFT vertex appearing in the on-shell action calculation \eqref{eq:OnShellS}, we obtain
\begin{align}
    g^\ast-g &= -\frac{\pi^2}{3}\bigg[(\lambda^\ast(y) )^3 \langle cV,cV\ast cV\rangle -  3(\lambda^\ast(y) )^4\Big\langle cV\ast cV,\frac{b_0}{L_0}\bar{P}_0\, (cV\ast cV)\Big\rangle\bigg]+\mathcal{O}(y^5)\,,
\end{align}
where the two terms appearing on the r.h.s.\ can be expanded as
\begin{align}
     &(\lambda^\ast(y) )^3 \langle cV,cV\ast cV\rangle=\nonumber\\
    &\hspace{1cm}=\bigg(\frac{y}{C_{VVV}} \bigg)^3 gC_{VVV}\bigg[1+\frac{y}{C_{VVV}}\bigg(\frac{\tilde{\mathcal{A}}_{VVVV}}{gC_{VVV}}+3C_{VVV}\log K\bigg)\bigg]\big(1+3y\log K\big)+\mathcal{O}(y^5)\,,\label{eq:term1}
\end{align}
and
\begin{align}
    &3(\lambda^\ast(y) )^4\Big\langle cV\ast cV,\frac{b_0}{L_0}\bar{P}_0\, (cV\ast cV)\Big\rangle=\frac{1}{2}\bigg(\frac{y}{C_{VVV}}\bigg)^4\big(\tilde{\mathcal{A}}_{VVVV}+12gC_{VVV}^2\log K\big)+\mathcal{O}(y^5)\,,\label{eq:term2}
\end{align}
respectively.
Here we have used the result \eqref{eq:AVVVVeval} to evaluate the 4-point Witten OSFT diagram $\mathcal{A}_{VVVV}$. We can quickly note that upon subtracting \eqref{eq:term1} and \eqref{eq:term2}, all dependence on the local coordinates of the SFT disappears and we finally obtain the result
\be
g^\ast-g = -\frac{\pi^2}{3}\Biggr[\bigg(\frac{y}{C_{VVV}}\bigg)^3\tilde{\mathcal{A}}_{VVV} + \frac{1}{2}\bigg(  \frac{y}{C_{VVV}}\bigg)^4\tilde{\mathcal{A}}_{VVVV} + \mathcal{O}(y^5)\Biggr]\,.
\ee


\section{Anomalous dimensions from shifted BRST cohomology}
\label{sec:anomalous_dimensions}

To compute the change in the conformal dimensions of primary operators as the BCFT flows from the UV to the IR fixed point, we implement the procedure originally proposed by Ashoke Sen (see \cite{Sen:2014dqa,Sen:2019jpm} as well as \cite{Maccaferri:2022yzy}). We start by embedding a boundary primary operator $\psi_i$ with a UV conformal dimension $h_i$ into the BRST cohomology of the worldsheet theory by dressing it with a primary field $v_i$ with dimension $1-h_i$ excited in the auxiliary sector $\text{BCFT}_\text{aux}$ so that the resulting (dressed) field $\tilde{\psi}_i=\psi_i v_i$ has dimension $1$. We choose the dressing field $v_i$ to be normalized to have unit 2-point function on the disk.
The open-string state
\begin{align}
    \mathcal{U}_i = c\tilde{\psi}_i
\end{align}
then belongs to the cohomology of the BRST charge $Q$ of the OSFT formulated for the worldsheet theory in the UV. While we will not need to make the exact form of the dressing state $v_i$ explicit at any point in the calculation, it may help to think of it as containing a non-compact free-boson subsector where the dressing may be realized in terms of momentum plane-wave operators. 

Given the on-shell open-string state $\mathcal{U}_i$, we will then look for a state $\mathcal{U}^\ast_i$ which solves the equation
\begin{align}
    Q_{\Psi^\ast(y)}\mathcal{U}^\ast_i \equiv Q\mathcal{U}^\ast_i + [\Psi^\ast(y),\mathcal{U}^\ast_i]=0\label{eq:shifted_cohomology}
\end{align}
which is a condition necessary for $\mathcal{U}^\ast_i$ to belong into the BRST cohomology of the OSFT around the vacuum given by the classical solution $\Psi^\ast(y)$. Assuming that the spectrum of the BCFT of interest is discrete and that $\psi_i$ does not mix with other primaries with dimension $h_i+\mathcal{O}(y)$ under the RG flow (below we will formulate conditions for this not to happen), we can expect that $\mathcal{U}^\ast_i$, as a state in the UV worldsheet theory, takes the form
\begin{align}
    \mathcal{U}^\ast_i = u^\ast_i +Y^\ast_i\,,
\end{align}
where $Y^\ast_i =(1- P_0) \mathcal{U}^\ast_i$ encodes contributions which are not close to $\mathrm{ker}\,L_0$ (perturbatively in $y$). On the other hand, $u^\ast=P_0\mathcal{U}^\ast_i$ can generally be expected to take the form
\begin{align}
    u^\ast_i=c\tilde{\psi}_i^\ast+ {u_i^\prime}^\ast\,,\label{eq:ustar}
\end{align}
where $\tilde{\psi}_i^\ast=\psi_i v_i^\ast$ results from dressing the original state of interest $\psi_i$ of the UV BCFT with a field $v_i^\ast$ from the auxiliary sector, which generally differs from the original UV dressing field $v_i$. For the purposes of this section, we choose to normalize $v_i^\ast$ so that it again has unit 2-point function on the disk. Finally, ${u^\prime}^\ast$ represents potential contributions in the (almost) $\mathrm{ker}\,L_0$ which are due to descendants of states constructed from matter primaries $\psi_j$ with $h_j<h_i$.

As the flow takes place purely within the BCFT of interest, and, since the spectrum of this BCFT is assumed to be discrete, it follows that the cohomology element $\mathcal{U}^\ast_i$ has to correspond to a state $\psi_i^\ast$ of the IR BCFT into which the original state $\psi_i$ flows. As a result, the dimension of the dressing field $v_i^\ast$ has to be $1-h_i^\ast$, where $h_i^\ast$ is the dimension of the state $\psi_i^\ast$ of the IR BCFT. Thus, assuming that we have found $\tilde{\psi}_i^\ast$, we can measure $h_i^\ast$ by computing the overlap
\begin{align}
    -\big\langle c\tilde{\psi}_i^\ast, Q\mathcal{U}^\ast_i\big\rangle = (h_i^\ast-h_i)\,g\,.\label{eq:h_overlap}
\end{align}
In practice, we will proceed by solving the equations
\begin{align}
    Q\mathcal{U}^{(k)}_i + [\Psi^{(k)},\mathcal{U}^{(k-1)}_i]=0\label{eq:cohom_approx}
\end{align}
for successive approximations $\mathcal{U}^{(k)}_i$ to the IR cohomology element $\mathcal{U}^\ast_i$. Similarly to the string fields $\Psi^{(k)}$ (see Section \ref{sec:rg}), the states $\mathcal{U}^{(k)}_i$ are defined to agree with $\mathcal{U}^\ast_i$ up to and including the order $y^k$ (we are intentionally dropping the superscript $\ast$ on the approximations $\mathcal{U}^{(k)}_i$ so as to lighten up the notation somewhat).

First, for $k=0$, the equation \eqref{eq:cohom_approx} reduces to the UV cohomology condition $Q\mathcal{U}^{(0)}_i=0$ which is simply solved by identifying
\begin{subequations}
    \begin{align}
        \tilde{\psi}_i^{(0)} &= \tilde{\psi}_i\,,\\
        {u^\prime}^{(0)}_i&=0\,,\\
        Y^{(0)}_i&=0\,.
    \end{align}
\end{subequations}
The l.h.s.\ of \eqref{eq:h_overlap} then evaluates to 0, signifying that $h_i^\ast$ only differs from $h_i$ by $\mathcal{O}(y)$ terms.

For $k=1$, we can consider acting on the equation \eqref{eq:cohom_approx} by the projector $P_0$ to obtain the condition
\begin{align}
    Qu^{(1)}_i+\frac{y}{C_{VVV}}P_0[cV,c\tilde{\psi}_i]=0\,.\label{eq:cohom_1}
\end{align}
First, in order for our ansatz \eqref{eq:ustar} for the $P_0$ projection of $\mathcal{U}^\ast_i$ to be valid, we note that \eqref{eq:cohom_1} tells us that the $V\psi_i$ collision may not produce any primaries with dimension $h_i$ (up to $\mathcal{O}(y)$ corrections) other than $\psi_i$. We shall henceforth adopt this assumption so that we do not have to deal with the corresponding mixing of operators along the RG flow. Second, given the analysis of boundary OPEs in Appendix \ref{app:opes}, we observe that if the collision $V\psi_i$ produces primaries $\psi_j$ with $h_j<h_i$, $h_j-h_i\in\mathbb{Z}$, which would violate the restriction \eqref{eq:res},
the projected star-commutator $P_0[cV,c\tilde{\psi}_i]$ might be contaminated by the descendants of $\psi_j$. This would, in turn, require switching on a non-trivial state ${u^\prime_i}^{(1)}$ whose precise form would be found by algebraically inverting $Q$ on these descendants. In what follows, we will choose to assume, to simplify our analysis, that all states $\psi_j$ with $h_j<h_i$, $h_j-h_i\in\mathbb{Z}$ produced in the $V\psi_i$ collision satisfy \eqref{eq:res} so that we can set ${u^\prime_i}^{(1)}=0$. Although it will be crucial for the computations below to make sense (without the necessity of adding more terms corresponding to massless descendant propagation), this assumption seems to play no significant role in the final result for anomalous dimensions, leading us to believe that the results  \eqref{eq:Amp4ptB} and \eqref{eq:deltah} should hold more generally.

Given these provisions, we can solve \eqref{eq:cohom_1} by suitably adjusting the weight of the auxiliary-sector field $v_i^{(1)}$ by which we dress $\psi_i$. Computing the overlap on the l.h.s.\ of \eqref{eq:h_overlap} and substituting from \eqref{eq:cohom_approx} for $k=1$, we therefore find
\begin{align}
    (h_i^\ast - h_i)\,g =\frac{y}{C_{VVV}}\big\langle c\tilde{\psi}_i,[cV,c\tilde{\psi}_i] \big\rangle= 2\frac{y}{C_{VVV}}\tilde{\mathcal{A}}_{iiV}+\mathcal{O}(y^2)\,,
\end{align}
where the 3-point amplitude $\tilde{\mathcal{A}}_{iiV}$ was defined in \eqref{eq:Amp3pt} and is simply given by the structure constant $C_{iiV}$. On the other hand, projecting the equation \eqref{eq:cohom_approx} into the image of $1-P_0$ enables one to find 
\begin{align}
    Y^{(1)}_i = -\frac{y}{C_{VVV}}\frac{b_0}{L_0}\bar{P}_0[cV,c\tilde{\psi}_i]\label{eq:Y1}
\end{align}
upon fixing Siegel gauge. Altogether we can therefore write
\begin{align}
    \mathcal{U}^{(1)}_i = c\tilde{\psi}_i^{(1)}-\frac{y}{C_{VVV}}\frac{b_0}{L_0}\bar{P}_0[cV,c\tilde{\psi}_i]+\mathcal{O}(y^2)\label{eq:U1}
\end{align}
for the element of the shifted BRST cohomology which is exact up to the order $y^1$.

In order to compute the anomalous dimensions at next-to-leading order in $y$, let us consider solving the $P_0$ projection of the equation \eqref{eq:cohom_approx} up to the order $k=2$. This entails solving the equation
\begin{align}
   & Qu^{(2)}_i +P_0\big[\lambda^{(2)} (y)cV,c\tilde{\psi}_i^{(1)}\big]-\frac{1}{2}\bigg(\frac{y}{C_{VVV}}\bigg)^2P_0\bigg[\frac{b_0}{L_0}\bar{P}_0 [cV, cV],c\tilde{\psi}_i\bigg]+\nonumber\\
    &\hspace{7cm}-\bigg(\frac{y}{C_{VVV}}\bigg)^2P_0\bigg[cV,\frac{b_0}{L_0}\bar{P}_0 \big[cV,c\tilde{\psi}_i\big]\bigg]=0\,,\label{eq:cohom_2}
\end{align}
where the third term comes from the leading $\bar{P}_0$-component of the solution $\Psi^\ast(y)$, while the fourth term comes from the leading $\bar{P}_0$-part of $\mathcal{U}^\ast_i$ given by \eqref{eq:Y1}. Leaving aside the question whether one can solve \eqref{eq:cohom_2} without having to turn on non-zero ${u^\prime_i}^{(2)}$, we can overlap \eqref{eq:cohom_2} with $c\tilde{\psi}_i^{(2)}$ to find
\begin{align}
    (h_i^\ast - h_i)\,g =\lambda^{(2)}(y)\big\langle c\tilde{\psi}_i^{(1)},[cV,c\tilde{\psi}_i^{(1)}] \big\rangle -\bigg(\frac{y}{C_{VVV}}\bigg)^2 \mathcal{A}_{iViV}+\mathcal{O}(y^3)\,,\label{eq:AD1}
\end{align}
where by $\mathcal{A}_{iViV}$ we have denoted the 4-point Witten OSFT Feynman diagram
\begin{align}
    \mathcal{A}_{iViV}=\frac{1}{2}\Big\langle[c\tilde{\psi}_i,c\tilde{\psi}_i],\frac{b_0}{L_0}\bar{P}_0 [cV, cV]\Big\rangle+\Big\langle[cV,c\tilde{\psi}_i],\frac{b_0}{L_0}\bar{P}_0 [cV, c\tilde{\psi}_i]\Big\rangle\,.\label{eq:4pt_cohom}
\end{align}
Realizing that the state $c\tilde{\psi}_i^{(1)}$ appearing in the first term on the r.h.s.\ of \eqref{eq:AD1} is no longer exactly dimension 0, we can evaluate
\begin{align}
    \lambda^{(2)}(y)\big\langle c\tilde{\psi}_i^{(1)},[cV,c\tilde{\psi}_i^{(1)}] \big\rangle &= 2\frac{y}{C_{VVV}}gC_{iiV}\bigg[1+\frac{1}{3}\frac{y}{C_{VVV}}\bigg(\frac{\tilde{\mathcal{A}}_{VVVV}}{gC_{VVV}}+3C_{VVV}\log K\bigg)\bigg]\times\nonumber\\
    &\hspace{3.4cm}\times\bigg[1+\frac{y}{C_{VVV}}\big(C_{VVV}+4{C_{iiV}}\big)\log K\bigg]+\mathcal{O}(y^3)\,,
\end{align}
while the 4-point diagram \eqref{eq:4pt_cohom} can be evaluated using the results of Appendix \ref{app:feyn} as
\begin{align}
    \mathcal{A}_{iViV} = \tilde{\mathcal{A}}_{iViV}+4g\big(2C_{iiV}+C_{VVV}\big)C_{iiV}\log K\,.
\end{align}
Here the zero-momentum amplitude $\tilde{\mathcal{A}}_{iViV}$ can be expressed as an integrated correlator
\begin{align}
    \tilde{\mathcal{A}}_{iViV}&=\int\limits_{\substack{-\infty \\ \text{a.c.}}}^\infty d\xi\, \bigg[\big\langle\mathsf{T}\big\{\psi_i(\infty)\,V(1)\,\psi_i(0)\,V(\xi)\big\}\big\rangle-\frac{1}{|\xi|}gC_{iiV}C_{iiV}+\nonumber\\[-3mm]
    &\hspace{4.9cm}-\frac{1}{2}\bigg(\frac{1}{|\xi-1|} + \frac{1}{|\xi||\xi-1|}-\frac{1}{|\xi|}\bigg)gC_{VVV}C_{iiV}\bigg]\,,\label{eq:Amp4ptB}
\end{align}
where the (by now) usual prescription for making sense of power-divergences is implemented, while logarithmic divergences are taken care of by means of the explicit subtractions. At this point it is manifest that all dependence on the local coordinates of the Witten vertex cancels between the first and the second term on the r.h.s.\ of \eqref{eq:AD1}. One then finally obtains the result
  \begin{align}
        (h_i^\ast -h_i)\,g= 2\,\bigg(\frac{y}{C_{VVV}}\bigg)\tilde{\mathcal{A}}_{iiV}-\bigg(\frac{y}{C_{VVV}}\bigg)^2\bigg(\tilde{\mathcal{A}}_{iViV}-\frac{2}{3}\frac{C_{iiV}}{C_{VVV}}\tilde{\mathcal{A}}_{VVVV}\bigg)+\mathcal{O}(y^3)\label{eq:deltah}
    \end{align}
    for the anomalous dimensions at next-to-leading order in $y$.

As a consistency check, let us show that the boundary identity field $\mathbf{1}$ does not receive anomalous dimension under short RG flows of the type considered here. Indeed, we note that $C_{\mathbf{1}\mathbf{1}V}=0$, so the only possible contribution to the r.h.s.\ of \eqref{eq:deltah} can come from
\begin{align}
    \tilde{\mathcal{A}}_{\mathbf{1}V\mathbf{1}V}=\int\limits_{\substack{-\infty \\ \text{a.c.}}}^\infty d\xi\, \big\langle\mathsf{T}\big\{\mathbf{1}(\infty)\,V(1)\,\mathbf{1}(0)\,V(\xi)\big\}\big\rangle= \int\limits_{\substack{-\infty \\ \text{a.c.}}}^\infty d\xi\, \frac{1}{(1-\xi)^2}\,.
\end{align}
The function $(1-\xi)^{-2}$ is integrable everywhere except for intervals containing the point $\xi=1$. This means that for any $\alpha$, $\beta$ such that $\alpha < 1 < \beta$, we can evaluate
\begin{align}
    \tilde{\mathcal{A}}_{\mathbf{1}V\mathbf{1}V}=\underbrace{\int\limits_{-\infty }^\alpha d\xi\, \frac{1}{(1-\xi)^2}}_{=\frac{1}{1-\alpha}}+\underbrace{\int\limits_{\substack{\alpha \\ \text{a.c.}}}^\beta d\xi\, \frac{1}{(1-\xi)^2}}_{=\frac{1}{1-\beta}-\frac{1}{1-\alpha}}+\underbrace{\int\limits_{\beta}^\infty d\xi\, \frac{1}{(1-\xi)^2}}_{=-\frac{1}{1-\beta}}=0\,,\label{eq:ac_comp}
\end{align}
which finally gives $h_\mathbf{1}^\ast=h_\mathbf{1}$.



\section{Boundary state from the Ellwood invariant}
\label{sec:ellwood}

To obtain the change in the boundary state coefficient corresponding to a bulk scalar primary $\phi_a(z,\bar{z})$ with dimensions $h_a=\bar{h}_a$ (and scaling dimension $\Delta_a=2h_a$) as the BCFT flows from the UV to the IR, we will implement the idea of \cite{Kudrna:2012re} and compute the corresponding shift in the disk one-point amplitude 
\begin{align}
  \mathcal{B}_a=  \big\langle  c_0^{-}\, c\bar{c}\tilde{\phi}_a(0,0)\big\rangle_\text{disk}\,.
\end{align}
Here $c\bar{c}\tilde{\phi}_a(z,\bar{z})$ is an on-shell closed string state, where $\tilde{\phi}_a(z,\bar{z})$ is a dimension $(1,1)$ bulk primary which is obtained from $\phi_a(z,\bar{z})$ by dressing it with a suitable primary field $w_a(z,\bar{z})$ from the auxiliary sector $\text{BCFT}_\text{aux}$ with dimensions $(1-h_a,1-h_a)$. According to the Ellwood conjecture \cite{Ellwood:2008jh}, this shift in the disk amplitude is measured by a gauge-invariant observable \cite{Hashimoto:2001sm} derived from the Shapiro-Thorn vertex \eqref{eq:ShapiroThorn}, namely
\begin{align}
    \mathcal{B}_a^\ast-\mathcal{B}_a  = -4\pi i\,\mathcal{V}_{1,1}\big(\tilde{\phi}_a,\Psi^\ast(y)\big)\,.\label{eq:ell_conj}
\end{align}
Assuming that we take the one-point function of the dressing field $w_a(z,\bar{z})$ to be (conventionally) normalized to unity, we can write\footnote{Here we make use of the disk correlator
\begin{align}
    \big\langle c_0^- \, c\bar{c}(0)\big\rangle_\text{disk} =-2
\end{align}
in the ghost sector.
}
\begin{align}
    \mathcal{B}_a = -2\,{\langle\phi_a\| B\rangle\!\rangle}\label{eq:ampa}
\end{align}
for the UV boundary state $\| B\rangle\!\rangle$ in the BCFT of interest and similarly for the relation between the amplitude $\mathcal{B}_a^\ast$ and the IR boundary state $\|B^\ast\rangle\!\rangle$.
Altogether, combining \eqref{eq:ell_conj} with \eqref{eq:ampa}, we can therefore compute the shift in the boundary state as the BCFT flows from the UV to the IR as
\begin{align}
    \langle\phi_a\| B^\ast\rangle\!\rangle-\langle\phi_a\| B\rangle\!\rangle=2\pi i\, \big\langle c\bar{c}\tilde{\phi}_a(i,-i) I, \Psi^\ast(y)\big\rangle\,.\label{eq:Bdiff}
\end{align}
It remains to substitute for the solution $\Psi^\ast(y)$ into the r.h.s.\ of \eqref{eq:Bdiff}. Using the earlier result \eqref{eq:PsiStar}, we can write
\begin{align}
    \big\langle c\bar{c}\tilde{\phi}_a(i,-i) I, \Psi^\ast(y)\big\rangle=\lambda^\ast (y)\, \big\langle c\bar{c}\tilde{\phi}_a(i,-i) I,cV\big\rangle - \bigg(\frac{y}{C_{VVV}}\bigg)^2i\mathcal{B}_{aVV}+\mathcal{O}(y^3)\,,\label{eq:ell_eval}
\end{align}
where $\mathcal{B}_{aVV}$ denotes the open-closed diagram
\begin{align}
    i\mathcal{B}_{aVV}=\frac{1}{2}\Big\langle c\bar{c}\tilde{\phi}_a(i,-i) I,\frac{b_0}{L_0}\bar{P}_0  [cV, cV]\Big\rangle
\end{align}
with one bulk and two boundary insertions. Using the results of Appendix \ref{app:feyn}, this can be evaluated as
\begin{align}
   i \mathcal{B}_{aVV} = i\tilde{\mathcal{B}}_{aVV}+igB_{aV}C_{VVV}\big(\log K-2\log 2\big)\,,\label{eq:termE1}
\end{align}
in terms of an integrated bulk-boundary correlator 
\begin{align}
    \tilde{\mathcal{B}}_{aVV}=2^{\Delta_a-2}\int\limits_{\substack{-\infty \\ \text{a.c.}}}^\infty d\xi\,\bigg[\big\langle \mathsf{T}\big\{\phi_a(i,-i)\,V(0)\,V(\xi)\big\}\big\rangle-2^{1-\Delta_a} gC_{VVV}B_{aV}\bigg(\frac{1}{|\xi|}-\frac{1}{|\xi+i|}\bigg)\bigg]\,.
\end{align}
Consistently with \eqref{eq:ampa}, here we have ignored the normalization of the 1-point function of the dressing state $w_a(z,\bar{z})$ on the upper-half plane.
At the same time, the first term on the r.h.s.\ of \eqref{eq:ell_eval} is readily computed as
\begin{align}
    &\lambda^\ast (y)\, \big\langle c\bar{c}\tilde{\phi}_a(i,-i) I,cV\big\rangle = \nonumber\\
    &\hspace{1cm}=igB_{aV}\frac{y}{C_{VVV}}\big(1-2y\log 2\big)\bigg[1+\frac{1}{3}\bigg(\frac{y}{C_{VVV}}\bigg)\bigg(\frac{\tilde{\mathcal{A}}_{VVVV}}{g C_{VVV}}+3C_{VVV}\log K\bigg)\bigg]+\mathcal{O}(y^3)\,.\label{eq:termE2}
\end{align}
Combining \eqref{eq:termE1} with \eqref{eq:termE2}, it is clear that all dependence on local coordinates drops out and one ends up with
\begin{align}
    &\langle\phi_a\| B^\ast\rangle\!\rangle-\langle\phi_a\| B\rangle\!\rangle=\nonumber\\
    &\hspace{1cm}=-2\pi \bigg[\bigg(\frac{y}{C_{VVV}}\bigg)\tilde{\mathcal{B}}_{aV}- \bigg(\frac{y}{C_{VVV}}\bigg)^2\bigg(\tilde{\mathcal{B}}_{aVV}-\frac{1}{3}\frac{B_{aV}}{C_{VVV}}\tilde{\mathcal{A}}_{VVVV}\bigg)+\mathcal{O}(y^3)\bigg]\,.\label{eq:BSdiff}
\end{align}
To perform a sanity check, let us consider setting $\phi_a(z,\bar{z})$ to be the bulk identity field ${\mathbf{1}}(z,\bar{z})$. Then \eqref{eq:BSdiff} should measure the change in the $g$-function which, as we have seen in Section \ref{sec:g_function}, only starts receiving corrections at the order $y^3$. Indeed, we have $B_{\mathbf{1}V}=0$ as well as 
\begin{align}
      \tilde{\mathcal{B}}_{{\mathbf{1}}VV}=\frac{1}{4}\int\limits_{\substack{-\infty \\ \text{a.c.}}}^\infty d\xi\,\big\langle \mathsf{T}\big\{{\mathbf{1}}(i,-i)\,V(0)\,V(\xi)\big\}\big\rangle=\frac{1}{4}\int\limits_{\substack{-\infty \\ \text{a.c.}}}^\infty d\xi\,\frac{1}{\xi^2}=0\,,
\end{align}
where we have recalled the computation \eqref{eq:ac_comp}.

\section{Structure constants from on-shell string amplitudes}
\label{sec:structure}

At this point, the only BCFT data whose change under the boundary RG flow has not yet been discussed, are the boundary OPE coefficients $C_{ijk}$, as well as the bulk-boundary OPE coefficients $B_{ai}$. Recall that we are normalizing the boundary primaries to have unit 2-point function on the disk so that there is no relative factor between the OPE coefficients and the corresponding chiral 3-point correlators (except for the usual $g$-factor arising from the 1-point function of the boundary identity).

To extract the change of $C_{ijk}$ and $B_{ai}$ from the classical solution $\Psi^\ast(y)$, we will compute \cite{Mazel:2024alu} the change in the 3-point boundary amplitude and 2-point bulk-boundary amplitude involving on-shell open and closed string states constructed from the boundary primaries $\psi_i$, $\psi_j$, $\psi_k$ and the bulk primary $\phi_a$. While the closed-string state does not undergo any change as the background is shifted by $\Psi^\ast(y)$, the on-shell open-string states which are to be put on the external legs have to be taken from the shifted BRST cohomology, as was discussed in detail in Section \ref{sec:anomalous_dimensions}. Since we will only aim for leading order corrections to $C_{ijk}$ and $B_{ai}$ (as those will already involve integrated correlators), it will be sufficient to consider the $\mathcal{O}(y)$ result \eqref{eq:U1} for the shifted open-string cohomology states.
The 3-point open-string amplitude around the background shifted by $\Psi^\ast(y)$ is then computed by simply plugging these IR cohomology states into the Witten 3-vertex $ \mathcal{V}_{0,3}(\Psi_1,\Psi_2,\Psi_3)$
of the OSFT around the UV background: this is a consequence of the fact that the Witten OSFT does not feature any higher elementary vertices so that when expanding around the classical solution $\Psi^\ast(y)$, no additional contributions to the 3-vertex are acquired (unlike in the case of the kinetic term). Similarly, the open-closed amplitude around the IR background involving one on-shell closed string and one on-shell open-string is obtained by substituting an open-string state from the shifted cohomology into the UV SFT Shapiro-Thorn vertex $\mathcal{V}_{1,1}(\Phi;\Psi)$, where $\Phi$ is forced to be a weight $(0,0)$ primary closed-string state. Again, since in the Witten OSFT there is no need for higher elementary open-closed vertices to cover moduli spaces of disk amplitudes with open and closed punctures, no additional contributions to the open-closed vertex are produced upon expanding the action in small fluctuations around $\Psi^\ast(y)$.

\subsection{Normalizing the IR cohomology}

A somewhat subtle point concerns the normalization of the shifted cohomology elements: even if we ensure that the UV cohomology elements can be normalized\footnote{By taking the auxiliary BCFT which we use to dress the states $\psi_i$ to contain e.g.\ multiple non-compact free bosons, \eqref{eq:UVnorm} can always be kinematically achieved by dressing the states $\psi_i$ with states $v_i$ which can be chosen as suitable momentum plane-waves. We will conventionally take the corresponding boundary OPE structure constant in the auxiliary sector to be unity.} so that
\begin{align}
    \mathcal{V}_{0,3}\big( c\tilde{\psi}_i,c\tilde{\psi}_j, c\tilde{\psi}_k\big) = g C_{ijk}\,,\label{eq:UVnorm}
\end{align}
there is no reason to expect that the particular normalization chosen for the shifted cohomology elements \eqref{eq:U1} in Section \ref{sec:anomalous_dimensions} should directly give $g^\ast C_{ijk}^\ast$ upon computing the corresponding on-shell 3-point scattering amplitude around $\Psi^\ast(y)$. We will therefore start by assuming a more general normalization
\begin{align}
     \mathcal{U}^{(1)}_i =\chi_i^{(1)}(y)\bigg( c\tilde{\psi}_i^{(1)}-\frac{y}{C_{VVV}}\frac{b_0}{L_0}\bar{P}_0[cV,c\tilde{\psi}_i]+\mathcal{O}(y^2)\bigg)
\end{align}
for the shifted cohomology states
and we will look for $\chi^{(1)}(y)=1+\eta_i y+\mathcal{O}(y^2)$ so that scattering two $\mathcal{U}^\ast_i$ states with the on-shell state $\mathcal{U}^\ast_\mathbf{1}$ constructed from the boundary identity $\mathbf{1}$ in the UV BCFT of interest results in the correctly normalized OPE structure constant $C_{ii\mathbf{1}}^\ast =1$. In other words, to fix $\eta_i$, we will demand
\begin{align}
   \mathcal{V}_{0,3}\big(  \mathcal{U}^{(1)}_\mathbf{1},\mathcal{U}^{(1)}_i, \mathcal{U}^{(1)}_i \big)=g+\mathcal{O}(y^2)\,.\label{eq:norm_crit}
\end{align}
Here we have noted that since $g^\ast$ differs from $g$ only at third order in $y$, we can simply replace $g^\ast$ by $g$ in the $\mathcal{O}(y)$ calculations of this section.

Starting with the case $i=\mathbf{1}$ and assuming the normalization \eqref{eq:UVnorm} for the states in the UV cohomology, we can expand
\begin{align}
    \mathcal{V}_{0,3}\big(\mathcal{U}^{(1)}_\mathbf{1},\mathcal{U}^{(1)}_\mathbf{1}, \mathcal{U}^{(1)}_\mathbf{1} \big) &= g+3yg\eta_\mathbf{1}-\frac{3}{2}\frac{y}{C_{VVV}}\Big\langle[c\tilde{\psi}_\mathbf{1},c\tilde{\psi}_\mathbf{1}] ,\frac{b_0}{L_0}\bar{P}_0[cV,c\tilde{\psi}_\mathbf{1}]\Big\rangle +\mathcal{O}(y^2)\,,\label{eq:111}
\end{align}
where we have noticed that since $C_{\mathbf{1}\mathbf{1}V}=0$ in any theory, the corrected state $c\tilde{\psi}_\mathbf{1}^{(1)}$ remains dimension 0 even at $\mathcal{O}(y)$. Furthermore,
we can readily see that the propagator term on the r.h.s.\ of \eqref{eq:111} vanishes as the corresponding 4-point correlator in the UV BCFT of interest is clearly zero. We therefore obtain
\begin{align}
    \eta_\mathbf{1} =0\,,
\end{align}
namely that the state $\mathcal{U}^{(1)}_\mathbf{1}$ does not get renormalized with respect to the conventions of Section \ref{sec:anomalous_dimensions}.

More generally, we can compute the 3-point scattering amplitude
\begin{align}
    \mathcal{V}_{0,3}\big(\mathcal{U}^{(1)}_\mathbf{1},\mathcal{U}^{(1)}_i, \mathcal{U}^{(1)}_i \big) &= g+2yg\eta_i+4\frac{y}{C_{VVV}}gC_{iiV}\log K-\frac{y}{C_{VVV}}\mathcal{A}_{ii\mathbf{1}V}\,,\label{eq:ii1}
\end{align}
where $\mathcal{A}_{ii\mathbf{1}V}$ denotes the Feynman diagram
\begin{align}
\mathcal{A}_{ii\mathbf{1}V}=\frac{1}{2}\Big\langle[c\tilde{\psi}_i,c\tilde{\psi}_i],\frac{b_0}{L_0}\bar{P}_0[cV,c\tilde{\psi}_\mathbf{1}] \Big\rangle+\Big\langle[cV,c\tilde{\psi}_i],\frac{b_0}{L_0}\bar{P}_0[c\tilde{\psi}_\mathbf{1},c\tilde{\psi}_i] \Big\rangle\,.
\end{align}
Applying the results of Appendix \ref{app:feyn} and recalling that $C_{\mathbf{1}\mathbf{1}V}=0$, $C_{ii\mathbf{1}}=1$ we can evaluate
\begin{align}
    \mathcal{A}_{ii\mathbf{1}V}&=8gC_{iiV}\log K+  \!\int_{-\infty}^\infty \! d\xi\, \bigg[\big\langle\mathsf{T}\big\{\psi_i(\infty)\,\psi_i(1)\,\mathbf{1}(0)\,V(\xi)\big\}\big\rangle- \frac{1}{|\xi-1|} gC_{iiV}\bigg]=8gC_{iiV}\log K\,,
\end{align}
where the last equality follows from observing that the 4-point correlation function under the integral contains one insertion of the boundary identity and can therefore be evaluated as the 3-point function
\begin{align}
    \big\langle\mathsf{T}\big\{\psi_i(\infty)\,\psi_i(1)\,V(\xi)\big\}\big\rangle= \frac{1}{|\xi-1|} gC_{iiV}\,.
\end{align}
Altogether, substituting back into the calculation \eqref{eq:ii1} and enforcing the normalization condition \eqref{eq:norm_crit}, we learn that 
\begin{align}
    \eta_i = 2\frac{C_{iiV}}{C_{VVV}}\log K\,.
\end{align}
The correctly normalized elements of the shifted open-string cohomology (to be compatible with demanding $C_{ii\mathbf{1}}^\ast$) therefore read
\begin{align}
     \mathcal{U}^{(1)}_i = \bigg(1+2\frac{y}{C_{VVV}}C_{iiV}\log K\bigg)c\tilde{\psi}_i^{(1)}-\frac{y}{C_{VVV}}\frac{b_0}{L_0}\bar{P}_0[cV,c\tilde{\psi}_i]+\mathcal{O}(y^2)\,.
\end{align}

\subsection{Boundary 3-point coupling}

Having suitably normalized the elements of the shifted open-string cohomology, we can now simply compute the 3-point scattering amplitude
\begin{align}
    \mathcal{V}_{0,3}\big( \mathcal{U}^\ast_i,\mathcal{U}^\ast_j, \mathcal{U}^\ast_k \big) = g^\ast C_{ijk}^\ast
\end{align}
around the shifted background up to $\mathcal{O}(y)$. In particular, taking into account the $\mathcal{O}(y)$ anomalous dimensions of the fields $\psi_i$, $\psi_j$ and $\psi_k$ when evaluating the Witten 3-vertex, we can expand
\begin{align}
    \mathcal{V}_{0,3}\big(\mathcal{U}^{(1)}_i,\mathcal{U}^{(1)}_j, \mathcal{U}^{(1)}_k \big) &= gC_{ijk}\bigg(\!1+4\frac{y}{C_{VVV}}(C_{iiV}+C_{jjV}+C_{kkV})\log K\!\bigg)+\nonumber\\
    &\hspace{7cm}-\frac{y}{C_{VVV}}\mathcal{A}_{ijkV}+\mathcal{O}(y^2)\,,
\end{align}
where the last term on the r.h.s.\ is expressed in terms of the 4-point diagram
\begin{align}
    \mathcal{A}_{ijkV}&=\Big\langle c\tilde{\psi}_i\ast c\tilde{\psi}_{j},\frac{b_0}{L_0}\bar{P}_0[cV,c\tilde{\psi}_k]\Big\rangle+\Big\langle c\tilde{\psi}_j\ast c\tilde{\psi}_{k},\frac{b_0}{L_0}\bar{P}_0[cV,c\tilde{\psi}_i]\Big\rangle+\nonumber\\
    &\hspace{7cm}+\Big\langle c\tilde{\psi}_k\ast c\tilde{\psi}_{i},\frac{b_0}{L_0}\bar{P}_0[cV,c\tilde{\psi}_j]\Big\rangle\,.
\end{align}
Finally, using the result \eqref{eq:AijkVApp} and recalling that $g^\ast$ differs from $g$ only by $\mathcal{O}(y^3)$ terms, we arrive at the formula
\begin{align}
    \big(C_{ijk}^\ast - C_{ijk}\big)\,g = -\frac{y}{C_{VVV}}\tilde{\mathcal{A}}_{ijkV}+\mathcal{O}(y^2)\,,
\end{align}
where the amplitude $\tilde{\mathcal{A}}_{ijkV}$ was defined in \eqref{eq:Amp4pt}.

\subsection{Bulk-boundary coupling}

The open-closed amplitude which involves one on-shell open string state $\mathcal{U}_i=c\tilde{\psi}_i$ and one on-shell closed string $c\bar{c}\tilde{\phi}_a$ around the perturbative vacuum $\Psi=0$ is directly given by the Shapiro-Thorn vertex \eqref{eq:ShapiroThorn} and reads
\begin{align} \mathcal{V}_{1,1}\big(c\bar{c}\tilde{\phi}_a;\mathcal{U}_i\big)=\big\langle c\bar{c}\tilde{\phi}_a(i,-i)I,\mathcal{U}_i\big\rangle=igB_{ai}  \,,\label{eq:V11UV}
\end{align}
where $B_{ai}$ is the coefficient in front of the boundary primary $\psi_i$ in the bulk-boundary OPE of bulk primary $\phi_a$. Here we again assume that the fields $\phi_a$ and $\psi_i$ can be dressed with suitably normalized auxiliary-sector fields so that \eqref{eq:V11UV} holds. 
In particular, note that here the dressing $w_a$ of the bulk field $\tilde{\phi}_a$ is different from the one used in Section \ref{sec:ellwood} as we need to ensure that the bulk-boundary correlator $\langle w_a(i,-i)\,v_i(0)\rangle$ in the auxiliary BCFT is non-zero. We also assume that the corresponding auxiliary-field bulk-boundary structure constant is made equal to unity so that \eqref{eq:V11UV} holds.

To find the bulk-boundary coupling $B_{ai}^\ast$ in the IR BCFT, we compute the corresponding open-closed amplitude in the theory around $\Psi^\ast(y)$, namely
\begin{align}
    \mathcal{V}_{1,1}\big( c\bar{c}\tilde{\phi}_a;\mathcal{U}_i^\ast\big)=ig^\ast B_{ai}^\ast \,.
\end{align}
We can see that the change of the background only enters the amplitude through the change in the state $\mathcal{U}_i^\ast$ which is taken to be the element of the IR (shifted) cohomology.
Up to the first order in $y$, we can expand
\begin{align}
    \mathcal{V}_{1,1}\big(  c\bar{c}\tilde{\phi}_a;\mathcal{U}_i^{(1)}\big)=\bigg[1+2\frac{y}{C_{VVV}}C_{iiV}\big(\log K-2\log 2\big)\bigg]igB_{ai}-2\frac{y}{C_{VVV}}i\mathcal{B}_{aiV}+\mathcal{O}(y^2)\,,
\end{align}
where $i\mathcal{B}_{aiV}$ denotes the open-closed diagram
\begin{align}
    i\mathcal{B}_{aiV} = \frac{1}{2}\Big\langle c\bar{c}\tilde{\phi}_a(i,-i)I ,\frac{b_0}{L_0}\bar{P}_0[cV,c\tilde{\psi}_i]\Big\rangle\,.
\end{align}
Recalling the result \eqref{eq:AaiVApp} expressing $i\mathcal{B}_{aiV}$ in terms of the integrated bulk-boundary correlator \eqref{eq:Amp3ptBB}, we observe that all dependence on off-shell SFT data again disappears, leaving us with the formula
\begin{align}
    (B_{ai}^\ast-B_{ai})\,g = -2\frac{y}{C_{VVV}}\tilde{\mathcal{B}}_{aiV}+\mathcal{O}(y^2)\label{eq:BBdiff}
\end{align}
for the change in the bulk-boundary structure constant where $\tilde{\mathcal{B}}_{aiV}$ can be computed using \eqref{eq:Amp3ptBB}.

As a consistency check, let us compute the r.h.s.\ of \eqref{eq:BBdiff} setting $\psi_i$ to be the boundary identity, that is $i=\mathbf{1}$. That is, we compute the change in the open-closed scattering amplitude which involves an on-shell open-string which corresponds to the boundary identity in the BCFT of interest. The result of such a calculation should agree with the leading-order part of the result \eqref{eq:BSdiff} for the change in the boundary state coefficient $\langle \phi_a \| B\rangle\!\rangle$ along the RG flow which was derived from the Ellwood conjecture. In particular, in the strict limit $y\to 0$, we have
\begin{align}
    \big\langle\mathsf{T}\big\{\phi_a(i,-i)\mathbf{1}(0)V(\xi)\big\}\big\rangle = 2^{-\Delta_a+1}\frac{ gB_{aV}}{1+\xi^2}\,,
\end{align}
so that no logarithmic or power divergences plague the integral over $\xi$ appearing in $\tilde{\mathcal{B}}_{a\mathbf{1}V}$. (Indeed, the logarithmic subtractions \eqref{eq:Sigmaai} vanish as we have $C_{\mathbf{1}\mathbf{1}V}=0$.) We can therefore compute the integrated bulk-boundary correlator as
\begin{align}
    \tilde{\mathcal{B}}_{a\mathbf{1}V} = g B_{aV}\int_{-\infty}^\infty d\xi\,\frac{1}{1+\xi^2} = \pi gB_{aV}\,,
\end{align}
which, upon substituting into \eqref{eq:BBdiff}, gives
\begin{align}
    (B_{ai}^\ast-B_{ai})\,g = -2\pi \frac{y}{C_{VVV}}gB_{aV}\,.
\end{align}
Realizing that
\begin{subequations}
    \begin{align}
        \langle\phi_a\| B\rangle\!\rangle&=gB_{a\mathbf{1}}\,,\\
        \langle\phi_a\| B^\ast\rangle\!\rangle&=g^\ast B_{a\mathbf{1}}^\ast\,,
    \end{align}
\end{subequations}
where no difference between $g$ and $g^\ast$ is felt at $\mathcal{O}(y)$, we indeed obtain agreement with the leading-order part of \eqref{eq:BSdiff}. We note in passing that this constitutes a non-trivial perturbative check of the Ellwood conjecture.

\section{Example: $(1,3)$-flows in diagonal minimal models with boundary}
\label{sec:mm}

In order to test the formulae for IR BCFT data derived from SFT in previous sections, we consider a specific deformation of Cardy boundary states in unitary minimal models $\mathcal{M}_m$ with central charge close to 1 i.e. in the large $m$ limit (see appendix \ref{app:virasoroMM} for a recap of this setup). The deformation of interest is triggered by the nearly marginal relevant operator $V \equiv \psi_{(1,3)}$, which is present on every non-identity Cardy boundary, and is distinguished by the property that the only nearly marginal operator appearing in the OPE of $V$ with itself is again $V$. It therefore respects the property \eqref{eq:VVOPE} which we required so that the deformation does not excite other nearly marginal operators at higher order.
Note that this deformation was already studied in the context of BCFT in \cite{Lesage:1998qf,Recknagel:2000ri,Graham:2000si}.

\subsection{Perturbed $g$-function}

We now compute the next-to-leading order shift $\Delta g_{(\alpha_1, \alpha_2)} \equiv g_{(\alpha_1, \alpha_2)}^* - g_{(\alpha_1, \alpha_2)}$ for the RG flow triggered by $V$ (we now specialize $g \to g_{(\alpha_1, \alpha_2)}$, which is given by (\ref{gFunMM})), living on a generic Cardy boundary with Kac label $(\alpha_1, \alpha_2)$. The RG eigenvalue for the above perturbation is $y = 1 - h_{(1,3)} = \frac{2}{m+1}$ and the relevant Runkel's structure constant of appendix \ref{app: boundaryOPE} takes the form
\be
\label{CVVV}
C_{(1,3)(1,3)(1,3)} = \sqrt{\frac{8}{\alpha_2^2 - 1}} \bigg(1 - \frac{3}{2} \frac{1}{m+1}\bigg) + \mathcal{O}\bigg(\frac{1}{m^2}\bigg)\,.
\ee
Using the formula (\ref{gFinal}) and specializing the integrated four-point correlator (\ref{integrated4ptCorrelator}) to that of four $V$ insertions, giving
\begin{align}
    \tilde{\mathcal{A}}_{VVVV} = -\frac{3}{2} g C_{VVV}^2\,,\label{eq:AVVVV13}
\end{align}
it is then simple to obtain the result
\be
\frac{\Delta g_{(\alpha_1, \alpha_2)}}{g_{(\alpha_1, \alpha_2)}}\equiv \frac{g_{(\alpha_1, \alpha_2)}^\ast-g_{(\alpha_1, \alpha_2)}}{g_{(\alpha_1, \alpha_2)}} = -\frac{\pi^2}{3}(\alpha_2^2-1) \Biggr[\frac{1}{(m+1)^3} + \frac{3}{2} \frac{1}{(m+1)^4}\Biggr] +  \mathcal{O}\bigg(\frac{1}{m^5}\bigg).
\label{eq:mmGResult}
\ee
Following \cite{Recknagel:2000ri}, we now use this result to derive constraints on a generic ansatz 
\begin{align}
    (\alpha_1, \alpha_2) \longrightarrow \bigoplus_l\, \big(\beta_1^l, \beta_2^l\big)\label{eq:assumedRG}
\end{align}
for the RG flow triggered by $V$. As we can write
\be 
\log \frac{\sum_l g_{(\beta^l_1, \beta^l_2)}}{g_{(\alpha_1,\alpha_2)}} =\log\bigg(1+\frac{\sum_l g_{(\beta^l_1, \beta^l_2)}-g_{(\alpha_1,\alpha_2)}}{g_{(\alpha_1,\alpha_2)}}\bigg) = \frac{\sum_l g_{(\beta^l_1, \beta^l_2)}-g_{(\alpha_1,\alpha_2)}}{g_{(\alpha_1,\alpha_2)}} + \mathcal{O}\big((\Delta g_{(\alpha_1, \alpha_2)})^2\big)\,,
\ee 
then, to next-to-leading order, we have to equate the $\Delta g_{(\alpha_1, \alpha_2)}/g_{(\alpha_1, \alpha_2)}$ observable we computed in \eqref{eq:mmGResult} with the log of the ratio of the final and initial $g$-functions for the assumed RG flow \eqref{eq:assumedRG}. Using the expressions for $g$-functions of Cardy boundaries (\ref{gFunMM}), the latter gives us the expression
\begin{align}
    & \log \frac{\sum_l g_{(\beta^l_1, \beta^l_2)}}{g_{(\alpha_1,\alpha_2)}} =\nonumber\\
    &\hspace{0.2cm}=\log \sigma + \frac{\pi^2}{6}\bigg(\alpha_1^2+\alpha_2^2-\sum\limits_l \frac{s_l}{\sigma} \big[(\beta_1^l)^2+(\beta_2^l)^2\big]\bigg)\frac{1}{(m+1)^2} + \frac{\pi^2}{3} \big(\alpha_1^2 - \sum\limits_l \frac{s_l}{\sigma}(\beta_1^l)^2\big)\frac{1}{(m+1)^3} + \nonumber\\
    &\hspace{1cm}+\biggr[\frac{\pi^2}{2}\big(\alpha_1^2 - \sum\limits_l\frac{s_l}{\sigma} (\beta_1^l)^2\big) +  \frac{\pi^4}{180} \biggr(\alpha_1^4+\alpha_2^4 - \sum\limits_l\frac{(\beta_1^l)^2(\beta_2^l)^2((\beta_1^l)^4+(\beta_2^l)^4)}{(\sigma \alpha_1\alpha_2)^2} +\nonumber\\
    &\hspace{2cm}+\frac{1}{2 (\sigma \alpha_1\alpha_2)^2} \sum\limits_{\substack{k,l \\ k\neq l}} \beta_1^l \beta_2^l \beta_1^k \beta_2^k \big(3 \big[(\beta_1^l)^4 +  (\beta_2^l)^4\big] + 10 (\beta_1^l)^2(\beta_2^l)^2+\nonumber\\[-4mm]
    &\hspace{5cm}- 5 \big[(\beta_2^l \beta_2^k)^2 + (\beta_1^l \beta_1^k)^2 + 2(\beta_1^l \beta_2^k)^2\big]\big)\biggr)\biggr]\frac{1}{(m+1)^4} + \mathcal{O}\bigg(\frac{1}{m^5}\bigg)
\end{align}
with $\sigma = \sum_l s_l$, where $s_l \equiv \frac{\beta_1^l \beta_2^l}{\alpha_1 \alpha_2}$. Equating this expression with the explicit computation (\ref{eq:mmGResult}), gives the following constraints
\begin{subequations}
    \begin{align}
           1 &= \sigma\,,\\[3.4mm]
            \alpha_1^2 + \alpha_2^2&= \sum\limits_l s_l \big[(\beta_1^l)^2+(\beta_2^l)^2\big]\,, \\ 
           1 &= \sum\limits_l s_l (\beta_2^l)^2 \,,\\[-1mm]
\alpha_1^2\alpha_2^2(\alpha_1^4+\alpha_2^4)&=\sum\limits_l(\beta_1^l)^2(\beta_2^l)^2((\beta_1^l)^4+(\beta_2^l)^4)  -\frac{1}{2} \sum\limits_{\substack{k,l \\ k\neq l}} \beta_1^l \beta_2^l \beta_1^k \beta_2^k \Big(3 \big((\beta_1^l)^4 +  (\beta_2^l)^4\big) +\nonumber\\
     &\hspace{4cm}+ 10 (\beta_1^l)^2(\beta_2^l)^2- 5 \big((\beta_2^l \beta_2^k)^2 + (\beta_1^l \beta_1^k)^2 + 2(\beta_1^l \beta_2^k)^2\big)\Big)\,.
    \end{align}
\end{subequations}
The first three constraints were found by \cite{Recknagel:2000ri}, but the fourth one is new and obviously not linearly dependent on the rest. In \cite{Recknagel:2000ri}, the generic solution
\be
\beta_1^l = \alpha_1 + \alpha_2 + 1 - 2l, \, \, \beta_2^l = 1
\label{kacSolution}
\ee
with $l = 1, \ldots, \min(\alpha_1, \alpha_2)$ was proposed. We have verified that this solution satisfies even the fourth constraint.
 It was further discussed that constraints from higher loops can rule out some non-generic solutions to these constraints.
An example of a non-generic flow that our new constraint rules out is the flow $(4,9) \to (11,1) \oplus 2 (10,1) \oplus (5,1)$, which satisfies the first three constraints but not the fourth one.

\subsection{Anomalous dimensions}
\label{anomalousDimComputation}
In this section, using the formula \eqref{anomalDim}, we compute the boundary spectrum of the IR endpoint of a few selected RG flows triggered by $V$. To facilitate straightforward application of \eqref{anomalDim}, the flows are selected so that no operator mixing occurs. To perform this computation, in addition to three-point boundary contact interactions (\ref{eq:Amp3pt}), we need the result \eqref{integrated4ptCorrelator} for the integrated four-point correlators involving two insertions of $V$ and two insertions of the UV BCFT operator whose change in the anomalous dimension we wish to compute. The input to both of these are the boundary OPE structure constants \eqref{eq:Cboundary}, which we evaluate on a case-by-case basis.

\subsubsection*{The flow $(1,3) \to (3,1)$}
We compute the spectrum of the deformed $(1,3)$ boundary, which according to the proposed RG flow \eqref{kacSolution} flows to the $(3,1)$ boundary. 
On the $(1,3)$ boundary, there are three Virasoro primaries, namely 
\begin{align}
    \mathbf{1}\,,\qquad  \psi_{(1,3)}\quad \text{and}\quad \psi_{(1,5)}\,,
\end{align}
which do not mix in perturbation theory since their weights ($0+\mathcal{O}(\tfrac{1}{m})$, $1+\mathcal{O}(\tfrac{1}{m})$ and $4+\mathcal{O}(\tfrac{1}{m})$, respectively) differ by $\mathcal{O}(1)$.  
On the $(3,1)$ boundary, which is supposed to represent the IR fixed point, there are again three different Virasoro primaries 
\begin{align}
    \mathbf{1}\,,\qquad \psi_{(3,1)}\quad \text{and}\quad \psi_{(5,1)}\,,
\end{align}
so that we expect that along the RG trajectory triggered by the $\psi_{(1,3)}$ deformation in the UV, the primaries flow as 
\begin{subequations}
    \begin{align}
       \mathbf{1} &\to \mathbf{1} \,,\\
       \psi_{(1,3)} &\to \psi_{(3,1)}\,,\\
       \psi_{(1,5)} &\to \psi_{(5,1)}\,.
    \end{align}
\end{subequations}
To confirm these operator flows, we will now apply the formula \eqref{anomalDim} to compute the anomalous dimensions and compare with the Kac table for the UV and IR BCFTs. In particular, as we have shown around \eqref{eq:ac_comp}, the boundary identity does not acquire an anomalous dimension, so it is sufficient to study the anomalous dimensions of $\psi_{(1,3)}$ and $\psi_{(1,5)}$. From now on, we will generally use $y=\tfrac{2}{m+1}$ as an expansion parameter.

First, plugging in $\tilde{\mathcal{A}}_{VVV} = g_{(1,3)} C_{VVV}$, as well as the result \eqref{eq:AVVVV13} for $\tilde{\mathcal{A}}_{VVVV}$ into \eqref{anomalDim}, we get that $\psi_{(1,3)}$ has the anomalous dimension
\be
h^*_{(1,3)} - h_{(1,3)} = 2 y + \frac{1}{2} y^2 + \mathcal{O}(y^3)\,,
\label{anomal13}
\ee
which, up to this order in $y$, indeed agrees with the difference $h_{(3,1)} - h_{(1,3)}$ obtained from the Kac table, thus confirming the oprator flow $\psi_{(1,3)} \to \psi_{(3,1)}$. It is not difficult to see that this is the case for any flow $(1, \beta) \to (\beta,1)$ of Cardy boundaries, as the factor of $C_{VVV}$ cancels out.

Similarly, we can express $\tilde{\mathcal{A}}_{(1,5)(1,5) V} = g_{(1,3)} C_{(1,5)(1,5)V}$ and use \eqref{integrated4ptCorrelator} to find
\be
\tilde{\mathcal{A}}_{(1,5) V (1,5) V} = -\frac{1}{4} g_{(1,3)} C_{(1,5)(1,5)V}^2 - \frac{23}{12} g_{(1,3)} C_{VVV} C_{(1,5)(1,5) V}\,,
\label{15V15V}
\ee
where on the $(1,3)$ boundary, the boundary OPE structure constant $C_{(1,5)(1,5) V}$ evaluates to
\be
C_{(1,5)(1,5) V} = 3 - 4y + \mathcal{O}(y^2)\,.
\ee
Plugging this into (\ref{anomalDim}) gives
\be
h_{(1,5)}^* - h_{(1,5)} = 6y + \frac{3}{2} y^2 + \mathcal{O}(y^3)\,,
\ee
which agrees with $h_{(5,1)} - h_{(1,5)}$ computed from the Kac table.
This confirms the flow $\psi_{(1,5)} \to \psi_{(5,1)}$.

\subsubsection*{The flow $(1,5) \to (5,1)$}
We now proceed to compute the spectrum of the deformed $(1,5)$ boundary which, according to the solution \eqref{kacSolution}, flows to the $(5,1)$ boundary.
On the $(1,5)$ boundary, there are five Virasoro primaries 
\begin{align}
    \mathbf{1}\,,\qquad \psi_{(1,3)}\,,\qquad \psi_{(1,5)}\,, \qquad \psi_{(1,7)}\quad \text{and}\quad \psi_{(1,9)}\,,
\end{align}
which do not mix in perturbation theory since their weights are close to $0, 1, 4, 9$ and $16$. We can thus use \eqref{anomalDim} and proceed as on the $(1,3)$ boundary. On the $(5,1)$ boundary, there are five Virasoro primaries with mirrored Kac labels, and we expect operator flows $\psi_{(r,s)} \to \psi_{(s,r)}$, as on the $(1,3)$ boundary. We only need to check three of them, as for the identity and $\psi_{(1,3)}$, the flows are immediate (for the $\psi_{(1,3)}$ case, see the comment below \eqref{anomal13}).

Using the general form \eqref{15V15V} and that on the $(1,5)$ boundary, we have
\be
C_{(1,5)(1,5)V} = \sqrt{3} - \frac{4}{\sqrt{3}} y + \mathcal{O}(y^2)\,,
\ee
we obtain 
\be
h_{(1,5)}^* - h_{(1,5)} = 6y + \frac{3}{2} y^2 + \mathcal{O}(y^3)\,,
\ee
which, up to second order in $y$, agrees with $h_{(5,1)} - h_{(1,5)}$ from the Kac table. This confirms the flow $\psi_{(1,5)} \to \psi_{(5,1)}$.

To compute the anomalous dimensions of the remaining two primaries $\psi_{(1,7)}$ and $\psi_{(1,9)}$, we compute the integrated correlators
\begin{subequations}
\begin{align}
    \tilde{\mathcal{A}}_{(1,7)V(1,7)V} &= -\frac{1}{9} g_{(1,5)} C_{(1,7)(1,7)V}^2 - \frac{41}{15} g_{(1,5)} C_{VVV} C_{(1,7)(1,7)V}\,, \\
    \tilde{\mathcal{A}}_{(1,9)V(1,9)V} &= -\frac{1}{16} g_{(1,5)} C_{(1,9)(1,9)V}^2 - \frac{927}{280} g_{(1,5)} C_{VVV} C_{(1,9)(1,9)V}\,,
\end{align}
\end{subequations}
and the structure constants
\begin{subequations}
    \begin{align}
     C_{(1,7)(1,7)V} &= 2\sqrt{3} - \frac{17\sqrt{3}}{5} y + \mathcal{O}(y^2)\,, \\
     C_{(1,9)(1,9)V} &= \frac{10}{\sqrt{3}} - \frac{551}{28 \sqrt{3}} y + \mathcal{O}(y^2)\,. 
    \end{align}
\end{subequations}
Upon substituting into \eqref{anomalDim}, these give
\begin{subequations}
    \begin{align}
     h^*_{(1,7)} - h_{(1,7)} &= 12 y + 3 y^2 + \mathcal{O}(y^3)\,, \\
     h^*_{(1,9)} - h_{(1,9)} &= 20y + 5y^2 + \mathcal{O}(y^3)\,,
    \end{align}
\end{subequations}
which, to this order in $y$, agrees with $h_{(7,1)} - h_{(1,7)}$ and $h_{(9,1)} - h_{(1,9)}$, as expected.

\subsection{Boundary-state coefficients}
\label{bstateCoeffComputation}
In this section, we compute shifts in a few boundary state coefficients 
of a couple select RG flows triggered by $V$ using \eqref{bStateShift}. In contrast to the case of anomalous dimensions, here there is no operator mixing problem to attend to (the bulk scalars do not flow under the boundary RG flow), so that we can also treat the case in which a single Cardy boundary flows to a superposition of multiple Cardy boundaries. To compute the shift in the boundary state, in addition to the bulk-boundary correlator \eqref{eq:Amp2ptBB}, we need to compute the integrated bulk-boundary-boundary correlators \eqref{eq:BBIntegrated}. The input to both are the bulk-boundary structure constants \eqref{eq:Bbulk_boundary}.

First, using (\ref{eq:bs_mm}), we can expand general boundary state coefficients in Virasoro Minimal Models in powers of $y$ as
\begin{align}
\langle \phi_{(r,s)} \| B_{(\alpha, \beta)}\rangle\!\rangle &=g_{(\alpha, \beta)} (-1)^{(r+s)(\alpha+\beta)} \sqrt{r s} \bigg(1-\frac{1}{48}\pi^2 \Big[r^2(2\alpha^2-1)+s^2(2\beta^2-1)+\nonumber\\
&\hspace{7cm}-2(\alpha^2+\beta^2-1)\Big]y^2 + \mathcal{O}(y^3) \bigg)\,.
\label{boundaryExpansion}
\end{align}
We will use this expansion to make comparison with our perturbative results. For the generic solution \eqref{kacSolution} for the IR fixed point, \eqref{boundaryExpansion} then gives the prediction
    \begin{align}
        \langle \phi_{(r,s)} \| B^*\rangle\!\rangle 
&= (-1)^{(r+s)(\alpha_1+\alpha_2)} \sqrt{rs}\sum\limits_{l=1}^{\min(\alpha_1, \alpha_2)} g_{(\alpha_1 + \alpha_2 + 1 - 2l, 1)}\, \big[1 + \mathcal{O}(y^2)\big]\,,
    \end{align}
for the boundary state coefficients of the IR fixed point.
This, in turn, yields the expected scaling
\be
y^{-\frac{3}{2}}\big(\langle \phi_{(r,s)} \| B^*\rangle\!\rangle  - \langle \phi_{(r,s)} \| B_{(\alpha_1, \alpha_2)}\rangle\!\rangle\big) = \mathcal{O}(y^2)\,,
\ee
where we used that $g_{(\alpha, \beta)}$ scales as $\mathcal{O}(y^\frac{3}{2})$, and that the relative change $(g^* - g_{(\alpha, \beta)})/g_{(\alpha,\beta)}$ of the $g$-function  scales like $\mathcal{O}(y^3)$. This means that the shift of the boundary state coefficients divided by the $g$-function of the UV BCFT is expected to be $\mathcal{O}(y^2)$. Correspondingly, we should expect that the perturbative result \eqref{bStateShift} will give non-zero change in the boundary state coefficient at $\mathcal{O}(y^2)$. This is indeed the case as we can show that the bulk-boundary OPE coefficients involving the boundary insertion of $V=\psi_{(1,3)}$ satisfy
\begin{align}
    B_{(r,s)(1,3)} = \mathcal{O}(y)\,.
\end{align}

\subsubsection*{The flow $(1,3) \to (3,1)$}
Using \eqref{boundaryExpansion}, we expect to generally obtain the change
\be
\langle \phi_{(r,s)} \| B_{(3,1)}\rangle\!\rangle - \langle \phi_{(r,s)} \| B_{(1,3)}\rangle\!\rangle = g_{(1,3)} \bigg[\frac{\pi^2}{3} \sqrt{rs}(s^2-r^2)y^2 + \mathcal{O}(y^3)\bigg]\,,
\label{1331boundaryState}
\ee
in the boundary-state coefficients for a generic bulk scalar $\phi_{(r,s)}$. 

We can first test this expectation on $(r,s) = (1,3)$, for which we have
\begin{subequations}
\label{eq:BBs13}
\begin{align}
B_{(1,3)(1,3)} &= -\frac{8\pi}{\sqrt{3}} y +\mathcal{O}(y^2)\,,\\
B_{(1,3)(1,1)} &= \sqrt{3} + \mathcal{O}(y)\,,
\end{align}
\end{subequations}
and, using the result \eqref{eq:BBIntegrated} for the integrated three-point bulk-boundary correlator, we also get
\be
\tilde{\mathcal{B}}_{(1,3)(1,3)(1,3)} = -\frac{4\pi}{\sqrt{3}}g_{(1,3)}\,.
\ee
The equation \eqref{bStateShift} then gives
\be
\langle \phi_{(1,3)} \| B_{(1,3)}^*\rangle\!\rangle - \langle \phi_{(1,3)} \| B_{(1,3)}\rangle\!\rangle = g_{(1,3)}\bigg[\frac{8\pi^2}{\sqrt{3}}y^2 + \mathcal{O}(y^3)\bigg]\,,
\ee
which indeed agrees with (\ref{1331boundaryState}) for $(r,s)=(1,3)$.

Another example is that of $(r,s) = (2,3)$, for which we can determine
\begin{align}
  B_{(2,3)(1,3)} &= -4 \sqrt{\frac{2}{3}} y + \mathcal{O}(y^2)\,,\\
  B_{(2,3)(1,1)} &= \sqrt{6}\,,
\end{align}
and
\be
\tilde{\mathcal{B}}_{(2,3)(1,3)(1,3)} = -\sqrt{\frac{3}{2}}\pi g_{(1,3)}\,.
\ee
Given these results, \eqref{bStateShift} yields
\be
\langle \phi_{(2,3)} \| B_{(1,3)}^*\rangle\!\rangle - \langle \phi_{(2,3)} \| B_{(1,3)}\rangle\!\rangle = g_{(1,3)}\bigg[5\sqrt{\frac{2}{3}}\pi^2 y^2 + \mathcal{O}(y^3)\bigg]\,,
\ee
which again with \eqref{1331boundaryState} for $(r,s) = (2,3)$.

One can also try to compute the change in the boundary state coefficient for $(r,s) = (3,3)$. From \eqref{1331boundaryState}, we expect no shift in the boundary state at $\mathcal{O}(y^2)$. From the perturbation theory result \eqref{bStateShift}, we can see this by realizing that $B_{(3,3)(1,3)} = \mathcal{O}(y^2)$ and that $\tilde{\mathcal{B}}_{(3,3)(1,3)(1,3)} = 0$ as the formula \eqref{eq:BBIntegrated} for the integrated 3-point bulk-boundary correlator is proportional to $|r-s|$.

\subsubsection*{The flow $(2,3) \to (4,1) \oplus (2,1)$}
As in the previous case, we first compute the expected difference in the boundary state coefficients using \eqref{boundaryExpansion}. We obtain
\begin{align}
\langle \phi_{(r,s)} \| B_{(4,1)}\rangle\!\rangle + \langle \phi_{(r,s)} \| B_{(2,1)}\rangle\!\rangle - \langle \phi_{(r,s)} \| B_{(2,3)}\rangle\!\rangle  
&= g_{(2,3)}\bigg[(-1)^{r+s}\frac{\pi^2}{3}\sqrt{rs}(s^2-r^2)y^2 + \mathcal{O}(y^3)\bigg]\,.
\label{23boundaryState}
\end{align}

The test for $(r,s) = (1,3)$ proceeds as for the $(1,3)$ boundary because the relevant bulk-boundary structure constants are again given by the expression \eqref{eq:BBs13}.
As before, we also work out the boundary state coefficient for $(r,s) = (2,3)$, for which
\begin{subequations}
    \begin{align}
        B_{(2,3)(1,3)} &= 4 \sqrt{\frac{2}{3}} y + \mathcal{O}(y^2)\,,\\
        B_{(2,3)(1,1)} &= -\sqrt{6}+\mathcal{O}(y)\,,
    \end{align}
\end{subequations}
and
\be
\tilde{\mathcal{B}}_{(2,3)(1,3)(1,3)} = \sqrt{\frac{3}{2}}\pi g_{(2,3)}\,.
\ee
Using these results (as well as the result that the boundary structure constant $C_{VVV}$ on the $(2,3)$ boundary takes the same value as on the $(1,3)$ boundary), \eqref{bStateShift} yields
\be
\langle \phi_{(2,3)} \| B_{(2,3)}^*\rangle\!\rangle - \langle \phi_{(2,3)} \| B_{(2,3)}\rangle\!\rangle = g_{(2,3)}\bigg[-5\sqrt{\frac{2}{3}}\pi^2 y^2 + \mathcal{O}(y^3)\bigg]\,,
\ee
which agrees with (\ref{23boundaryState}) for $(r,s) = (2,3)$.
Finally, for the same reasons as for the $(1,3)$ boundary, the boundary state coefficient for $(r,s) = (3,3)$ does not get shifted at $\mathcal{O}(y^2)$.

\subsubsection*{The flow $(2,4) \to (5,1) \oplus (3,1)$}
Using (\ref{boundaryExpansion}), we get the expected shift in the boundary state coefficients
\be
\begin{aligned}
\langle \phi_{(r,s)} \| B_{(5,1)}\rangle\!\rangle + \langle \phi_{(r,s)} \| B_{(3,1)}\rangle\!\rangle - \langle \phi_{(r,s)} \| B_{(2,4)}\rangle\!\rangle  
&= g_{(2,4)}\bigg[\frac{5}{8}\pi^2\sqrt{rs}(s^2-r^2)y^2 + \mathcal{O}(y^3)\bigg]\,.
\label{24boundaryState}
\end{aligned}
\ee
We again perform two tests, one for $(r,s) = (1,3)$ and the other for $(r,s) = (2,3)$. For both of these calculations, we need to compute the structure constant $C_{VVV}  = 2\sqrt{{2}/{15}} + \mathcal{O}(y)$ on the $(2,4)$ boundary.

To make the test for $(r,s) = (1,3)$, we first compute
\begin{subequations}
    \begin{align}
        B_{(1,3)(1,3)} &= -2\sqrt{10}\pi y + \mathcal{O}(y^2)\,,\\
        B_{(1,3)(1,3)} &= \sqrt{3} + \mathcal{O}(y)\,,
    \end{align}
\end{subequations}
which gives
\be
\tilde{\mathcal{B}}_{(1,3)(1,3)(1,3)} = - g_{(2,4)} \frac{4\pi}{\sqrt{3}} y\,.
\ee
Formula \eqref{bStateShift} then yields
\be
\langle \phi_{(1,3)} \| B_{(2,4)}^*\rangle\!\rangle - \langle \phi_{(1,3)} \| B_{(2,4)}\rangle\!\rangle = g_{(2,4)}\bigg[5\sqrt{3}\pi^2 y^2 + \mathcal{O}(y^3)\bigg]
\ee
in accordance with \eqref{23boundaryState} for $(r,s) = (1,3)$.

The other test for $(r,s) = (2,3)$ is performed by first computing 
\begin{subequations}
    \begin{align}
        B_{(2,3)(1,3)} &= -2\sqrt{5}\pi y + \mathcal{O}(y^2)\,,\\
        B_{(2,3)(1,1)} &= \sqrt{6}+ \mathcal{O}(y)\,,
    \end{align}
\end{subequations}
so that \eqref{eq:BBIntegrated} can be evaluated as
\be
\tilde{\mathcal{B}}_{(2,3)(1,3)(1,3)} = - g_{(2,4)} \sqrt{\frac{3}{2}} \pi y\,,
\ee
from this we obtain
\be
\langle \phi_{(2,3)} \| B_{(2,4)}^*\rangle\!\rangle - \langle \phi_{(2,3)} \| B_{(2,4)}\rangle\!\rangle = g_{(2,4)}\bigg[\frac{25}{4}\sqrt{\frac{3}{2}}\pi^2 y^2 + \mathcal{O}(y^3)\bigg]\,,
\ee
in accordance with \eqref{23boundaryState} for $(r,s) = (2,3)$.

\subsection{Boundary OPE structure constants}
\label{boundaryOPEcomp}
In this subsection, we use the formula \eqref{boundaryOPE} to compute a few selected boundary-boundary-boundary OPE structure constants of the IR endpoint for the same flows as considered in Subsection \ref{anomalousDimComputation}. 
The input of this computation will be the integrated correlators \eqref{integrated4ptCorrelator} and \eqref{17131513amp}, into which we plug in the structure constants \eqref{eq:Cboundary}.

\subsubsection*{The flow $(1,3) \to (3,1)$}
For this RG flow, we compute the shifts in the boundary structure constants of the form $C_{(i, j)(i,j) V}$. On the $(1,3)$ boundary, such nontrivial structure constants are 
\be
C_{VVV} = 1 - \frac{3}{4}y + \mathcal{O}(y^2)\,,  \qquad C_{(1,5)(1,5)V} = 3 - 4y + \mathcal{O}(y^2)\,,
\ee
and we expect these to flow to
\be
C_{(3,1)(3,1)(3,1)} = 1 + \frac{3}{4}y + \mathcal{O}(y^2)\,, \qquad C_{(5,1)(5,1)(3,1)} = 3 + 4y + \mathcal{O}(y^2)\,,
\label{structureShift13}
\ee
on the $(3,1)$ boundary. The formula \eqref{boundaryOPE} then instructs us to compute the integrated four point correlators \eqref{integrated4ptCorrelator}, which can be evaluated as
\be
\tilde{\mathcal{A}}_{VVVV} = - \frac{3}{2} g_{(1,3)} + \mathcal{O}(y)\,, \qquad \tilde{\mathcal{A}}_{(1,5)V(1,5)V} = -8 g_{(1,3)} + \mathcal{O}(y)\,,
\ee
giving us
\be
C^*_{VVV} - C_{VVV} = \frac{3}{2} y + \mathcal{O}(y^2)\,,  \quad C_{(1,5)(1,5)V}^* - C_{(1,5)(1,5)V}  = 8 y + \mathcal{O}(y^2)\,,
\ee
in accordance with \eqref{structureShift13}.

\subsubsection*{The flow $(1,5) \to (5,1)$}
For this flow, we again compute the shifts of the structure constants of the form $C_{(i, j)(i,j) V}$. In addition, we also compute the shift of $C_{V(1,5)(1,7)}$ using the integrated correlator (\ref{17131513amp}). The UV structure constants of interest are thus
\be
\begin{aligned}
    &C_{VVV} = \frac{1}{\sqrt{3}} - \frac{\sqrt{3}}{4} y + \mathcal{O}(y^2)\,, \quad C_{(1,5)(1,5)V} = \sqrt{3} - \frac{4}{\sqrt{3}} y + \mathcal{O}(y^2)\,, \\
    &C_{(1,7)(1,7)V} =2\sqrt{3} - \frac{17 \sqrt{3}}{5} y + \mathcal{O}(y^2) \,,\quad C_{(1,9)(1,9)V} = \frac{10}{\sqrt{3}} - \frac{551}{28\sqrt{3}} y + \mathcal{O}(y^2)\,, \\
    & C_{V(1,5)(1,7)} = \sqrt{\frac{2}{5}} - \frac{101}{60\sqrt{10}} y + \mathcal{O}(y^2)\,,
\end{aligned}
\ee
which are expected to flow to
\be
\begin{aligned}
    &C_{(3,1)(3,1)(3,1)} = \frac{1}{\sqrt{3}} + \frac{\sqrt{3}}{4} y + \mathcal{O}(y^2)\,, \quad C_{(5,1)(5,1)(3,1)} = \sqrt{3} + \frac{4}{\sqrt{3}} y + \mathcal{O}(y^2)\,, \\
    &C_{(7,1)(7,1)(3,1)} =2\sqrt{3} + \frac{17 \sqrt{3}}{5} y + \mathcal{O}(y^2) \,,\quad C_{(9,1)(9,1)(3,1)} = \frac{10}{\sqrt{3}} + \frac{551}{28\sqrt{3}} y + \mathcal{O}(y^2), 
    \\ & C_{(3,1)(5,1)(7,1)} = \sqrt{\frac{2}{5}} + \frac{101}{60\sqrt{10}} y + \mathcal{O}(y^2)\,,\label{structureShift15}
\end{aligned}
\ee
on the $(5,1)$ boundary. The relevant integrated correlators, needed in \eqref{boundaryOPE}, are given by \eqref{integrated4ptCorrelator} and \eqref{17131513amp}. These yield the results
\be
\begin{aligned}
&\tilde{\mathcal{A}}_{VVVV} = - \frac{1}{2} g_{(1,5)} + \mathcal{O}(y)\,, \quad \tilde{\mathcal{A}}_{(1,5)V(1,5)V} = - \frac{8}{3} g_{(1,5)} + \mathcal{O}(y)\,, \\
&\tilde{\mathcal{A}}_{(1,7)V(1,7)V} = -\frac{34}{5} g_{(1,5)} + \mathcal{O}(y)\,, \quad \tilde{\mathcal{A}}_{(1,9)V(1,9)V} = -\frac{551}{42} g_{(1,5)} + \mathcal{O}(y)\,, \\
& \tilde{\mathcal{A}}_{(1,5)V(1,9)V} = -\frac{101}{30\sqrt{30}}g_{(1,5)} + \mathcal{O}(y)\,,
\end{aligned}
\ee
which, in turn, give
\be
\begin{aligned}
&C^*_{VVV} - C_{VVV} = \frac{\sqrt{3}}{2} y + \mathcal{O}(y^2)\,,  \quad C_{(1,5)(1,5)V}^* - C_{(1,5)(1,5)V}  = \frac{8}{\sqrt{3}}y + \mathcal{O}(y^2)\,, \\
&C^*_{(1,7)(1,7)V} - C_{(1,7)(1,7)V} = \frac{34\sqrt{3}}{5} y + \mathcal{O}(y^2)\,,  \quad C^*_{(1,9)(1,9)V} - C_{(1,9)(1,9)V} = \frac{551}{14\sqrt{3}} y + \mathcal{O}(y^2)\,,\\
&C^*_{V(1,5)(1,7)} - C_{V(1,5)(1,7)} = \frac{101}{30\sqrt{10}} y + \mathcal{O}(y^2)\,,
\end{aligned}
\ee
in accordance with \eqref{structureShift15}.

\subsection{Bulk-boundary OPE structure constants}
In this subsection, we use the formula \eqref{bulkBoundaryOPE} to compute some of the bulk-boundary structure constants of the IR endpoint for the flows considered in subsections \eqref{anomalousDimComputation} and \eqref{boundaryOPEcomp}.
As an input, we take the integrated correlators \eqref{eq:BBIntegrated}, into which we plug in the bulk-boundary structure constants \eqref{eq:Bbulk_boundary}.

\subsubsection*{The flow $(1,3) \to (3,1)$}

We compute the shifts of $B_{(r,s)V}$ (the boundary insertion is $V$, so that we can use \eqref{eq:BBIntegrated}) for $(r,s) = (1,3)$ and $(r,s) = (2,3)$. In the UV, we have the bulk-boundary structure constants 
\be
B_{(1,3)V} = -\frac{8\pi}{\sqrt{3}}y + \mathcal{O}(y^2)\,, \qquad B_{(2,3)V} = -4\sqrt{\frac{2}{3}}\pi y + \mathcal{O}(y^2)\,,
\label{bb1}
\ee
which are predicted to flow to
\be
B_{(1,3)(3,1)} = \mathcal{O}(y^2)\,, \qquad B_{(2,3)(3,1)} = -\sqrt{\frac{2}{3}}\pi y + \mathcal{O}(y^2)\,.
\label{bb2}
\ee
Using \eqref{eq:BBIntegrated} and $B_{(1,3)(1,1)} = \sqrt{3} + \mathcal{O}(y)$, $B_{(2,3)(1,1)} = \sqrt{6} + \mathcal{O}(y)$, we obtain
\be
\tilde{\mathcal{B}}_{(1,3) VV} = -\frac{4\pi}{\sqrt{3}} g_{(1,3)} + \mathcal{O}(y^2)\,, \qquad \tilde{\mathcal{B}}_{(2,3)VV} = -\sqrt{\frac{3}{2}}\pi g_{(1,3)} + \mathcal{O}(y^2)\,,
\ee
which, when plugged into \eqref{bulkBoundaryOPE}, gives using $C_{VVV} = 1 + \mathcal{O}(y)$ the shifts
\be
B^*_{(1,3)V} - B_{(1,3)V} = \frac{8\pi}{\sqrt{3}} y + \mathcal{O}(y^2)\,, \quad B^*_{(2,3)V}-B_{(2,3)V} = \sqrt{6}\pi y + \mathcal{O}(y^2),
\ee
which agree with the difference between (\ref{bb2}) and (\ref{bb1}).

\subsubsection*{The flow $(1,5) \to (5,1)$}
We compute the shifts of bulk-boundary structure constants with the same Kac labels as we did above for the flow $(1,3)\to (3,1)$, i.e. $B_{(r,s)V}$ for $(r,s) = (1,3)$ and $(r,s) = (2,3)$. The bulk-boundary structure constants
\be
B_{(1,3)V} = -8\pi y + \mathcal{O}(y^2)\,, \qquad B_{(2,3)V} = -4\sqrt{2}\pi y + \mathcal{O}(y^2)\,,
\label{bb3}
\ee
should flow to
\be
B_{(1,3)(3,1)} = \mathcal{O}(y^2)\,, \qquad B_{(2,3)(3,1)} = -\sqrt{2}\pi y + \mathcal{O}(y^2)\,.
\label{bb4}
\ee
The corresponding values of the integrated correlators \eqref{eq:BBIntegrated} are 
\be
\tilde{\mathcal{B}}_{(1,3) VV} = -\frac{4\pi}{\sqrt{3}} g_{(1,5)} + \mathcal{O}(y^2)\,, \qquad \tilde{\mathcal{B}}_{(2,3)VV} = -\sqrt{\frac{3}{2}}\pi g_{(1,5)} + \mathcal{O}(y^2)\,,
\ee
which we evaluated using $B_{(1,3)(1,1)} = \sqrt{3} + \mathcal{O}(y)$ and $B_{(2,3)(1,1)} = \sqrt{6} + \mathcal{O}(y)$.
Using that $C_{VVV} = \frac{1}{\sqrt{3}} + \mathcal{O}(y)$, we obtain
\be
B^*_{(1,3)V} - B_{(1,3)V} = 8\pi y + \mathcal{O}(y^2)\,, \qquad B^*_{(2,3)V}-B_{(2,3)V} = 3\sqrt{2}\pi y + \mathcal{O}(y^2)\,,
\ee
in agreement with the difference of \eqref{bb4} and \eqref{bb3}.

\section{Discussion and outlook}
\label{sec:disc}

In this paper we have demonstrated the utility of string field theory in studying RG flows between boundary conformal field theories in two dimensions by deriving a set of simple formulae which allow one to compute all data of the IR BCFT up to subleading order in perturbation (leading only in the case of structure constants). The SFT approach played an instrumental role in deriving the correct form of the logarithmic subtractions in integrated correlators by recognizing that these would come from propagation of nearly on-shell states in the corresponding string theory amplitudes. Having evaluated the string propagator by explicitly inverting the SFT kinetic operator, we were also able to show that power divergences in integrated correlators can be consistently avoided by analytically continuing in the dimensions of states propagating in the collisions with the deforming operators.

The apparent structural simplicity of our formulae provides optimism towards pursuing the effort of deriving the IR BCFT data to an arbitrary order in the perturbation. Repeating the strategy outlined in this paper to a few more orders in $y$ (possibly choosing a different gauge to facilitate the rewriting of SFT Feynman diagrams in terms of matter BCFT correlators on the upper-half plane), it seems plausible that one should be able to guess an all-order expansion of the IR BCFT data in terms of ``amplitudes'' involving insertions of the perturbing operator at the moving punctures. It would be also intriguing to test whether our formulae (particularly the form of the log subtractions) extrapolate to higher dimensions, as suggested by the apparent similarity with the results of \cite{Komargodski:2016auf}.

Having utilizable results for IR BCFT data to a high order in $y$ promises a plethora of interesting applications. For one, there are known examples of boundary RG flows where one lands at an unknown BCFT in the IR. See \cite{Affleck:2000ws,1996cond.mat..2099Y,Schnabl:2019oom,honeycomb} for examples of such BCFTs which occur in as simple bulk theories as compactified free bosons. Perturbation theory (whenever applicable) is often one of the only options of studying these BCFTs as they typically break most of the bulk CFT chiral algebra which makes them difficult to find by other means. In many of such applications, one will encounter mixing between boundary operators as the theory flows to the IR. We expect that the results of this paper should extend rather straightforwardly to the cases where mixing occurs.

More abstractly, having established a direct link between the classical solution $\Psi^\ast(y)$ and the IR BCFT data, one should be able to study (at least perturbatively in $y$) the relation between conformal bootstrap and string field theory equations of motion. This has a potential of shedding more light on the fundamental mechanism behind the Sen's conjecture, which claims that classical solutions of OSFT are in one-to-one correspondence (modulo gauge transformation) with consistent worldsheet BCFTs.

In this paper we have made provisions for the IR BCFT data to be expandable as an integer power series in the parameter $y$ which measures the departure of the perturbing operator $V$ from marginality. In particular, we have see that this property is guaranteed by assuming that $V$ experiences cubic potential, namely that $C_{VVV}\neq 0$. However, there are instances of fixed points which are only seen once quartic (or even higher) terms in the effective potential for $V$ are taken into account. Prime examples of such scenaria are perturbations by boundary-condition-changing operators (BCCOs) \cite{Graham:2001pp,Gaberdiel:2008fn}. In the SFT framework, these can be naturally implemented by endowing the string field with a matrix (Chan-Paton) degree of freedom. The systematics of perturbation theory is then expected to be qualitatively different from what was presented in this paper and merits a project of its own. It would be interesting to see applications of such results on the condensation of barely tachyonic stretched strings on a pair of D-branes (or a $\mathrm{D}p$-$\overline{\mathrm{D}p}$ brane pair in the case of superstring) in suitable backgrounds which might allow for the corresponding RG flow to be short.

Finally, it is to be expected that the strategy outlined in this paper will allow for a direct derivation of analogous results for short RG flows in bulk CFT using closed string field theory.\footnote{See the recent progress \cite{Scheinpflug:2023lfn,Mazel:2024alu} as well as the older works \cite{Mukherji:1991tb,GHOSHAL1991295} in this direction.} The qualitatively different feature appearing in the case of closed strings is the universal presence of the zero-momentum ghost dilaton, which often introduces obstructions to solving CSFT equations of motion. These can be avoided, for instance, by coupling a linear-dilaton sector into the worldsheet theory which has the effect of BRST-trivializing the ghost dilaton state. At the same time, this allows for the central charge of the CFT of interest to change as the theory flows from the UV to the IR because the linear dilaton sector is capable of compensating for it, so that the total central charge can remain fixed (as required by BRST consistency). While it should be relatively straightforward to repeat the derivation of anomalous dimensions and OPE structure constants presented in this paper,\footnote{Some care might be needed to account for the induced change in the string coupling constant $g_\mathrm{s}$ which was recently observed in \cite{Maccaferri:2022yzy,Maccaferri:2024cjb}.} it remains unclear what the best strategy is for tackling the computation of the change in central charge. The most direct method seems to be the evaluation of the on-shell action, as proposed by \cite{Scheinpflug:2023lfn}.\footnote{See also \cite{Okawa:2004rh} for an older calculation of the on-shell CSFT action in the specific case of twisted tachyon condensation.} However, it remains somewhat unclear how exactly (if at all) the on-shell action is related to the change in the central charge. In particular, it appears that a careful treatment of boundary terms (arising due to the presence of the non-compact linear dilaton CFT factor) in the SFT action will be essential, as it was shown \cite{Erler:2022agw} that non-zero on-shell CSFT action can only be sourced by boundary terms.\footnote{See \cite{Stettinger:2024uus,Firat:2024kxq,Maccaferri:2025orz,Maccaferri:2025onc,Bernardes:2025uzg,Bernardes3,Bernardes2} for recent progress on boundary terms and conserved charges in string field theory.} Alternatively, one could compute the change in the central charge from the induced change in the background charge in the linear dilaton sector (as in \cite{Mazel:2024alu}). While, at given order in perturbation theory, this method seems to require one more integration compared to the computation from the on-shell action, it is amenable to an efficient computation using the flat vertices introduced in \cite{Mazel:2024alu}. Other observables for bulk RG flows can be provided by coupling the closed-string theory to D-brane probes \cite{Maccaferri:2021ksp,Maccaferri:2022yzy,Maccaferri:2024cjb} and studying the corresponding bulk-induced boundary flows \cite{Fredenhagen:2006dn,Fredenhagen:2009tn} on the string worldsheet.\footnote{See also \cite{Lauria:2023uca,Antunes:2024hrt} for an intriguing recent investigation of bulk flows in a presence of conformal boundary where the bulk theory is put into $\mathrm{AdS}_2$.} As in the case of boundary flows, working out the IR CFT data for bulk flows to a high order in perturbation theory would open up avenues to many interesting applications \cite{Antunes:2022vtb,Antunes:2024mfb,Nakayama:2024msv,Ambrosino:2025yug}.

\section*{Acknowledgements}
We are grateful to Davide Gaiotto and Xi Yin for useful conversations.
We thank the International Centre for Theoretical Physics (ICTP), Trieste, for providing a stimulating environment while this work was in progress. The work of JS was supported by DOE grant DE-SC0007870. The work of MS was supported by Grant Agency of Czech Republic under the grant
EXPRO 20-25775X, and in its final stages  also by the Czech Ministry of Education, Youth and Sports, project No. FORTE – CZ.02.01.01/00/22{\_}008/0004632, co-funded by the European Union.
The work of JV is supported by the ERC Starting Grant 853507.



\appendix


\section{General structure of boundary OPEs}
\label{app:opes}

In this appendix, we aim to review some general features of OPEs of boundary primary fields. 

\subsection{Quasi-primary expansion}

Let us start by considering a collision of two quasi-primary fields $\psi_i$ and $\psi_j$ along the boundary. We will assume that boundary 2-point functions of all quasi-primary fields is normalized so that we do not have to make any distinction between the OPE coefficients $\tensor{C}{_{ij}^k}$ and the 3-point structure constants $C_{ijk}$. For $x>y$, we can evaluate the collision at $y$ as
\begin{align}
    \psi_i (x)\,\psi_j(y) = \sum_{k} \sum_{n=0}^\infty \frac{C_{ijk}}{(x-y)^{h_i+h_j-h_k}} \frac{1}{n!}\frac{(h_i-h_j+h_k)^{\overline{n}}}{(2h_k)^{\overline{n}}} (x-y)^n \partial^n\psi_k(y)\,,\label{eq:coll_xy}
\end{align}
where the sum on the r.h.s.\ runs over quasi-primaries $\psi_k$ and 
\begin{align}
    x^{\overline{n}} \equiv x(x+1)\ldots (x+n-1)
\end{align}
denotes the rising factorial (Pochhammer function). In order to evaluate the collision at $y$ also in the case $x<y$, let us start with the collision 
\begin{align}
    \psi_j(y)\,\psi_i (x) = \sum_{k} \sum_{n=0}^\infty \frac{C_{jik}}{(y-x)^{h_j+h_i-h_k}} \frac{1}{n!}\frac{(h_j-h_i+h_k)^{\overline{n}}}{(2h_k)^{\overline{n}}} (y-x)^n \partial^n\psi_k(x)\,,\label{eq:yx_coll}
\end{align}
for $y>x$ (which is evaluated at $x$) and Taylor-expand it around $y$. To this end, we can write
\begin{subequations}
  \begin{align}
  &\sum_{n=0}^\infty  \frac{1}{n!}\frac{(h_j-h_i+h_k)^{\overline{n}}}{(2h_k)^{\overline{n}}} (y-x)^n \partial^n\psi_k(x)=\nonumber\\
  &\hspace{3cm}= \sum_{n=0}^\infty\sum_{l=0}^\infty  \frac{1}{n!}\frac{1}{l!}\frac{(h_j-h_i+h_k)^{\overline{n}}}{(2h_k)^{\overline{n}}}(-1)^n (x-y)^{n+l} \partial^{n+l}\psi_k(y)\\
   &\hspace{3cm}= \sum_{m=0}^\infty \frac{1}{m!}(x-y)^{m} \partial^m \psi_k (y)\sum_{n=0}^m (-1)^n {m\choose n}\frac{(h_j-h_i+h_k)^{\overline{n}}}{(2h_k)^{\overline{n}}}  \,,
\end{align}  
\end{subequations}
where, to perform the sum over $n$ in the last line, we make use of the ``umbral'' binomial theorem in the form
\begin{align}
    \sum_{n=0}^{m}  {m\choose n } (-1)^n \,\frac{x^{\overline{n}}}{y^{\overline{n}}} = \frac{(y-x)^{\overline{m}}}{y^{\overline{m}}}\,.
\end{align}
We have therefore arrived at the equality
\begin{align}
    \sum_{n=0}^\infty  \frac{1}{n!}\frac{(h_j-h_i+h_k)^{\overline{n}}}{(2h_k)^{\overline{n}}} (y-x)^n \partial^n\psi_k(x)=\sum_{m=0}^\infty \frac{1}{m!}
    \frac{(h_i-h_j+h_k)^{\overline{m}}}{(2h_k)^{\overline{m}}}
    (x-y)^{m} \partial^m \psi_k (y)\,,
\end{align}
which enables us to rewrite the collision in \eqref{eq:yx_coll} as
\begin{align}
    \psi_j(y)\,\psi_i (x) = \sum_{k} \sum_{n=0}^\infty \frac{C_{jik}}{(y-x)^{h_j+h_i-h_k}} \frac{1}{n!}\frac{(h_i-h_j+h_k)^{\overline{n}}}{(2h_k)^{\overline{n}}} (x-y)^n \partial^n\psi_k(y)\,.
\end{align}
Altogether, we can therefore write down the OPE
\begin{align}
    \mathsf{T}\{ \psi_i(x)\,\psi_j(y)\}= \sum_{k} \sum_{n=0}^\infty \frac{C_{ijk}(x-y)}{|x-y|^{h_i+h_j-h_k}} \frac{1}{n!}\frac{(h_i-h_j+h_k)^{\overline{n}}}{(2h_k)^{\overline{n}}} (x-y)^n \partial^n\psi_k(y)\,,
\end{align}
where we define
\begin{align}
    C_{ijk}(x-y)  =\bigg\{\begin{array}{ll}
        C_{ijk} & \text{if $x>y$} \\
        C_{jik} & \text{if $y<x$}
    \end{array}\,.
\end{align}

\subsection{Symmetric OPE}

When evaluating 2-products in a twist-symmetric open string field theory, it comes in handy to have an explicit form of a \textsl{symmetric} OPE at our disposal. By this we mean an OPE of two operators inserted at $x$ and $-x$ for $x>0$, which is evaluated at $0$. We will derive this OPE by taking the collision \eqref{eq:coll_xy}, setting $y=-x$ and Taylor-expanding the r.h.s.\ (which is evaluated at $-x$) around 0. Doing this, we first obtain
\begin{align}
    \psi_i (x)\,\psi_j(-x) = \sum_{k} \sum_{n=0}^\infty \frac{C_{ijk}}{(2x)^{h_i+h_j-h_k}} \frac{1}{n!}\frac{(h_i-h_j+h_k)^{\overline{n}}}{(2h_k)^{\overline{n}}} 2^n x^n \partial^n\psi_k(-x)\,,
\end{align}
where we can manipulate
\begin{subequations}
    \begin{align}
       & \sum_{n=0}^\infty\frac{1}{n!}\frac{(h_i-h_j+h_k)^{\overline{n}}}{(2h_k)^{\overline{n}}} 2^n x^n \partial^n\psi_k(-x)=\nonumber\\
       &\hspace{2cm}= \sum_{n=0}^\infty\sum_{l=0}^\infty \frac{1}{n!}\frac{1}{l!}\frac{(h_i-h_j+h_k)^{\overline{n}}}{(2h_k)^{\overline{n}}} (-1)^l 2^n x^{n+l} \partial^{n+l}\psi_k(0)\\
       &\hspace{2cm}= \sum_{m=0}^\infty \frac{1}{m!}(-1)^{m}x^m \partial^m\psi_k(0)\sum_{n=0}^m{m\choose n}(-1)^n \,2^n \,\frac{(h_i-h_j+h_k)^{\overline{n}}}{(2h_k)^{\overline{n}}}  \,.
    \end{align}
\end{subequations}
In the final line, we can recognize
\begin{align}
    \sum_{n=0}^m{m\choose n}(-1)^n \,2^n \,\frac{(h_i-h_j+h_k)^{\overline{n}}}{(2h_k)^{\overline{n}}} = {_2 F_1}(-m,h_i-h_j+h_k;2h_k;2)\,.
\end{align}
This gives the symmetric OPE
\begin{align}
    \psi_i (x)\,\psi_j(-x) = \sum_{k}\sum_{m=0}^\infty \frac{(-1)^{m} C_{ijk}}{(2x)^{h_i+h_j-h_k}} \frac{1}{m!}\,{_2 F_1}(-m,h_i-h_j+h_k;2h_k;2) \, x^m\partial^m\psi_k(0) \,.\label{eq:symmetric_ope}
\end{align}
When computing the graded commutators, we frequently encounter the symmetrized OPE 
\begin{align}
    \psi_i (x)\,\psi_j(-x)+\psi_j (x)\,\psi_i(-x)\,.
\end{align}
Here we can conveniently make use of the identity
\begin{align}
    {_2 F_1}(-m,y-x;y;2)=(-1)^m{_2 F_1}(-m,x;y;2)\,,
\end{align}
which holds for all non-negative integers $k$, to write
\begin{align}
    &\psi_i (x)\,\psi_j(-x)+\psi_j (x)\,\psi_i(-x)=\nonumber\\
    &\hspace{2cm}= \sum_{k}\sum_{m=0}^\infty \frac{(-1)^{m} C_{ijk}+C_{jik}}{(2x)^{h_i+h_j-h_k}} \frac{1}{m!}\,{_2 F_1}(-m,h_i-h_j+h_k;2h_k;2) \, x^m\partial^m\psi_k(0)\,.\label{eq:symmetrized}
\end{align}
Hence, if the structure constant $C_{ijk}$ is symmetric / anti-symmetric in its first two indices only even / odd derivatives of quasi-primaries will propagate in the symmetrized collision in \eqref{eq:symmetrized}. Also note that in the particular case when $\psi_i=\psi_j$, the terms in the sum on the r.h.s.\ of \eqref{eq:symmetrized} become proportional to
\begin{align}
    (-1)^{m} C_{ijk}+C_{jik} = C_{iik}\big[1+(-1)^{m}\big]\,,
\end{align}
so that the sum over $m$ is projected to $m\in 2\mathbb{Z}$. That is, we obtain the result
\begin{align}
    \psi_i (x)\,\psi_i(-x)= \sum_{k}\sum_{r=0}^\infty \frac{ C_{iik}}{(2x)^{2h_i-h_k}} \frac{1}{(2r)!}\,\frac{(\tfrac{1}{2})^{\bar{r}}}{(h_k+\tfrac{1}{2})^{\bar{r}}} \, x^{2r}\partial^{2r}\psi_k(0)\,,\label{eq:symmetrizedEQ}
\end{align}
where we have noticed the relation
\begin{align}
    {_2 F_1}(-2r,h_k;2h_k;2)=\frac{(\tfrac{1}{2})^{\bar{r}}}{(h_k+\tfrac{1}{2})^{\bar{r}}} 
\end{align}
for any integer $r\geqslant 0$.

\subsection{Collisions with $V$}
\label{app:collV}

Let us now focus specifically on the case where one of the boundary operators entering the collision is the deforming operator $V$. Let us work in the strict $y=0$ limit where $V$ has dimension $h_V=1$. 

Starting with the OPE
\begin{align}
    \mathsf{T}\{ V(\xi)\,\psi_j(0)\}= \sum_{k} \sum_{n=0}^\infty \frac{C_{Vjk}(\xi)}{|\xi|^{1+h_j-h_k}} \frac{1}{n!}\frac{(1-h_j+h_k)^{\overline{n}}}{(2h_k)^{\overline{n}}} \xi^n \partial^n\psi_k(0)\,,\label{eq:Vpsij}
\end{align}
where
\begin{align}
    C_{Vjk}(\xi) = \left\{
    \begin{array}{ll}
     C_{Vjk}    &  \text{for $\xi>0$}\\
     C_{jVk}    &  \text{for $\xi<0$}
    \end{array}
    \right.\,,
\end{align}
we notice that for quasi-primaries $\psi_k$ with $0<h_j-h_k\in \mathbb{Z}$, the coefficient of $\partial^{h_j-h_k}\psi_k(0)$ \textsl{vanishes} because
\begin{align}
    (1-h_j+h_k)^{\overline{h_j-h_k}} = (1-h_j+h_k)(2-h_j+h_k)\ldots (1-h_j+h_k+h_j-h_k-1)=0\,.\label{eq:poch0}
\end{align}
This ensures that derivatives of quasi-primaries which are more relevant than $\psi_j$ do \textsl{not} contribute to the simple pole in the $V\psi_j$ OPE \eqref{eq:Vpsij}. Furthermore, assuming that the only quasi-primary of dimension $h_j$ which appears in the $V\psi_j$ collision is again $\psi_j$ guarantees that the singular part of the $V\psi_j$ OPE takes the form \eqref{eq:Vpsi_coll}. This, in particular, ensures that the subtractions \eqref{eq:Sigmaijk} and \eqref{eq:Sigmaai} correctly cancel the logarithmic divergences which would otherwise appear in the amplitudes \eqref{eq:Amp4pt} and \eqref{eq:Amp3ptBB}.

Let us continue by discussing the symmetrized collision of $V$ with $\tilde{\psi}_j=\psi_j v_j$ where $v_j$ is a primary from the auxilliary theory $\text{BCFT}_\text{aux}$ with dimension $1-h_j$ so that the dressed field $\tilde{\psi}_j$ has conformal dimension 1. Using \eqref{eq:symmetric_ope}, we can first write
\begin{align}
    V (x)\,\tilde{\psi}_j(-x) = \sum_{k}\sum_{m,l=0}^\infty \frac{(-1)^{m+l} C_{Vjk}}{(2x)^{1+h_j-h_k}} \frac{1}{m!l!}\,{_2 F_1}(-m,1-h_j+h_k;2h_k;2) \, x^{m+l}\partial^m\psi_k \partial^l v_j (0)\,,
\end{align}
so that after symmetrizing, we obtain
\begin{align}
    &V (x)\,\tilde{\psi}_j(-x)+\tilde{\psi}_j (x)\,V(-x)=\nonumber\\
    &\hspace{0.8cm}= \sum_{k}\sum_{m,l=0}^\infty \frac{(-1)^{m+l} C_{Vjk}+C_{jVk}}{(2x)^{1+h_j-h_k}} \frac{1}{m!l!}\,{_2 F_1}(-m,1-h_j+h_k;2h_k;2) \, x^{m+l}\partial^m\psi_k\partial^l\psi_j(0)\,.\label{eq:Vpsi_symm}
\end{align}
When computing the star-commutator $[cV,c\tilde{\psi}_j]$ of the states $cV$ and $c\tilde{\psi}_j$, in addition, one has to multiply this by the symmetric OPE $c(x)c(-x)$, which is known to contain only odd powers of $x$, and also act with $U_3^\ast$, which contains only even modes of the total worldsheet stress-energy tensor. 

Altogether, we can see that the star-commutator $[cV,c\tilde{\psi}_j]$ will generally produce states in the $\mathrm{ker}\,L_0$ provided that the collision of $V$ with $\psi_j$ propagates a quasi-primary $\psi_k$ with weight $h_k$ such that 
\begin{align}
    0\leqslant h_j-h_k\in\mathbb{Z}\,.
\end{align}
However, at the same time, to keep the analysis in this paper manageable, we would like to ensure that the only state produced by $P_0[cV,c\tilde{\psi}_j]$ is proportional to $c\partial c \tilde{\psi}_j$. As explained in Sections \ref{sec:rg} and \ref{sec:anomalous_dimensions}, this is because 1.\ appearance of other quasi-primaries $\psi_k\neq \psi_j$ with weight $h_k=h_j$ in the projected star-product would cause operator mixing along the RG trajectory, 2.\ contamination of $P_0[cV,c\tilde{\psi}_j]$ by descendants of quasi-primaries $\psi_k$ with $h_k<h_j$ would necessitate adding algebraic terms to the open-string propagator strip.

To ensure that $P_0[cV,c\tilde{\psi}_j]$ is proportional to $c\partial c \tilde{\psi}_j$, we can see from \eqref{eq:Vpsi_symm} that when $0<h_j-h_k$ is odd, one should require that the structure constant $C_{Vjk}$ is symmetric in the first two indices: then the sums over $m$ and $l$ in \eqref{eq:Vpsi_symm} are projected to $m+l\in 2\mathbb{Z}$ so that only even derivatives propagate, ensuring that no $\mathrm{ker}\,L_0$ states in the module of $\psi_k$ are produced in $[cV,c\tilde{\psi}_j]$ because only odd powers of $k$ survive after multiplying with the symmetric $cc$ OPE. Similarly, when $0<h_j-h_k$ is even, we can prevent the states from the $\psi_k$-module from appearing in $P_0[cV,c\tilde{\psi}_j]$ by assuming the anti-symmetry $C_{Vjk}=-C_{jVk}$. This allows only odd derivatives on the r.h.s.\ of \eqref{eq:Vpsi_symm} so that when combined with the symmetric $cc$ OPE, again, only odd powers of $x$ survive. Altogether, we conclude that assuming that the structure constants satisfy
\begin{align}
    C_{Vjk}=(-1)^{h_j-h_k+1}C_{jVk}\,,\qquad\text{for $0< h_j-h_k\in \mathbb{Z}$}\label{eq:cond_app}
\end{align}
ensures that no states from the $\psi_k$-module ($h_k<h_j$) contaminate $P_0 [cV,c\tilde{\psi}_j]$. Finally, when $h_k=h_j$, we have to assume that the only quasi-primary appearing in the $V\psi_j$ collision is again $\psi_j$ so that no operator mixing occurs.

In the particular case when $\tilde{\psi}_j=V$ which, in the strict limit $y=0$ has already dimension $1$, no dressing is needed so that the symmetrized OPE \eqref{eq:Vpsi_symm} is given by \eqref{eq:symmetrizedEQ} for $h_i=1$. Then, since only even derivatives appear in the OPE \eqref{eq:symmetrizedEQ}, it follows that in order for $P_0[cV,cV]$ not to be contaminated by the descendants of $\psi_k$, one has to disallow quasi-primary states $\psi_k$ with $h_k$ odd in the $VV$ collision. On the other hand, no restriction is placed on the appearance of states with $h_k$ even. These remarks indeed follow as a special case of the condition \eqref{eq:cond_app} which, for $\psi_j=V$, reads $C_{VVk} = (-1)^{h_k}C_{VVk}$. For $h_k$ odd, this forces $C_{VVk}=0$ while being automatically satisfied for $h_k$ even.

\section{Some SFT Feynman diagrams}
\label{app:feyn}

In this appendix, we will go through the details of the computation of two archetypal tree-level diagrams in Witten's open SFT  which frequently appear in the calculations of this paper.

\subsection{Boundary 4-point diagram $\mathcal{A}_{ijkV}$}
\label{app:feyn_4pt}

The first diagram we will evaluate reads
\begin{align}
    \mathcal{A}_{ijkV}&=\Big\langle c\tilde{\psi}_i\ast c\tilde{\psi}_{j},\frac{b_0}{L_0}\bar{P}_0[cV,c\tilde{\psi}_k]\Big\rangle+\text{cyc.}\,,\label{eq:Aijk_start}
\end{align}
where ``$\text{cyc.}$'' denotes the cyclic permutation of the indices $i$, $j$ and $k$.
It involves tree-level scattering of three on-shell zero-momentum states $c\tilde{\psi}_i$, $c\tilde{\psi}_j$ and $c\tilde{\psi}_k$, which were constructed from three generic primaries $\psi_i$, $\psi_j$, $\psi_k$ in the UV BCFT by dressing with suitable states in the auxiliary sector, as well as the state $cV$ constructed from the deforming primary $V$. We will evaluate $\mathcal{A}_{ijkV}$ in the strict limit $y\to 0$ in which $cV$ also becomes exactly on-shell.

Let us first note that we can expand the $\ast$-products to write
\begin{align}
    \Big\langle c\tilde{\psi}_i\ast c\tilde{\psi}_{j},\frac{b_0}{L_0}\bar{P}_0[cV,c\tilde{\psi}_k]\Big\rangle &= \langle 0| c\tilde{\psi}_i(-\sqrt{3}) c\tilde{\psi}_{j}(\sqrt{3})\, U_3 \frac{b_0}{L_0}\bar{P}_0  U_3^\ast \, \Big[c\tilde{\psi}_k(\tfrac{1}{\sqrt{3}}) cV(-\tfrac{1}{\sqrt{3}})+\nonumber\\
    &\hspace{6cm}+cV(\tfrac{1}{\sqrt{3}}) c\tilde{\psi}_k(-\tfrac{1}{\sqrt{3}})\Big]|0\rangle\,.\label{eq:expU}
\end{align}
In particular, in \eqref{eq:expU}, we have used the fact that in terms of the operators $U_r$, one may express the $\ast$-product of two dimension-0 states $\varphi_i(0)|0\rangle$, $\varphi_j(0)|0\rangle$ as
\begin{align}
    \varphi_i(0)|0\rangle\ast \varphi_j(0)|0\rangle = U_3^\ast \varphi_i(\tfrac{1}{\sqrt{3}})\varphi_j(-\tfrac{1}{\sqrt{3}})|0\rangle\,.
\end{align}
By inserting the Hodge-Kodaira decomposition od the identity
\begin{align}
    1 = P_0 +Q\frac{b_0}{L_0}\bar{P}_0 + \frac{b_0}{L_0}\bar{P}_0 Q\,, \label{eq:HK}
\end{align}
as well as using the gluing theorem in the form \cite{Schnabl:2002gg}
\begin{align}
    U_r U_s^\ast = U^\ast_{2+\frac{2}{r}(s-2)} U_{2+\frac{2}{s}(r-2)}\,,\label{eq:gluing}
\end{align}
it is then possible to rewrite the propagator-part of \eqref{eq:expU} as
\begin{align}
    U_3 \frac{b_0}{L_0}\bar{P}_0  U_3^\ast = P_0 U_3 \frac{b_0}{L_0}\bar{P}_0  U_3^\ast P_0 +\frac{b_0}{L_0}\bar{P}_0 U_\frac{8}{3}^\ast U_\frac{8}{3}  \frac{1+P_0}{2}+\frac{1+P_0}{2} U_\frac{8}{3}^\ast U_\frac{8}{3}\frac{b_0}{L_0}\bar{P}_0\,,\label{eq:flattenize}
\end{align}
up to $Q$-exact terms. Those can be ignored because all four external legs of the diagram are on-shell. 

Using the results of \cite{Scheinpflug:2023osi} and assuming the projected $\ast$-commutator $P_0[cV,c\tilde{\psi}_k]$ is proportional to $c\partial c\tilde{\psi}_k$ (as we have seen above, this happens as long as \eqref{eq:cond_app} holds and when there is no mixing of $\psi_k$ with other quasi-primaries under the deformation by $V$), 
the first term of \eqref{eq:flattenize} together with the $P_0$-proportional parts of the second and third term in \eqref{eq:flattenize} contribute to the r.h.s.\ of \eqref{eq:expU} with
\begin{align}
    2gC_{ijk}C_{kkV}\log \frac{27a\sqrt{3}}{16} \,,
\end{align}
where $a=\sqrt{2}-1$. Here we also recall that we assume that the dressing fields $v_i$ can be normalized so that 
\begin{align}
    \big\langle v_i(\infty)\, v_j(1) \,v_k(0)\big\rangle = 1\,.\label{eq:dressing_C}
\end{align}
Such dressing factors can always be realized in terms of non-compact free bosons.

On the other hand, to evaluate the parts of the second and the third term in \eqref{eq:flattenize} which are proportional to the identity, we explicitly invert $L_0$ by replacing \cite{Larocca:2017pbo}
    \begin{align}
        \frac{1}{L_0}\longrightarrow \frac{1}{L_0+\varepsilon}=\int_0^1 ds\, s^{L_0+\varepsilon-1}\,,
    \end{align}
where the Schwinger integral is evaluated in a range of $\varepsilon$ where it converges and then it is analytically continued to $\varepsilon=0$. In this way we avoid the power divergences of the Schwinger parametrization caused by the propagation of tachyonic states.
Altogether, we can show that \eqref{eq:expU} can be rewritten as
\begin{align}
    &\Big\langle c\tilde{\psi}_i\ast c\tilde{\psi}_{j},\frac{b_0}{L_0}\bar{P}_0[cV,c\tilde{\psi}_k]\Big\rangle =\nonumber\\
    &\hspace{0.4cm}=
    \lim_{\varepsilon\to 0}\,\int_0^1 ds\, s^\varepsilon\bigg[\frac{4}{a}\bigg(\frac{1}{a^2}-\mu(s)^2\bigg)\frac{d\mu}{ds}\bigg(\big\langle \tilde{\psi}_i(-\tfrac{1}{a})\,\tilde{\psi}_j(+\tfrac{1}{a})\,\tilde{\psi}_k(+\mu(s))\,V(-\mu(s))\big\rangle+\nonumber\\
    &\hspace{4cm}+\big\langle \tilde{\psi}_i(-\tfrac{1}{a})\,\tilde{\psi}_j(+\tfrac{1}{a})\,V(+\mu(s))\,\tilde{\psi}_k(-\mu(s))\big\rangle\bigg)-2gC_{ijk}C_{kkV}\frac{1}{s}\bigg]+\nonumber\\
    &\hspace{10cm}+2gC_{ijk}C_{kkV}\log \frac{27a\sqrt{3}}{16}\,,\label{eq:schwinger_frame}
\end{align}
where we have defined
\begin{align}
    \mu(s) = f_{\frac{8}{3}}(\tfrac{s}{\sqrt{2}}) = \tan\big(\tfrac{3}{4}\arctan\tfrac{s}{\sqrt{3}}\big)\,.
\end{align}
The degeneration limit $s\to 0$ of the open-string propagator in \eqref{eq:schwinger_frame} clearly gives the two sides of the $V\psi_k$ collision. We also note that the $\frac{1}{s}$ subtraction in the integrand comes from the action of the projector $\bar{P}_0=1-P_0$ outside of the $\mathrm{ker}\,L_0$ and as such it serves to remove the would-be logarithmic divergences due to massless propagation in the intermediate channel of the diagram. Finally, we can apply the two $SL(2;\mathbb{R})$ maps
\begin{subequations}
    \begin{align}
         w_1(z) &= \frac{z+\mu(s)}{z+\tfrac{1}{a}}\frac{\tfrac{2}{a}}{\tfrac{1}{a}+\mu(s)}\,,\\
        w_6(z) &= \frac{z-\mu(s)}{z+\tfrac{1}{a}}\frac{\tfrac{2}{a}}{\tfrac{1}{a}-\mu(s)}\,,
    \end{align}
\end{subequations}
respectively to the two correlators in the integrand of \eqref{eq:schwinger_frame}. This has the effect of mapping the insertions of $\tilde{\psi}_i$, $\tilde{\psi}_j$ and $\tilde{\psi}_k$ to $\infty$, $1$ and $0$, while the insertion of the deforming operator $V$ is mapped to
\begin{subequations}
    \begin{align}
     \xi_1(s)&=+\frac{4a\mu(s)}{(1+a\mu(s))^2}\in (0,\tfrac{1}{2})\,,\\
        \xi_6(s)&=-\frac{4a\mu(s)}{(1-a\mu(s))^2} = \frac{\xi_1(s)}{\xi_1(s)-1}\in (-1,0)\,,
    \end{align}
\end{subequations}
respectively. Accounting for the conformal transformations of the individual insertions as well as for the change of measure, we arrive at the result
\begin{align}
    &\Big\langle c\tilde{\psi}_i\ast c\tilde{\psi}_{j},\frac{b_0}{L_0}\bar{P}_0[cV,c\tilde{\psi}_k]\Big\rangle =\nonumber\\
    &\hspace{1cm}= \lim_{\varepsilon\to 0}\,\int_{-1}^{0} d\xi\, s_6(\xi)^\varepsilon \,\bigg[\big\langle {\psi}_i(\infty)\,{\psi}_j(1)\,{\psi}_k(0) V(\xi)\big\rangle+gC_{ijk}C_{kkV}\frac{1}{s_6(\xi)}\frac{ds_6}{d\xi}\bigg]
   +\nonumber\\
   &\hspace{2cm}+\lim_{\varepsilon\to 0}\,\int_{0}^{\frac{1}{2}} d\xi\, s_1(\xi)^\varepsilon \,\bigg[\big\langle {\psi}_i(\infty)\,{\psi}_j(1)\,V(\xi)\,{\psi}_k(0)\big\rangle-gC_{ijk}C_{kkV}\frac{1}{s_1(\xi)}\frac{ds_1}{d\xi}\bigg]+\nonumber\\
   &\hspace{10.7cm}+2gC_{ijk}C_{kkV}\log \frac{27a\sqrt{3}}{16}\,,\label{eq:moving_puncture}
\end{align}
where we have already implemented \eqref{eq:dressing_C} to get rid of the dressing factors $v_i$, $v_j$ and $v_k$. We have also introduced functions $s_1(\xi)$ and $s_6(\xi)$ to be the inverses of $\xi_1(s)$ and $\xi_6(s)$, respectively. As such, they map the moving-puncture coordinate $\xi$ to the Feynman modulus $s$ on the two respective intervals. Recalling the form of the OPE \eqref{eq:Vpsij}, we also note that the subtractions in \eqref{eq:moving_puncture} correctly take care of the potential logarithmic divergences due to the propagation of $\psi_k$ in the $V\psi_k$ collision.

Similarly, insisting that the fields $\psi_i$, $\psi_j$ and $\psi_k$ are inserted at $\infty$, $1$ and $0$, respectively, we can express the diagram for the remaining two orderings as
\begin{align}
    &\Big\langle c\tilde{\psi}_k\ast c\tilde{\psi}_{i},\frac{b_0}{L_0}\bar{P}_0[cV,c\tilde{\psi}_j]\Big\rangle =\nonumber\\
    &\hspace{1cm}=
   \lim_{\varepsilon\to 0}\,\int_{\frac{1}{2}}^{1} d\xi\, s_2(\xi)^\varepsilon \,\bigg[\big\langle {\psi}_i(\infty)\,{\psi}_j(1)\,V(\xi)\,{\psi}_k(0)\big\rangle+gC_{ijk}C_{jjV}\frac{1}{s_2(\xi)}\frac{ds_2}{d\xi}\bigg]+\nonumber\\
   &\hspace{2cm}+ \lim_{\varepsilon\to 0}\,\int_{1}^{2} d\xi\, s_3(\xi)^\varepsilon \,\bigg[\big\langle {\psi}_i(\infty)\,V(\xi)\,{\psi}_j(1)\,{\psi}_k(0)\big\rangle-gC_{ijk}C_{jjV}\frac{1}{s_3(\xi)}\frac{ds_3}{d\xi}\bigg]+\nonumber\\
   &\hspace{10.7cm}+2gC_{ijk}C_{jjV}\log \frac{27a\sqrt{3}}{16}\,,
\end{align}
and
\begin{align}
    &\Big\langle c\tilde{\psi}_j\ast c\tilde{\psi}_{k},\frac{b_0}{L_0}\bar{P}_0[cV,c\tilde{\psi}_i]\Big\rangle =\nonumber\\
    &\hspace{1cm}=
   \lim_{\varepsilon\to 0}\,\int_{2}^{\infty} d\xi\, s_4(\xi)^\varepsilon \,\bigg[\big\langle {\psi}_i(\infty)\,V(\xi)\,{\psi}_j(1)\,{\psi}_k(0)\big\rangle+gC_{ijk}C_{iiV}\frac{1}{s_4(\xi)}\frac{ds_4}{d\xi}\bigg]+\nonumber\\
   &\hspace{2cm}+ \lim_{\varepsilon\to 0}\,\int_{-\infty}^{-1} d\xi\, s_5(\xi)^\varepsilon \,\bigg[\big\langle {\psi}_i(\infty)\,{\psi}_j(1)\,{\psi}_k(0)\,V(\xi)\big\rangle-gC_{ijk}C_{iiV}\frac{1}{s_5(\xi)}\frac{ds_5}{d\xi}\bigg]+\nonumber\\
   &\hspace{10.7cm}+2gC_{ijk}C_{iiV}\log \frac{27a\sqrt{3}}{16}\,,
\end{align}
where the the map between the moving puncture $\xi$ and the Feynman modulus $s$ is provided by the functions (for $0<s<1$, where $s\to 0$ corresponds to the open-string degeneration)
\begin{subequations}
    \begin{align}
        \xi_2(s)&=\bigg(\frac{1-a\mu(s)}{1+a\mu(s)}\bigg)^2=1-\xi_1(s)\in (\tfrac{1}{2},1)\,,\\
        \xi_3(s)&=\bigg(\frac{1+a\mu(s)}{1-a\mu(s)}\bigg)^2=\frac{1}{1-\xi_1(s)}\in (1,2)\,,\\
        \xi_4(s)&=\frac{(1+a\mu(s))^2}{4a\mu(s)}=\frac{1}{\xi_1(s)}\in (2,\infty)\,,\\
        \xi_5(s)&=-\frac{(1-a\mu(s))^2}{4a\mu(s)}=1-\frac{1}{\xi_1(s)}\in (-\infty,-1)\,.
    \end{align}
\end{subequations}
As before, the function $s_2(\xi)$, $s_3(\xi)$, $s_4(\xi)$, $s_5(\xi)$ are defined as inverses of $\xi_2(s)$, $\xi_3(s)$, $\xi_4(s)$, $\xi_5(s)$.

Furthermore, we would like to provide a more natural form for the log-subtractions by expressing them in terms of simple poles in $\xi$ around the collisions at $0$, $1$ and $\infty$ in a way which would allow to rewrite \eqref{eq:Aijk_start} as a single integral over $\xi$ from $-\infty$ to $\infty$.
In particular, using the results
\begin{subequations}
    \begin{align}
        \int_0^\frac{1}{2} d\xi\,\bigg[C_{kkV}\bigg(\frac{1}{\xi}-\frac{1}{s_1(\xi)}\frac{ds_1}{d\xi}\bigg)+C_{jjV}\frac{1}{1-\xi}\bigg]&=C_{kkV}\log \frac{1+\sqrt{2}}{2\sqrt{3}}+C_{jjV}\log 2\,,\\
        \int_\frac{1}{2}^1 d\xi\,\bigg[C_{jjV}\bigg(\frac{1}{1-\xi}+\frac{1}{s_2(\xi)}\frac{ds_2}{d\xi}\bigg)+C_{kkV}\frac{1}{\xi}\bigg]&=C_{jjV}\log \frac{1+\sqrt{2}}{2\sqrt{3}}+C_{kkV}\log 2\,,\\
        \int_1^2 d\xi\,\bigg[C_{jjV}\bigg(\frac{1}{\xi(\xi-1)}-\frac{1}{s_3(\xi)}\frac{ds_3}{d\xi}\bigg)+C_{iiV}\frac{1}{\xi}\bigg]&=C_{jjV}\log \frac{1+\sqrt{2}}{2\sqrt{3}}+C_{iiV}\log 2\,,\\
        \int_2^\infty d\xi\,\bigg[C_{iiV}\bigg(\frac{1}{\xi}+\frac{1}{s_4(\xi)}\frac{ds_4}{d\xi}\bigg)+C_{jjV}\frac{1}{\xi(\xi-1)}\bigg]&=C_{iiV}\log \frac{1+\sqrt{2}}{2\sqrt{3}}+C_{jjV}\log 2\,,\\
         \int_{-\infty}^{-1} d\xi\,\bigg[C_{iiV}\bigg(\frac{1}{1-\xi}-\frac{1}{s_5(\xi)}\frac{ds_5}{d\xi}\bigg)+C_{kkV}\frac{1}{\xi(\xi-1)}\bigg]&=C_{iiV}\log \frac{1+\sqrt{2}}{2\sqrt{3}}+C_{kkV}\log 2\,,\\
         \int_{-1}^{0} d\xi\,\bigg[C_{kkV}\bigg(\frac{1}{\xi(\xi-1)}+\frac{1}{s_6(\xi)}\frac{ds_6}{d\xi}\bigg)+C_{iiV}\frac{1}{1-\xi}\bigg]&=C_{kkV}\log \frac{1+\sqrt{2}}{2\sqrt{3}}+C_{iiV}\log 2\,,
    \end{align}
\end{subequations}
we can express \eqref{eq:Aijk_start} as
\begin{subequations}
\begin{align}
    \mathcal{A}_{ijkV} &=  \lim_{\varepsilon\to 0}\,\int_{-\infty}^{+\infty} d\xi\, s(\xi)^\varepsilon \,\bigg[\big\langle \mathsf{T}\big\{{\psi}_i(\infty)\,{\psi}_j(1)\,{\psi}_k(0) V(\xi)\big\}\big\rangle-\Sigma_{ijk}(\xi)\bigg]+\nonumber\\
    &\hspace{2cm}+2g(C_{iiV}+C_{jjV}+C_{kkV})C_{ijk}\bigg(\log 2+\log \frac{1+\sqrt{2}}{2\sqrt{3}}+\log \frac{27a\sqrt{3}}{16}\bigg)\\
     &=  \lim_{\varepsilon\to 0}\,\int_{-\infty}^{+\infty} d\xi\, s(\xi)^\varepsilon \,\bigg[\big\langle \mathsf{T}\big\{{\psi}_i(\infty)\,{\psi}_j(1)\,{\psi}_k(0) V(\xi)\big\}\big\rangle-\Sigma_{ijk}(\xi)\bigg]+\nonumber\\
    &\hspace{8cm}+4g(C_{iiV}+C_{jjV}+C_{kkV})C_{ijk}\log K\label{eq:amp4S}
\end{align}
\end{subequations}
where the log-subtractions are now all subsumed in the function
\begin{align}
       \Sigma_{ijk}(\xi)&\equiv \frac{1}{2}\bigg(\frac{1}{|\xi|}+\frac{1}{|\xi-1|} - \frac{1}{|\xi||\xi-1|}\bigg)gC_{iiV}C_{ijk}+\nonumber\\
    &\hspace{4cm} +\frac{1}{2}\bigg(\frac{1}{|\xi-1|} + \frac{1}{|\xi||\xi-1|}-\frac{1}{|\xi|}\bigg)gC_{jjV}C_{ijk}+\nonumber\\
    &\hspace{4cm}+ \frac{1}{2}\bigg(  \frac{1}{|\xi||\xi-1|}+\frac{1}{|\xi|}-\frac{1}{|\xi-1|}\bigg)gC_{kkV}C_{ijk}
\end{align}
and we have defined the function $s(\xi)$ which correctly maps the moving puncture coordinate $\xi$ to the Feynman modulus $s$ in each interval over which we integrate. That is, we set
\begin{align}
    s(\xi) = \left\{\begin{array}{cl}
       s_1(\xi)  & \xi\in (0,\tfrac{1}{2}) \\
       s_2(\xi)  &  \xi\in (\tfrac{1}{2},1)\\
       \vdots &
    \end{array}\right.\,.\label{eq:sxi}
\end{align}
We also have recognized the mapping radius $K = \frac{3\sqrt{3}}{4}$ of the Witten vertex. 

Finally, as we discuss in detail in Appendix \ref{app:plumbing}, avoiding the power (tachyon) divergences by multiplying the integrand by\footnote{In general, the function $s(\xi)$ mapping the moving-puncture modulus $\xi$ to the Feynman moduli near open-string degenerations, can be taken for any twist-symmetric OSFT 3-vertex.} $s(\xi)^\varepsilon$ and analytically continuing to $\varepsilon=0$ is equivalent to analytically continuing in the dimension of the states propagating in the collisions of $V(\xi)$ with the insertions at $0$, $1$ and $\infty$. This allows us to finally write
\begin{align}
   \mathcal{A}_{ijkV}&= \tilde{\mathcal{A}}_{ijkV}+4g\big(C_{iiV}+C_{jjV}+C_{kkV}\big)C_{ijk}\log K\,,\label{eq:AijkVApp}
\end{align}
where we define 
\begin{align}
    \tilde{\mathcal{A}}_{ijkV} = \int\limits_{\substack{-\infty\\ \text{a.c.}} }^{+\infty} d\xi\,\bigg[\big\langle \mathsf{T}\big\{{\psi}_i(\infty)\,{\psi}_j(1)\,{\psi}_k(0) V(\xi)\big\}\big\rangle-\Sigma_{ijk}(\xi)\bigg]\,,
\end{align}
and where ``a.c.'' stands for the analytic continuation in the dimensions of the states propagating in the collisions at $0$, $1$ and $\infty$. Somewhat expectedly, we will see below in Appendix \ref{app:equiv_from_sign} that the transition from \eqref{eq:amp4S} to \eqref{eq:AijkVApp} works only if the the collisions of $V$ with $\psi_i$ at $0$, $1$ and $\infty$ satisfy the restriction \eqref{eq:res}. Indeed, if \eqref{eq:res} were violated, there would have been additional contributions to the amplitude coming from propagating the descendants appearing in $P_0[cV,c\psi_i]$. These contributions would need to be canceled by the resonant contributions encountered in Appendix \ref{app:equiv_from_sign} so that one would still arrive at the result \eqref{eq:AijkVApp}. Since the expression \eqref{eq:AijkVApp} makes sense irrespective of whether the restriction \eqref{eq:res} is in place or not, we think that it is plausible that such cancellation mechanism occurs.

\subsection{Bulk-boundary 3-point diagram $\mathcal{B}_{aiV}$}
\label{app:feyn_ellwood}

For the observables which involve a bulk primary insertion, we will also need to compute the diagram
\begin{align}
    i\mathcal{B}_{aiV} = \frac{1}{2}\Big\langle c\bar{c}\tilde{\phi}_a(i,-i)I ,\frac{b_0}{L_0}\bar{P}_0[cV,c\tilde{\psi}_i]\Big\rangle\,,
\end{align}
where $c\bar{c}\tilde{\phi}_a$ is an element of the closed-string BRST cohomology which was constructed from the bulk CFT primary $\phi_a$ by dressing it with a suitable field $w_a(z,\bar{z})$ in the auxiliary sector so as to make it dimension $(1,1)$ and so that the bulk-boundary structure constant with the auxiliary field $v_i$ (which is used to dress $\psi_i$) is non-zero and is normalized to one. Also, $I$ denotes the identity string field.

To start with, let us make the vertex geometry explicit by writing
\begin{align}
    i\mathcal{B}_{aiV} = \frac{1}{2}\langle 0| c\bar{c} \tilde{\phi}_a(i,-i) \, U_1\frac{b_0}{L_0} \bar{P}_0 U_3^\ast \,\Big[c\tilde{\psi}_i(\tfrac{1}{\sqrt{3}})\, cV(-\tfrac{1}{\sqrt{3}})+cV(\tfrac{1}{\sqrt{3}})\, c\tilde{\psi}_i(-\tfrac{1}{\sqrt{3}})\Big]|0\rangle\,.
\end{align}
Again by suitably inserting the Hodge-Kodaira decomposition \eqref{eq:HK} and using the gluing theorem \eqref{eq:gluing}, we can rewrite
\begin{align}
    U_1\frac{b_0}{L_0} \bar{P}_0 U_3^\ast =P_0 U_1 \frac{b_0}{L_0}\bar{P}_0 U_3^\ast P_0 +\frac{1+P_0}{2}U_4^\ast U_\frac{4}{3} \frac{b_0}{L_0}\bar{P}_0+  \frac{b_0}{L_0}\bar{P}_0 U_4^\ast U_\frac{4}{3}\frac{1+P_0}{2}\label{eq:ellwood_decomp}
\end{align}
up to $Q$-exact terms, which, can again be ignored since all insertions are assumed to be on-shell. Taking into account the assumption \eqref{eq:cond_app}, as well as the assumption that the only quasi-primary with dimension $h_i$ propagating in the $V\psi_i$ collision is again $\psi_i$, we ensure that the projected commutator $P_0[cV,c\tilde{\psi}_i]$ is proportional to $c\partial c \tilde{\psi}_i$. Employing a procedure similar to the one described in \cite{Scheinpflug:2023osi}, we can then compute the contributions of the $P_0$-terms in \eqref{eq:ellwood_decomp} to $i\mathcal{B}_{aiV}$ as
\begin{align}
    &\frac{1}{2}\langle 0| c\bar{c} \tilde{\phi}_a(i,-i) 
    P_0 U_1 \frac{b_0}{L_0}\bar{P}_0 U_3^\ast P_0
    \Big[c\tilde{\psi}_i(\tfrac{1}{\sqrt{3}})\, cV(-\tfrac{1}{\sqrt{3}})+cV(\tfrac{1}{\sqrt{3}})\, c\tilde{\psi}_i(-\tfrac{1}{\sqrt{3}})\Big]|0\rangle =\nonumber\\
    &\hspace{11cm}=-ig C_{iiV}B_{ai}\log \frac{16}{9}\,,
\end{align}
as well as
\begin{subequations}
\begin{align}
    &\frac{1}{2}\langle 0| c\bar{c} \tilde{\phi}_a(i,-i) 
   \frac{b_0}{L_0}\bar{P}_0 U_4^\ast U_\frac{4}{3}\frac{P_0}{2}
    \Big[c\tilde{\psi}_i(\tfrac{1}{\sqrt{3}})\, cV(-\tfrac{1}{\sqrt{3}})+cV(\tfrac{1}{\sqrt{3}})\, c\tilde{\psi}_i(-\tfrac{1}{\sqrt{3}})\Big]|0\rangle =\nonumber\\
    &\hspace{11cm}=\frac{1}{2}ig C_{iiV}B_{ai}\log 2\,,\\
      &\frac{1}{2}\langle 0| c\bar{c} \tilde{\phi}_a(i,-i)\frac{P_0}{2} U_4^\ast U_\frac{4}{3}
   \frac{b_0}{L_0}\bar{P}_0 
    \Big[c\tilde{\psi}_i(\tfrac{1}{\sqrt{3}})\, cV(-\tfrac{1}{\sqrt{3}})+cV(\tfrac{1}{\sqrt{3}})\, c\tilde{\psi}_i(-\tfrac{1}{\sqrt{3}})\Big]|0\rangle =\nonumber\\
    &\hspace{11cm}=\frac{1}{2}ig C_{iiV}B_{ai}\log \frac{2}{\sqrt{3}}\,.
\end{align}
\end{subequations}
At the same time, avoiding the power divergences due to tachyons propagating in the $V\tilde{\psi}_i$ collision by the $\varepsilon$-prescription, as in Appendix \ref{app:feyn_4pt},
the terms in \eqref{eq:ellwood_decomp} containing no $P_0$ on either side of the propagator can be shown to translate into the contributions
\begin{align}
     &\frac{1}{4}\langle 0| c\bar{c} \tilde{\phi}_a(i,-i) 
   \frac{b_0}{L_0}\bar{P}_0 U_4^\ast U_\frac{4}{3}
    \Big[c\tilde{\psi}_i(\tfrac{1}{\sqrt{3}})\, cV(-\tfrac{1}{\sqrt{3}})+cV(\tfrac{1}{\sqrt{3}})\, c\tilde{\psi}_i(-\tfrac{1}{\sqrt{3}})\Big]|0\rangle =\nonumber\\
    &\hspace{1cm}=\frac{i}{4}\lim_{\varepsilon\to 0}\int_0^1 ds\,s^\varepsilon\,\bigg[4\big(1+\mu_4(s)^2\big)\,\frac{d\mu_4}{ds}\bigg(\big\langle \tilde{\phi}_a(i\mu_4(s),-i\mu_4(s))\,\tilde{\psi}_i(-1)\,V(1)\big\rangle+\nonumber\\
    &\hspace{6cm}+\big\langle \tilde{\phi}_a(i\mu_4(s),-i\mu_4(s))\,V(-1)\,\tilde{\psi}_i(1)\big\rangle\bigg)-\frac{2}{s}gB_{ai}C_{iiV}\bigg]\label{eq:schwinger_ellwood1}
\end{align}
and
\begin{align}
     &\frac{1}{4}\langle 0| c\bar{c} \tilde{\phi}_a(i,-i) 
   U_4^\ast U_\frac{4}{3}\frac{b_0}{L_0}\bar{P}_0 
    \Big[c\tilde{\psi}_i(\tfrac{1}{\sqrt{3}})\, cV(-\tfrac{1}{\sqrt{3}})+cV(\tfrac{1}{\sqrt{3}})\, c\tilde{\psi}_i(-\tfrac{1}{\sqrt{3}})\Big]|0\rangle =\nonumber\\
    &\hspace{1cm}=\frac{i}{4}\lim_{\varepsilon\to 0}\int_0^1 ds\,s^\varepsilon\,\bigg[4\big(1+\mu_\frac{4}{3}(s)^2\big)\,\frac{d\mu_\frac{4}{3}}{ds}\bigg(\big\langle \tilde{\phi}_a(i,-i)\,\tilde{\psi}_i(\mu_\frac{4}{3}(s))\,V(-\mu_\frac{4}{3}(s))\big\rangle+\nonumber\\
    &\hspace{6cm}+\big\langle \tilde{\phi}_a(i,-i)\,V(\mu_\frac{4}{3}(s))\,\tilde{\psi}_i(-\mu_\frac{4}{3}(s))\big\rangle\bigg)-\frac{2}{s}gB_{ai}C_{iiV}\bigg]\,,\label{eq:schwinger_ellwood2}
\end{align}
where we define
\begin{subequations}
    \begin{align}
        \mu_4(s)&= \tanh\big(\tfrac{1}{2}\mathrm{arctanh}\,s\big)\,,\\
        \mu_\frac{4}{3}(s)&=\tan\big(\tfrac{3}{2}\arctan \tfrac{s}{\sqrt{3}}\big)\,.
    \end{align}
\end{subequations}
The $\frac{1}{s}$ subtractions in \eqref{eq:schwinger_ellwood1} and \eqref{eq:schwinger_ellwood2}, which originate from the insertion of $\bar{P}_0=1-P_0$ in front of the propagator, indeed correctly cancel the dangerous logarithmic divergences due to propagation of $\tilde{\psi}_i$ in the $V\tilde{\psi}_i$ collision.

Furthermore, we can apply the $SL(2;\mathbb{R})$ maps
\begin{align}
    w_\frac{4}{3}^{(\pm)}(z) = \frac{z\pm\mu_\frac{4}{3}(s)}{1\mp \mu_\frac{4}{3}(s)z}
\end{align}
to the two correlators in the integrand of \eqref{eq:schwinger_ellwood2} to map the insertion of $\tilde{\psi}_i(x)$ to $x=0$, fix the insertion of $\tilde{\phi}_a(z,\bar{z})$ at $z=i$, $\bar{z}=-i$ and map the insertion of $V(x)$ to
\begin{align}
    \xi_\frac{4}{3}^{(\pm)}(s) = \pm \frac{2\mu_\frac{4}{3}(s)}{1-\mu_\frac{4}{3}(s)^2}\,.
\end{align}
Carefully accounting for all conformal factors and change of variables, one can show that
\begin{align}
      &\frac{1}{4}\langle 0| c\bar{c} \tilde{\phi}_a(i,-i) 
   U_4^\ast U_\frac{4}{3}\frac{b_0}{L_0}\bar{P}_0 
    \Big[c\tilde{\psi}_i(\tfrac{1}{\sqrt{3}})\, cV(-\tfrac{1}{\sqrt{3}})+cV(\tfrac{1}{\sqrt{3}})\, c\tilde{\psi}_i(-\tfrac{1}{\sqrt{3}})\Big]|0\rangle =\nonumber\\
    &\hspace{1cm}=\frac{i}{2}\lim_{\varepsilon\to 0} \int_0^\infty d\xi\, S_\frac{4}{3}^{(+)}(\xi)^\varepsilon\,\bigg[\big\langle \tilde{\phi}_a(i,-i)\,V(\xi)\,\tilde{\psi}_i(0)\big\rangle-\frac{1}{2} gB_{ai}C_{iiV}\frac{1}{S_\frac{4}{3}^{(+)}(\xi)}\frac{dS_\frac{4}{3}^{(+)}}{d\xi}\bigg]+\nonumber\\
    &\hspace{2cm}+\frac{i}{2}\lim_{\varepsilon\to 0} \int_{-\infty}^0 d\xi\, S_\frac{4}{3}^{(-)}(\xi)^\varepsilon\,\bigg[\big\langle \tilde{\phi}_a(i,-i)\,\tilde{\psi}_i(0)\,V(\xi)\big\rangle+\frac{1}{2} gB_{ai}C_{iiV}\frac{1}{S_\frac{4}{3}^{(-)}(\xi)}\frac{dS_\frac{4}{3}^{(-)}}{d\xi}\bigg]\,,\label{eq:ellwood_moving1}
\end{align}
where the functions $S_{4/3}^{(\pm)}(\xi)$ can be found as inverses of $\xi_{4/3}^{(\pm)}(s)$. Similarly, applying the $SL(2;\mathbb{R})$ maps
\begin{align}
   w_4^{(\pm)}(z)= \mu_4(s)\frac{z\mp 1}{\mu_4(s)^2\pm z}
\end{align}
to the correlators in the integrand of \eqref{eq:schwinger_ellwood1}, we can map the position where $\tilde{\psi}_i$ is inserted to $0$, the position of $\tilde{\phi}_a(z,\bar{z})$ to $z=i$, $\bar{z}=-i$ and the position of the deforming operator $V$ to
\begin{align}
    \xi_4^{(\pm)} = \pm \frac{2\mu_4(s)}{1-\mu_4(s)^2}\,.
\end{align}
This eventually gives
\begin{align}
      &\frac{1}{4}\langle 0| c\bar{c} \tilde{\phi}_a(i,-i) 
   \frac{b_0}{L_0}\bar{P}_0 U_4^\ast U_\frac{4}{3}
    \Big[c\tilde{\psi}_i(\tfrac{1}{\sqrt{3}})\, cV(-\tfrac{1}{\sqrt{3}})+cV(\tfrac{1}{\sqrt{3}})\, c\tilde{\psi}_i(-\tfrac{1}{\sqrt{3}})\Big]|0\rangle =\nonumber\\
    &\hspace{1cm}=\frac{i}{2}\lim_{\varepsilon\to 0} \int_0^\infty d\xi\, S_4^{(+)}(\xi)^\varepsilon\,\bigg[\big\langle \tilde{\phi}_a(i,-i)\,V(\xi)\,\tilde{\psi}_i(0)\big\rangle-\frac{1}{2} gB_{ai}C_{iiV}\frac{1}{S_4^{(+)}(\xi)}\frac{dS_4^{(+)}}{d\xi}\bigg]+\nonumber\\
    &\hspace{2cm}+\frac{i}{2}\lim_{\varepsilon\to 0} \int_{-\infty}^0 d\xi\, S_4^{(-)}(\xi)^\varepsilon\,\bigg[\big\langle \tilde{\phi}_a(i,-i)\,\tilde{\psi}_i(0)\,V(\xi)\big\rangle+\frac{1}{2} gB_{ai}C_{iiV}\frac{1}{S_4^{(-)}(\xi)}\frac{dS_4^{(-)}}{d\xi}\bigg]\,.\label{eq:ellwood_moving2}
\end{align}
Furthermore, we would like to simplify the log-subtractions in \eqref{eq:ellwood_moving1} and \eqref{eq:ellwood_moving2} so that they are naturally written in terms of simple poles in $\xi$. To this end, we note the results
\begin{subequations}
    \begin{align}
        \int_0^\infty d\xi\,\bigg(\frac{1}{\xi}-\frac{1}{|\xi+i|}-\frac{1}{S_4^{(+)}(\xi)}\frac{dS_4^{(+)}}{d\xi}\bigg) &=-\log 2\,,\\
        \int_0^\infty d\xi\,\bigg(\frac{1}{\xi}-\frac{1}{|\xi+i|}-\frac{1}{S_{4/3}^{(+)}(\xi)}\frac{dS_{4/3}^{(+)}}{d\xi}\bigg) &=-\frac{1}{2}\log 3-\log 2\,,\\
         \int_{-\infty}^0 d\xi\,\bigg(-\frac{1}{\xi}-\frac{1}{|\xi+i|}+\frac{1}{S_4^{(-)}(\xi)}\frac{dS_4^{(-)}}{d\xi}\bigg) &=-\log 2\,,\\
        \int_{-\infty}^0 d\xi\,\bigg(-\frac{1}{\xi}-\frac{1}{|\xi+i|}+\frac{1}{S_{4/3}^{(-)}(\xi)}\frac{dS_{4/3}^{(-)}}{d\xi}\bigg) &=-\frac{1}{2}\log 3-\log 2\,.
    \end{align}
\end{subequations}
Altogether we can therefore write
\begin{subequations}
\begin{align}
    i\mathcal{B}_{aiV} &= i \lim_{\varepsilon\to 0} \int_{-\infty}^\infty d\xi\, S(\xi)^\varepsilon\,\bigg[\big\langle \mathsf{T}\big\{ \tilde{\phi}_a(i,-i)\, \tilde{\psi}_i(0)\,V(\xi)\big\}\big\rangle-\Sigma_{ai}(\xi)\bigg]+\nonumber\\
    &\hspace{1cm}-igB_{ai}C_{iiV}\bigg(\frac{1}{4}\log 3+\log 2\bigg)+igB_{ai}C_{iiV}\bigg(\frac{1}{2}\log 2 + \frac{1}{2}\log\frac{2}{\sqrt{3}}-\log \frac{16}{9}\bigg)\\
 &= \frac{i}{4} \lim_{\varepsilon\to 0} \int_{-\infty}^\infty d\xi\, \big[S_4(\xi)^\varepsilon+S_\frac{4}{3}(\xi)^\varepsilon\big]\,\bigg[2^{\Delta_a-h_i}\big\langle \mathsf{T}\big\{ {\phi}_a(i,-i)\, {\psi}_i(0)\,V(\xi)\big\}\big\rangle-\Sigma_{ai}(\xi)\bigg]+\nonumber\\
    &\hspace{9cm}+igB_{ai}C_{iiV}\big(\log K-2\log 2\big)\,,  \label{eq:amp3SEllwood} 
\end{align}
\end{subequations}
where now we can package the log-subtractions as
\begin{align}
    \Sigma_{ai}(\xi) = gB_{ai}C_{iiV}\bigg(\frac{1}{|\xi|}-\frac{1}{|\xi+i|}\bigg)
\end{align}
and we define
\begin{align}
    S_4(\xi)= \left\{\begin{array}{ll}
      S_4^{(+)}(\xi)  &  \xi>0\\
       S_4^{(-)}(\xi)    & \xi<0
    \end{array}\right.\qquad\text{and}\qquad  S_\frac{4}{3}(\xi) = \left\{\begin{array}{ll}
      S_{4/3}^{(+)}(\xi)   & \xi>0 \\
       S_{4/3}^{(-)}(\xi)    & \xi<0
    \end{array}\right.\,.\label{eq:sxiEllwood}
\end{align}
We have also performed the correlator in the auxiliary sector.
Finally, following the detailed explanation presented in Appendix \ref{app:plumbing}, we realize that the dealing with the power-divergences via the Feynman $\varepsilon$-regularization is equivalent to analytically continuing the $\xi$-integral in the conformal dimensions of the intermediate states propagating where the $\xi$ puncture collides with $0$, $1$ and $\infty$. We therefore obtain the result
\begin{align}
    \mathcal{B}_{aiV} = \tilde{\mathcal{B}}_{aiV}+gB_{ai}C_{iiV}\big(\log K-2\log 2\big)\,,\label{eq:AaiVApp}
\end{align}
where we define
\begin{align}
     \tilde{\mathcal{B}}_{aiV}=\frac{1}{2}\int\limits_{\substack{-\infty \\ \text{a.c.}}}^\infty d\xi\,\bigg[2^{\Delta_a-h_i}\, \big\langle \mathsf{T}\big\{\phi_a(i,-i)\,\psi_i(0)\,V(\xi)\big\}\big\rangle- \Sigma_{ai}(\xi)\bigg]\,.
\end{align}
Again, if the states produced in the collision of $V$ with $\psi_i$ at $0$ violate the assumption \eqref{eq:res}, we expect that one would still arrive at the final result \eqref{eq:AaiVApp}, as the condition \eqref{eq:res} seems to play no special role in 2d CFT. For this, one would have to show that the additional contributions to the amplitude coming from the propagating descendants in $\mathrm{ker}\,L_0$ can be canceled with the resonant terms encountered in Appendix \ref{app:equiv_from_sign}.

\section{Analytic continuation from the Feynman $\varepsilon$-prescription}
\label{app:plumbing}

In this appendix we will show, for any twist-symmetric OSFT, that the $s^\varepsilon$ regularization of the tachyon divergences of 4-point amplitudes expressed in terms of the Feynman modulus $s$ (see e.g.\ \eqref{eq:schwinger_frame}) is equivalent to analytically continuing in the conformal dimension of operators propagating in the collisions of the operator inserted at the moving puncture $\xi$ with the insertions at $0$, $1$ and $\infty$. The key result which facilitates this equivalence are the relations (which hold for any small enough $\alpha<0<\beta$ and $\mu\neq 1$)
\begin{subequations}
\label{eq:int_rels}
\begin{align}
    \lim_{\varepsilon\to 0} \int_0^\beta d\xi\, S(\xi)^\varepsilon\, \xi^{-\mu} &= +\frac{\beta^{-\mu+1}}{-\mu+1}+P(\mu)\,,\\
    \lim_{\varepsilon\to 0} \int_\alpha^0 d\xi\, S(\xi)^\varepsilon\, \xi^{-\mu} &= -\frac{\alpha^{-\mu+1}}{-\mu+1}-P(\mu)\,,
\end{align}
\end{subequations}
where $P(\mu)$ is given by \eqref{eq:Pmu} below. In particular, we have $P(\mu)=0$ unless $\mu\in \mathbb{Z}$, $\mu>1$. Here $S(\xi)$ stands for a function which maps the position $\xi$ of the moving puncture to the Feynman modulus $s$ for the amplitude in question near propagator degenerations. In the particular case of the Witten's OSFT, we have seen the examples of such functions above (\eqref{eq:sxi} for the 4-point boundary amplitude and \eqref{eq:sxiEllwood} for the 3-point bulk-boundary amplitude). 

We will first show in Appendix \ref{app:sign_from_plumbing} that in the particular case of the boundary 4-point amplitude, the plumbing fixture implies that around $\xi=0$, the function $S(\xi)$ (for a twist-symmetric cubic vertex) can be generally expressed in the form 
\begin{align}
    S(\xi)=f(\xi)\,\mathrm{sgn}\,\xi \,,\label{eq:S_sign}
\end{align}
where $f(\xi)$ is analytic around $\xi=0$. In the case of the bulk-boundary 3-point amplitudes discussed in Appendix \ref{app:feyn_ellwood}, the validity of \eqref{eq:S_sign} can be checked directly using the explicit form \eqref{eq:sxiEllwood}. We will then see in Appendix \ref{app:int_from_sign} how the integrals \eqref{eq:int_rels} follow from \eqref{eq:S_sign} and, finally, in Appendix \ref{app:equiv_from_sign}, we will show that they ensure the equivalence of the Feynman-region $\varepsilon$-regularization with the analytic continuation of the $\xi$-integral in the conformal dimensions of the intermediate states propagating in the collisions at $0$, $1$ and $\infty$.

\subsection{$S(\xi)=f(\xi)\,\mathrm{sgn}\,\xi $ from plumbing fixture}
\label{app:sign_from_plumbing}

Let us parametrize a generic cubic vertex by the coordinate maps $f_k(w)$ for $k=1,2,3$ mapping the local coordinate patches (unit upper-half disk with coordinate $w$) inside a the upper-half plane $f_1(0)=0$, $f_2(0)=1$ and $f_3(0)=\infty$. Cyclicity of the vertex relates the three coordinate maps $f_k(w)$ as
\begin{align}
    f_2(w) = \frac{1}{1-f_1(w)}\,,\qquad f_3(w) = \frac{1}{1-f_2(w)}\,.
\end{align}
For a healthy vertex, we will assume that the maps $f_k(w)$ are analytic on the unit upper-half disk. Let us also denote their inverses as $h_k\equiv f_k^{-1}$. Moreover, we will assume that the vertex is twist-symmetric. As we have already discussed in Section \ref{sec:sft_intro}, this means that there exist involutions $T_k$ preserving the 3-punctured upper-half plane such that 
\begin{align}
    T_k\circ f_k = f_k \circ t
\end{align}
holds for the involution $t(w)=-w^\ast$ of the local coordinate patch reversing the orientation of the open string. Realizing that the involution $T_1$ swaps the punctures at $1$ and $\infty$ (while it fixes 0), we conclude that for the coordinate map $f_1$, twist symmetry implies that
\begin{align}
    h_1(\infty) = -h_1(1)\,,\label{eq:twist_cons}
\end{align}
as both $1$ and $\infty$ are mapped onto the real line in the $w$ coordinate so that $t(w)$ simply acts by multiplying with the sign.

Let us now focus on deriving the map between the position $\xi$ of the moving puncture of a $4$-point amplitude and the Feynman modulus $s$ (with $0<s<1$) in the vicinity of $\xi=0$. For this, we will create a 4-punctured surface by connecting two 3-punctured UHPs via a strip of length $-\log s$ through their first puncture (the one equipped with local coordinate map $f_1$). Plumbing fixture then gives the relation
\begin{align}
    h_1(z) h_1(z^\prime) = -s \qquad \Longleftrightarrow\qquad z(z^\prime) = h_1^{-1}\big(\!-\!\tfrac{s}{h_1(z)}\big)
\end{align}
between two coordinates $z$, $z^\prime$ on the resulting 4-punctured surface. Hence, the remaining four punctures
\begin{align}
    z=1\,,\qquad z=\infty\,,\qquad z^\prime=1\,,\qquad z^\prime =\infty
\end{align}
which were not used for gluing the two UHPs via the strip, are in the $z$-coordinate positioned at
\begin{align}
    1\,,\qquad \infty\,,\qquad  h_1^{-1}\big(\!-\!\tfrac{s}{h_1(1)}\big)\,,\qquad h_1^{-1}\big(\!-\!\tfrac{s}{h_1(\infty)}\big)\,.
\end{align}
We can then apply the $SL(2;\mathbb{R})$ map
\begin{align}
    z\to \frac{z-h_1^{-1}\big(\!-\!\tfrac{s}{h_1(1)}\big)}{1-h_1^{-1}\big(\!-\!\tfrac{s}{h_1(1)}\big)}
\end{align}
to put the punctures at the positions\
\begin{align}
    1\,,\qquad \infty\,,\qquad 0\,,\qquad \xi_1(s)>0\,,
\end{align}
where
\begin{align}
    \xi_1(s) = \frac{h_1^{-1}\big(\!-\!\tfrac{s}{h_1(\infty)}\big)-h_1^{-1}\big(\!-\!\tfrac{s}{h_1(1)}\big)}{1-h_1^{-1}\big(\!-\!\tfrac{s}{h_1(1)}\big)}\equiv h(s)\,.\label{eq:xi1_plumbing}
\end{align}
This gives the map between the Feynman modulus $s$ and the position of the moving puncture inside the 4-point amplitude for small positive $\xi$.\footnote{The fact that $\xi_1(s)>0$ can be established by realizing that for $w$ real, $f_1(w)$ has to be monotonic (conventionally increasing) in order to provide a one-to-one mapping between the local coordinate patch and the boundary of the UHP.} Also note that even though the physical range of $s$ is $0<s<1$, the function $h(s)$ appearing on the r.h.s.\ of \eqref{eq:xi1_plumbing} is analytic for $-1<s<1$.

For small negative $\xi$, we instead obtain the map
\begin{align}
    \xi_6(s) = \frac{\xi_1(s)}{\xi_1(s)-1} = -\frac{h_1^{-1}\big(\!-\!\tfrac{s}{h_1(\infty)}\big)-h_1^{-1}\big(\!-\!\tfrac{s}{h_1(1)}\big)}{1-h_1^{-1}\big(\!-\!\tfrac{s}{h_1(\infty)}\big)}\,.\label{eq:xi6_init}
\end{align}
At this point, we can substitute the consequence \eqref{eq:twist_cons} of twist symmetry into the r.h.s.\ of \eqref{eq:xi6_init} to obtain
\begin{align}
     \xi_6(s) = -\frac{h_1^{-1}\big(\tfrac{s}{h_1(1)}\big)-h_1^{-1}\big(\tfrac{s}{h_1(\infty)}\big)}{1-h_1^{-1}\big(\tfrac{s}{h_1(1)}\big)} = h(-s)\,.
\end{align}
Hence, denoting by $f(\xi)$ the inverse of $h(s)$, we may write for the Feynman modulus $S(\xi)$ in terms of the position $\xi$ of the moving puncture
\begin{align}
    S(\xi) = f(\xi)\,\mathrm{sgn}\, \xi \,,
\end{align}
where $f(\xi)$ is analytic.

\subsection{\eqref{eq:int_rels} from $S(\xi)=f(\xi)\,\mathrm{sgn}\,\xi$}
\label{app:int_from_sign}

Let us now evaluate the integrals on the l.h.s.\ of \eqref{eq:int_rels}. To this end, let us first note that for small $s$ we can expand the analytic function $h(s)$ defined in \eqref{eq:xi1_plumbing} as
\begin{align}
    h(s) = -\frac{2}{h_1(\infty)h_1^\prime(0)} s +\mathcal{O}(s^2)\,.
\end{align}
Since we can assume that for a healthy 3-vertex, we have $h_1(\infty)h_1^\prime(0)\neq 0$,\footnote{We have $h_1(\infty)\neq 0$, because $h_1(0)=0$ and $h_1$ is one-to-one. We also have $h_1^\prime(0)\neq 0$, otherwise the vertex is ill-defined for states with non-zero weight.} we conclude that the inverse function $f(\xi)$ can be expanded as
\begin{align}
    f(\xi) = \sum_{k=1}^\infty \sigma_k\xi^k\,,
\end{align}
where
\begin{align}
    \sigma_1 = -\frac{1}{2}h_1(\infty)h_1^\prime(0) \neq 0\,.
\end{align}
This means that for $\xi>0$, we can expand
\begin{align}
&\int_0^\beta d\xi\, S(\xi)^\varepsilon\,\xi^{-\mu}=\nonumber\\
&\hspace{0.2cm} =\int_0^\beta d\xi\,\sigma_1^\varepsilon \,\sum_{l=0}^\infty \frac{\varepsilon(\varepsilon-1)\ldots(\varepsilon-l+1)}{l!}\sum_{k_1=2}^\infty\ldots\sum_{k_l=2}^\infty \frac{\sigma_{k_1}\ldots \sigma_{k_l}}{(\sigma_1)^l}\, \xi^{k_1+k_2+\ldots+k_l -l +\varepsilon-\mu}\,,\label{eq:intp}
\end{align}
while for $\xi<0$, we obtain
\begin{align}
&\int_\alpha^0 d\xi\, S(\xi)^\varepsilon\,\xi^{-\mu}=\nonumber\\
&\hspace{0.2cm} =\int_0^{|\alpha|} d\xi\,\sigma_1^\varepsilon \,\sum_{l=0}^\infty \frac{\varepsilon(\varepsilon-1)\ldots(\varepsilon-l+1)}{l!}\sum_{k_1=2}^\infty\ldots\sum_{k_l=2}^\infty \frac{\sigma_{k_1}\ldots \sigma_{k_l}}{(\sigma_1)^l}\,(-1)^\varepsilon (-\xi)^{k_1+k_2+\ldots+k_l -l +\varepsilon-\mu}\,.\label{eq:intm}
\end{align}
First, provided that $\mu\neq 1$ is chosen such that $\mu\notin \{2,3,\ldots\}$, then the power-law integrals in \eqref{eq:intp} and \eqref{eq:intm}
never produce a pole at $\varepsilon=0$. We can then safely take the $\varepsilon\to 0$ limit to establish the relations \eqref{eq:int_rels} in the cases when $\mu\notin \{2,3,\ldots\}$. Indeed, for such $\mu$, the $P(\mu)$ terms do not contribute non-trivially. On the other hand, for $\mu\in\{2,3,\ldots\}$, there are always $l\geqslant 0$ and $k_1,\ldots,k_l\geqslant 2$ such that
\begin{align}
    k_1+k_2+\ldots+k_l-l+1=\mu\,,\label{eq:resonance}
\end{align}
in which case we have
    \begin{align}
        \int_0^\beta d\xi\, \xi^{k_1+k_2+\ldots+k_l -l +\varepsilon-\mu} &= \frac{\beta^\varepsilon}{\varepsilon}\,.
    \end{align}
This has the effect of contributing an extra  ``resonant'' term to the integrals \eqref{eq:intp} and \eqref{eq:intm}. Realizing that thanks to the resonance condition \eqref{eq:resonance}, we can replace
\begin{align}
    (-1)^{k_1+k_2+\ldots+k_l-l-\mu}=-1\,,
\end{align}
in \eqref{eq:intm},
we readily obtain the relations \eqref{eq:int_rels} with
\begin{align}
    P(\mu) = \underset{k_1+\ldots+k_l-l+1=\mu}{\sum_{l=0}^{\infty}\sum_{k_1=2}^\infty\ldots\sum_{k_l=2}^\infty }\,\frac{\sigma_{k_1}\ldots\sigma_{k_l}}{(\sigma_1)^l}\frac{(-1)^{l-1}}{l}\,,\qquad \text{for $\mu\in\{2,3,\ldots\} $}\,.\label{eq:Pmu}
\end{align}
Note that $P(\mu)$ is manifestly finite because the sum on the r.h.s.\ of \eqref{eq:Pmu} is finite.

\subsection{Deriving analytic continuation for the collisions of $V(\xi)$}
\label{app:equiv_from_sign}

Finally, let us observe how the relations \eqref{eq:int_rels} imply that one can simply analytically continue the $\xi$-integral in the dimensions of states propagating in the collisions to avoid power-divergences. For the sake of simplicity, let us focus on the collision of $V(\xi)$ with $\psi_i(0)$, the collisions at $1$ and $\infty$ (which additionally appear in the case of the boundary 4-point amplitude) can be dealt with analogously. Recall that in general, this collision can be expanded as the OPE \eqref{eq:Vpsij}, namely
\begin{align}
    \mathsf{T}\{ V(\xi)\,\psi_j(0)\}= \sum_{k} \sum_{n=0}^\infty \frac{C_{Vjk}(\xi)}{|\xi|^{1+h_j-h_k}} \frac{1}{n!}\frac{(1-h_j+h_k)^{\overline{n}}}{(2h_k)^{\overline{n}}} \xi^n \partial^n\psi_k(0)\,.\label{eq:VpsiC}
\end{align}
Assuming that the simple poles have already been dealt with by explicit subtractions, let us concentrate on the terms in \eqref{eq:VpsiC} proportional to
\begin{align}
\frac{C_{Vjk}(\xi)}{|\xi|^{1+h_j-h_k}}\xi^{n} 
\end{align}
for $h_j-h_k-n>0$. These would lead to power divergences upon integrating over $\xi$ over an interval containing the collision point $\xi=0$. First, for $h_j-h_k\notin\mathbb{Z}$, we can drop the $P(\mu)$-terms in the relations \eqref{eq:int_rels} so that one obtains
\begin{subequations}
\begin{align}
      \lim_{\varepsilon\to 0} \int_0^\beta d\xi\, S(\xi)^\varepsilon\, \frac{C_{Vjk}(\xi)}{|\xi|^{1+h_j-h_k}}\xi^{n}  &= \frac{\beta^{n-h_j+h_k}}{n-h_j+h_k}\,C_{Vjk}\,,\\
      \lim_{\varepsilon\to 0} \int_\alpha^0 d\xi\, S(\xi)^\varepsilon\, \frac{C_{Vjk}(\xi)}{|\xi|^{1+h_j-h_k}}\xi^{n}&= \frac{\alpha^{n-h_j+h_k}}{n-h_j+h_k}\,e^{i\pi (h_k-h_j)} \,C_{jVk}\,.
\end{align}
\end{subequations}
This justifies the analytic-continuation prescription in the case $h_j-h_k\notin\mathbb{Z}$. When $h_j-h_k \in \mathbb{Z}$, more care is needed: applying \eqref{eq:int_rels}, one finds
\begin{align}
\lim_{\varepsilon\to 0} \int_\alpha^\beta d\xi\, S(\xi)^\varepsilon\, \frac{C_{Vjk}(\xi)}{|\xi|^{1+h_j-h_k}}\xi^{n}  &=
    C_{Vjk}\,\frac{\beta^{n-h_j+h_k}}{n-h_j+h_k}+(-1)^{h_j-h_k} \,C_{jVk}\,\frac{\alpha^{n-h_j+h_k}}{n-h_j+h_k}+\nonumber\\
    &\hspace{2cm}+\big[C_{Vjk}+(-1)^{h_j-h_k} C_{jVk}\big]P(1+h_j-h_k-n)\,.\label{eq:intab}
\end{align}
We can see that the simple analytic-continuation prescription appears to be obstructed by the resonant contribution on the second line of \eqref{eq:intab}
At this point, however, we need to recall that the Feynman $\varepsilon$-regularized results \eqref{eq:amp4S} and \eqref{eq:amp3SEllwood} were only derived assuming that the restriction \eqref{eq:res} on the boundary OPE structure constants was in place. This fortunately ensures that the prefactor of the resonant term in \eqref{eq:intab} vanishes and hence that the $\varepsilon$-prescription is indeed equivalent to analytically continuing the $\xi$-integral in the weights of intermediate states.
If the condition \eqref{eq:res} was not satisfied, the $\varepsilon$-regulated results \eqref{eq:amp4S} and \eqref{eq:amp3SEllwood} would have to be corrected by terms arising from (algebraically) propagating the $\mathrm{ker}\, L_0$ descendant states in the star product $cV\ast c\tilde{\psi}_j$ (which are generally present when \eqref{eq:res} is violated). It is then conceivable that these additional contributions would cancel with the resonant term in the second line of \eqref{eq:intab}, thus ensuring that the integrated correlator can be evaluated using analytic continuation in the intermediate conformal weights irrespective of whether the UV BCFT satisfies \eqref{eq:res} or not.

\section{Power-divergence subtractions}
\label{app:powers}

We will now demonstrate how the analytic continuation prescription (which was introduced above to deal with power divergences) can be implemented in practice by introducing suitable subtractions in the integrated correlators \eqref{eq:Amp4pt} and \eqref{eq:Amp3ptBB}. 

Considering first the 4-point boundary amplitude \eqref{eq:Amp4pt}, we recall that the collisions of $V(\xi)$ with the insertions of $\psi_i(\infty)$, $\psi_j(1)$ and $\psi_k(0)$ can be expressed in terms of the OPE \eqref{eq:Vpsij}. In particular, to isolate only the power-divergent terms produced by the collision, we may write
\begin{subequations}
\begin{align}
     \mathsf{T}\{ V(\xi)\,\psi_k(0)\}&\overset{\xi\sim 0}{=}\hspace{-1mm} \sum_{h_l<h_k}\hspace{-3mm} \sum_{n=0}^{n<h_k-h_l} \hspace{-1mm}\frac{C_{Vkl}(\xi)}{|\xi|^{1+h_k-h_l}} \frac{1}{n!}\frac{(1-h_k+h_l)^{\overline{n}}}{(2h_l)^{\overline{n}}} \xi^n \partial^n\psi_l(0)+\text{less singular}\,,\\
     \mathsf{T}\{ V(\xi)\,\psi_j(1)\}&\overset{\xi\sim 1}{=} \hspace{-1mm}\sum_{h_l<h_j}\hspace{-3mm} \sum_{n=0}^{n<h_j-h_l} \hspace{-1mm} \frac{C_{Vjl}(\xi-1)}{|\xi-1|^{1+h_j-h_l}} \frac{1}{n!}\frac{(1-h_j+h_l)^{\overline{n}}}{(2h_l)^{\overline{n}}} (\xi-1)^n \partial^n\psi_l(1)+\text{less singular}\,,\\
     \mathsf{T}\{ V(\xi)\,\psi_i(\infty)\}&\overset{\xi\sim \infty}{=}\hspace{-1mm} \sum_{h_l<h_i} \hspace{-3mm} \sum_{n=0}^{n<h_i-h_l} \hspace{-1mm} \frac{C_{Vil}(-\tfrac{1}{\xi})}{|\xi|^{-1-h_i+h_l}} \frac{1}{n!}\frac{(1-h_i+h_l)^{\overline{n}}}{(2h_l)^{\overline{n}}} (-\xi)^{-2-n} \partial^n\psi_l(\infty)+\text{less singular}\,.
\end{align}
\end{subequations}
Upon substituting these OPEs into the 4-point correlator of $\psi_i$, $\psi_j$, $\psi_k$ and $V$, one may expand
\begin{subequations}
\begin{align}
    &\big\langle\mathsf{T}\big\{\psi_i(\infty)\,\psi_j(1)\,\psi_k(0)\,V(\xi)\big\}\big\rangle =\nonumber\\
    &\hspace{0.2cm}\overset{\xi\sim 0}{=}  \sum_{h_l<h_k}\hspace{-3mm} \sum_{n=0}^{n<h_k-h_l} \hspace{-1mm}\frac{C_{Vkl}(\xi)C_{ijl}}{|\xi|^{1+h_k-h_l}} \frac{1}{n!}\frac{(1-h_k+h_l)^{\overline{n}} (h_l-h_i+h_j)^{\overline{n}}}{(2h_l)^{\overline{n}}} \xi^n +\text{less singular}\,,\\
    &\big\langle\mathsf{T}\big\{\psi_i(\infty)\,\psi_j(1)\,\psi_k(0)\,V(\xi)\big\}\big\rangle =\nonumber\\
    &\hspace{0.2cm}\overset{\xi\sim 1}{=}  \sum_{h_l<h_j}\hspace{-3mm} \sum_{n=0}^{n<h_j-h_l} \hspace{-1mm} \frac{C_{Vjl}(\xi-1)C_{ilk}}{|1-\xi|^{1+h_j-h_l}} \frac{1}{n!}\frac{(1-h_j+h_l)^{\overline{n}}(h_l-h_i+h_k)^{\overline{n}}}{(2h_l)^{\overline{n}}} (1-\xi)^n +\text{less singular}\,,\\
    &\big\langle\mathsf{T}\big\{\psi_i(\infty)\,\psi_j(1)\,\psi_k(0)\,V(\xi)\big\}\big\rangle =\nonumber\\
    &\hspace{0.2cm}\overset{\xi\sim \infty}{=}   \sum_{h_l<h_i} \hspace{-3mm} \sum_{n=0}^{n<h_i-h_l} \hspace{-1mm} \frac{C_{Vil}(-\tfrac{1}{\xi})C_{ljk}}{|\xi|^{-1-h_i+h_l}} \frac{1}{n!}\frac{(1-h_i+h_l)^{\overline{n}}(h_l-h_k+h_j)^{\overline{n}}}{(2h_l)^{\overline{n}}} \xi^{-2-n}+\text{less singular}\,,
\end{align}
\end{subequations}
where we have used the results
\begin{subequations}
    \begin{align}
        \big\langle \psi_i(\infty)\,\psi_j(1)\,\partial^n\psi_l(0)\big\rangle &=C_{ijl}\,(h_l-h_i+h_j)^{\overline{n}}\,,\\
        \big\langle 
        \psi_i(\infty)\,\partial^n\psi_l(1)\,\psi_k(0)\big\rangle &=C_{ilk}\,(h_l-h_i+h_k)^{\overline{n}}(-1)^n\,,\\
        \big\langle 
        \partial^n\psi_l(\infty)\,\psi_j(1)\,\psi_k(0)\big\rangle &=C_{ljk}(h_l-h_k+h_j)^{\overline{n}} (-1)^n\,.
    \end{align}
\end{subequations}
Let us then pick three points $a,b,c$ on the real line such that 
\begin{align}
    -\infty<a<0\,,\qquad 0<b<1\,, \qquad 1<c<\infty\,.
\end{align}
Introducing the total power subtraction
\begin{align}
    \Pi_{ijk}(\xi) &\equiv \sum_{h_l<h_k}\hspace{-3mm} \sum_{n=0}^{n<h_k-h_l} \hspace{-1mm}\frac{C_{Vkl}(\xi)C_{ijl}}{|\xi|^{1+h_k-h_l}} \frac{1}{n!}\frac{(1-h_k+h_l)^{\overline{n}} (h_l-h_i+h_j)^{\overline{n}}}{(2h_l)^{\overline{n}}} \xi^n+\nonumber\\
    &\hspace{1cm}+\sum_{h_l<h_j}\hspace{-3mm} \sum_{n=0}^{n<h_j-h_l} \hspace{-1mm} \frac{C_{Vjl}(\xi-1)C_{ilk}}{|1-\xi|^{1+h_j-h_l}} \frac{1}{n!}\frac{(1-h_j+h_l)^{\overline{n}}(h_l-h_i+h_k)^{\overline{n}}}{(2h_l)^{\overline{n}}} (1-\xi)^n +\nonumber\\
    &\hspace{1cm}+  \sum_{h_l<h_i} \hspace{-3mm} \sum_{n=0}^{n<h_i-h_l} \hspace{-1mm} \frac{C_{Vil}(-\tfrac{1}{\xi})C_{ljk}}{|\xi|^{-1-h_i+h_l}} \frac{1}{n!}\frac{(1-h_i+h_l)^{\overline{n}}(h_l-h_k+h_j)^{\overline{n}}}{(2h_l)^{\overline{n}}} \xi^{-2-n}\,,\label{eq:Piijk}
\end{align}
one may then rewrite the integrated correlator \eqref{eq:Amp4pt} as
\begin{align}
    \tilde{\mathcal{A}}_{ijkV} &= \int\limits_a^b d\xi\, \bigg[\big\langle\mathsf{T}\big\{\psi_i(\infty)\,\psi_j(1)\,\psi_k(0)\,V(\xi)\big\}\big\rangle-\Pi_{ijk}(\xi)-\Sigma_{ijk}(\xi)\bigg]+\int\limits_{\substack{a\\ \mathrm{a.c.}}}^b d\xi\, \Pi_{ijk}(\xi)+\nonumber\\
    &\hspace{1cm}+\int\limits_b^c d\xi\, \bigg[\big\langle\mathsf{T}\big\{\psi_i(\infty)\,\psi_j(1)\,\psi_k(0)\,V(\xi)\big\}\big\rangle-\Pi_{ijk}(\xi)-\Sigma_{ijk}(\xi)\bigg]+\int\limits_{\substack{b\\ \mathrm{a.c.}}}^c d\xi\, \Pi_{ijk}(\xi)+\nonumber\\
    &\hspace{1cm}+\int\limits_c^\infty d\xi\, \bigg[\big\langle\mathsf{T}\big\{\psi_i(\infty)\,\psi_j(1)\,\psi_k(0)\,V(\xi)\big\}\big\rangle-\Pi_{ijk}(\xi)-\Sigma_{ijk}(\xi)\bigg]+\int\limits_{\substack{c\\ \mathrm{a.c.}}}^\infty d\xi\, \Pi_{ijk}(\xi)+\nonumber\\
    &\hspace{1cm}+\int\limits_{-\infty}^a d\xi\, \bigg[\big\langle\mathsf{T}\big\{\psi_i(\infty)\,\psi_j(1)\,\psi_k(0)\,V(\xi)\big\}\big\rangle-\Pi_{ijk}(\xi)-\Sigma_{ijk}(\xi)\bigg]+\int\limits_{\substack{-\infty\\ \mathrm{a.c.}}}^a d\xi\, \Pi_{ijk}(\xi)\,,
\end{align}
where we have realized that we can drop the $\mathrm{a.c.}$ once the power divergences have been correctly subtracted from the correlators in the integrands. Moreover, using the prescription \eqref{eq:ac_prescription}, we can readily show that 
\begin{align}
    \int\limits_{\substack{a\\ \mathrm{a.c.}}}^b d\xi\, \Pi_{ijk}(\xi) +\int\limits_{\substack{b\\ \mathrm{a.c.}}}^c d\xi\, \Pi_{ijk}(\xi)+\int\limits_{\substack{c\\ \mathrm{a.c.}}}^\infty d\xi\, \Pi_{ijk}(\xi)+\int\limits_{\substack{-\infty\\ \mathrm{a.c.}}}^a d\xi\, \Pi_{ijk}(\xi)=0\,.
\end{align}
Altogether, we therefore find that analytically continuing in the dimensions of propagating states is equivalent to the formula
\begin{align}
    \tilde{\mathcal{A}}_{ijkV} &= \int\limits_{-\infty}^\infty d\xi\, \bigg[\big\langle\mathsf{T}\big\{\psi_i(\infty)\,\psi_j(1)\,\psi_k(0)\,V(\xi)\big\}\big\rangle-\Pi_{ijk}(\xi)-\Sigma_{ijk}(\xi)\bigg]
\end{align}
for the integrated 4-point function, with the power subtractions $\Pi_{ijk}(\xi)$ given by \eqref{eq:Piijk}.

Similarly, for the 3-point bulk-boundary amplitude \eqref{eq:Amp3ptBB}, we can first establish
\begin{align}
    &\big\langle \mathsf{T}\big\{\phi_a(i,-i)\,\psi_i(0)\,V(\xi)\big\}\big\rangle = \nonumber\\
    &\hspace{-0.2cm}\overset{\xi\sim 0}{=}\sum_{h_l<h_i}\hspace{-3mm} \sum_{m=0}^{2m<h_i-h_l} \hspace{-1mm}\frac{C_{Vil}(\xi)B_{al}}{|\xi|^{1+h_i-h_l}} \frac{(2m-1)!!}{(2m)!}\frac{(1-h_i+h_l)^{\overline{2m}}(h_l)^{\overline{m}}}{(2h_l)^{\overline{2m}}}(-1)^{m} 2^{h_l-\Delta_a+m} \xi^{2m} +\text{less singular} \,,
\end{align}
where we have made use of the result
\begin{subequations}
\begin{align}
    \big\langle \phi_a(i,-i)\, \partial^n \psi_l(0)\big\rangle &= B_{al} \,(\partial_x)^n \bigg(\frac{2^{h_l-\Delta_a}}{(1+x^2)^{h_l}}\bigg)\bigg|_{x=0}\nonumber \\ 
    &=\left\{
    \begin{array}{ll}
       B_{al}(-1)^{\frac{n}{2}} 2^{h_l-\Delta_a+\frac{n}{2}}(n-1)!!\,(h_l)^{\overline{\frac{n}{2}}}  &  \text{if $n\in 2\mathbb{Z}$}\\[2mm]
      0   & \text{if $n\in 2\mathbb{Z}+1$}
    \end{array}
    \right.\,.
\end{align}
Hence, introducing the subtraction
\begin{align}
    \Pi_{ai}(\xi) \equiv \sum_{h_l<h_i}\hspace{-3mm} \sum_{m=0}^{2m<h_i-h_l} \hspace{-1mm}\frac{C_{Vil}(\xi)B_{al}}{|\xi|^{1+h_i-h_l}} \frac{(2m-1)!!}{(2m)!}\frac{(1-h_i+h_l)^{\overline{2m}}(h_l)^{\overline{m}}}{(2h_l)^{\overline{2m}}}(-1)^{m} 2^{h_l-h_i+m} \xi^{2m}\,,\label{eq:Piai}
\end{align}
\end{subequations}
and realizing that
\begin{align}
    \int\limits_{\substack{-\infty \\ \text{a.c.}}}^\infty d\xi\, \Pi_{ai}(\xi)=0\,,
\end{align}
we may rewrite the integrated correlator \eqref{eq:Amp3ptBB} as
\begin{align}
    \tilde{\mathcal{B}}_{aiV} = \frac{1}{2}\int\limits_{-\infty}^\infty d\xi\,\bigg[2^{\Delta_a-h_i}\, \big\langle \mathsf{T}\big\{\phi_a(i,-i)\,\psi_i(0)\,V(\xi)\big\}\big\rangle-\Pi_{ai}(\xi)- \Sigma_{ai}(\xi)\bigg]\,.
\end{align}

\section{Virasoro Minimal Models}
\label{app:virasoroMM}

In this appendix, we aim to review some basic facts about the (unitary) A-series of Virasoro Minimal Models (VMMs) with Cardy boundaries. These are unitary CFTs $\mathcal{M}_m$ labeled by an integer $m\geqslant 3$ whose central charge takes values
\begin{align}
    c_m = 1-\frac{6}{m(m+1)}<1\,.
\end{align}
In this paper, we will be mostly interested in properties of VMMs in the limit $m\to \infty$, in which their central charge approaches $c=1$.

\subsection{Representations}

The spectrum of irreducible representations is labeled by a pair $(r,s)$ of integers which fall into the range
\begin{align}
    1\leqslant r\leqslant m-1\,,\qquad 1\leqslant s \leqslant m\,,
\end{align}
where we identify 
\begin{align}
(r,s)\sim (m-r,m+1-s)\label{eq:mm_id}\,.    
\end{align}
We call the pair $(r,s)$ the \textsl{Kac labels}. The corresponding conformal weights are given by the formula
\begin{align}
    h_{(r,s)} = \frac{[(m+1)r-ms]^2-1}{4m(m+1)}\overset{m\to\infty}{\sim} \frac{1}{4}(r-s)^2\,.\label{eq:Kac_table}
\end{align}
We can see that as $m\to\infty$, the spectrum of irreducible representations becomes infinite with accumulation points near weights $k^2/4$ for integer $k$.
The Verma modules $(r,s)$ are degenerate: in particular, they contain two primitive null states at level $rs$ and $(m-r)(m+1-s)$. Thus, for large $m$, the second null state becomes very heavy. 
The irreducible characters of the representations $(r,s)$ transform under the modular S-transformation $\tau \to -\frac{1}{\tau}$ as
\begin{align}
    \chi_{(r,s)}(-\tfrac{1}{\tau}) = \sum_{(r^\prime,s^\prime)} S_{(r,s)(r^\prime,s^\prime)}\,\chi_{(r^\prime,s^\prime)} (\tau)\,,
\end{align}
where the modular S-matrix reads
\begin{align}
    S_{(r,s)(r^\prime,s^\prime)} = \sqrt{\frac{8}{m(m+1)}}(-1)^{1+rs^\prime +r^\prime s}\sin \bigg(\pi rr^\prime \frac{m+1}{m}\bigg)\,\sin \bigg(\pi ss^\prime \frac{m}{m+1}\bigg)\,.\label{eq:S_mm}
\end{align}
Applying the Verlinde formula, one can derive the fusion rules
\begin{align}
    (r_1,s_2)\otimes (r_2,s_2)  =\bigoplus_{r_{3}{\overset {2}{=}}|r_{1}-r_{2}|+1}^{\min(r_{1}+r_{2},2q-r_{1}-r_{2})-1}\ \bigoplus _{s_{3}{\overset {2}{=}}|s_{1}-s_{2}|+1}^{\min(s_{1}+s_{2},2p-s_{1}-s_{2})-1}(r_3,s_3)\,,\label{eq:MMfusion}
\end{align}
where $\overset {2}{=}$ is to denote that the sums run with an increment of 2.
From now on, we will focus on the A-series of VMMs where the bulk spectrum pairs all left- and right-moving Kac modules diagonally. This yields a modular-invariant partition function as the modular S-matrix \eqref{eq:S_mm} is unitary.

\subsection{Cardy boundaries}

The conformal boundary conditions in the A-series of VMMs are classified \cite{Runkel:1998he} by the Cardy boundary states
\begin{align}
    \| B_{(\alpha,\beta)}\rangle\!\rangle = \sum_{(r,s)} \frac{S_{(\alpha,\beta)(r,s)}}{\sqrt{S_{(1,1),(r,s)}}}|(r,s)\rangle\!\rangle\,,\label{eq:bs_mm}
\end{align}
where $|(r,s)\rangle\!\rangle$ is the Ishibashi state constructed over the bulk primary state with Kac labels $(r,s)$.  For the $g$-function i.e. the boundary state coefficient on the identity, this gives
\be
\label{gFunMM}
g_{(\alpha, \beta)} = \frac{S_{(\alpha,\beta)(1,1)}}{\sqrt{S_{(1,1)(1,1)}}} = \Big(\frac{8}{m(m+1)}\Big)^{\frac{1}{4}} \frac{\sin \frac{\pi \alpha}{m} \sin \frac{\pi \beta }{m+1}}{(\sin \frac{\pi}{m} \sin \frac{\pi}{m+1})^\frac{1}{2}}.
\ee

The spectrum of boundary primaries $\psi_{(i,j)}$ living in the $\text{BCFT}_{(\alpha,\beta)}$ corresponding to the boundary state $\| (\alpha,\beta)\rangle\!\rangle$ can be obtained by looking at the boundary channel of the cylinder overlap involving two boundary states \eqref{eq:bs_mm}. One obtains the boundary partition function
\begin{align}
    Z_{(\alpha,\beta)}(t) = \sum_{(i,j)}\tensor{N}{_{(\alpha,\beta),(\alpha,\beta)}^{(i,j)}}\chi_{(i,j)}(t)\,,
\end{align}
where $\tensor{N}{_{(\alpha,\beta),(\alpha,\beta)}^{(i,j)}}$ encode the Verlinde fusion rules. In the limit of large $m$, where we keep the Kac labels $(\alpha,\beta)$ of the boundary state fixed, this translates into the spectrum 
\begin{align}
    \mathcal{H}_{(\alpha,\beta)}=\bigoplus_{i{\overset {2}{=}}1}^{2\alpha-1}\ \bigoplus _{j{\overset {2}{=}}1}^{2\beta-1}(i,j)\,.
\end{align}
Hence, provided that we keep $\alpha$ and $\beta$ finite as we take the large $m$ limit, the boundary spectrum of $\text{BCFT}_{(\alpha,\beta)}$, unlike the spectrum of the bulk theory, remains finite. Also, we notice that the Kac labels $(i,j)$ of boundary primaries are always odd, meaning that $i-j=2k_{(i,j)}\in 2\mathbb{Z}$ is always even, meaning that their dimensions
\begin{align}
    h_{(i,j)} \overset{m\to\infty}{\sim} \frac{1}{4}(i-j)^2 = k_{(i,j)}^2\label{eq:sq_int}
\end{align}
approach squares of integers as we take $m\to \infty$.

\subsubsection{Boundary OPE structure constants}
\label{app: boundaryOPE}
For the diagonal VMMs, the boundary OPE structure constants were found by Runkel in \cite{Runkel:1998he}. In the this paper we are interested in boundary fields which live on a fixed boundary with Kac labels $(\alpha,\beta)$. That is, we ignore boundary-condition-changing operators. Adopting the normalization such that
\begin{align}
    C_{(i,j)(i,j)(1,1)}=1
\end{align}
for all boundary primaries $\psi_{(i,j)}$, we can write the structure constants as
\begin{align}
    &C_{(i_1,j_1)(i_2,j_2)(i_3,j_3)}=\nonumber\\
    &\hspace{0.0cm}= \frac{\mathsf{F}_{(\alpha,\beta)(i_3,j_3)}\!\!\left[\!\begin{array}{cc}
      (\alpha,\beta)  & (\alpha,\beta)\\
      (i_1,j_1)  & (i_2,j_2)    
      \end{array}\!\right]}{\sqrt{\mathsf{F}_{(\alpha,\beta)(1,1)}\!\!\left[\!\begin{array}{cc}
      (\alpha,\beta)  & (\alpha,\beta)\\
      (i_1,j_1)  & (i_1,j_1)    
      \end{array}\!\right]\!\mathsf{F}_{(\alpha,\beta)(1,1)}\!\!\left[\!\begin{array}{cc}
      (\alpha,\beta)  & (\alpha,\beta)\\
      (i_2,j_2)  & (i_2,j_2)    
      \end{array}\!\right]\!\mathsf{F}_{(\alpha,\beta)(1,1)}\!\!\left[\!\begin{array}{cc}
      (\alpha,\beta)  & (\alpha,\beta)\\
      (i_3,j_3)  & (i_3,j_3)    
      \end{array}\!\right]}}\label{eq:Cboundary}
\end{align}
where the F-matrices can be computed using an algorithm which is described in Appendix A.4 of \cite{Runkel:thesis} (and also can be explicitly found in Appendix A of \cite{Kudrna:2019xnw}) and which can be readily implemented in MATHEMATICA.

\subsubsection{Bulk-boundary structure constants}

Furthermore, canonically normalizing the bulk OPE structure constants so as to have
\begin{align}
    \phi_{(a,b)}(z,\bar{z})\,\phi_{(a,b)}(w,\bar{w}) = \frac{\phi_{(1,1)}(w,\bar{w})}{|z-w|^{4h_{(a,b)}}}+\ldots\,,
\end{align}
the structure constants $B_{(a,b)(i,j)}$ appearing in the bulk-boundary OPE of the bulk field $\phi_{(a,b)}$ can be computed as
\begin{align}
    B_{(a,b)(i,j)} &= \sqrt{\mathsf{F}_{(\alpha,\beta)(1,1)}\!\!\left[\!\begin{array}{cc}
      (\alpha,\beta)  & (\alpha,\beta)\\
      (i,j)  & (i,j)    
      \end{array}\!\right]\frac{S_{(1,1)(a,b)}}{S_{(1,1)(1,1)}}}\sum_{(m,n)}e^{i\pi[2(h_{(m,n)}-h_{(\alpha,\beta)}-h_{(a,b)})+\frac{1}{2}h_{(i,j)}]}\times\nonumber\\
      &\hspace{4.4cm}\times \mathsf{F}_{(1,1)(m,n)}\!\!\left[\!\begin{array}{cc}
      (\alpha,\beta)  & (a,b)\\
      (\alpha,\beta)  & (a,b)    
      \end{array}\!\right] \mathsf{F}_{(m,n)(i,j)}\!\!\left[\!\begin{array}{cc}
      (a,b)  & (a,b)\\
      (\alpha,\beta)  & (\alpha,\beta)    
      \end{array}\!\right] \,,\label{eq:Bbulk_boundary}
\end{align}
where the Kac labels $(m,n)$ run over the representations appearing in the fusion of $(a,b)$ with $(\alpha,\beta)$.

\section{Correlation functions in diagonal VMMs with boundary}
\label{app:mm_correlators}

Let us now compute some correlation functions in diagonal Virasoro Minimal Models with Cardy boundaries. In particular, given the motivation of this paper, we will be interested in correlators involving one or more insertions of the boundary field $\psi_{(1,3)}$ with Kac labels $(1,3)$. This module contains a null state at level 3 whose existence can be exploited \cite{Belavin:1984vu} to write down a differential equation which is satisfied by correlation functions involving an insertion of $\psi_{(1,3)}$.

In particular, consider a chiral 4-point correlator
\begin{align}
    \mathcal{C}(z;z_1,z_2,z_4) \equiv \big\langle \psi_1(z_1)\,\psi_2(z_2)\,\psi_{(1,3)}(z)\,\psi_4(z_4) \big\rangle\,,
\end{align}
where $\psi_1$, $\psi_2$ and $\psi_4$ are some other primary-field insertions in the given minimal model, with conformal weights $h_1$, $h_2$ and $h_4$. The particular form
\begin{align} \mathcal{L}_{-3}|(1,3)\rangle \equiv \bigg[\big(h_{(1,3)}+2\big)L_{-3}-2L_{-1}L_{-2}+\frac{1}{h_{(1,3)+1}}L_{-1}^3\bigg]|(1,3)\rangle
\end{align}
of the level-3 null state in the $(1,3)$ module can then be leveraged to show that $\mathcal{C}$ solves the differential equation
\begin{align}
   0&= \frac{1}{h_{(1,3)}+1} \frac{\partial^3\mathcal{C}}{\partial z^3}
   -2\bigg(\frac{1}{z-z_1}\frac{\partial ^2\mathcal{C}}{\partial z\partial z_1}+\frac{1}{z-z_2}\frac{\partial^2\mathcal{C}}{\partial z\partial z_2}+\frac{1}{z-z_4}\frac{\partial^2\mathcal{C}}{\partial z\partial z_4}\bigg)+\nonumber\\
   &\hspace{1cm}
   -\frac{1}{(z-z_1)^2}\bigg(2h_1\frac{\partial \mathcal{C}}{\partial z}+h_{(1,3)}\frac{\partial \mathcal{C}}{\partial z_1}\bigg)-\frac{1}{(z-z_2)^2}\bigg(2h_2\frac{\partial \mathcal{C}}{\partial z}+h_{(1,3)}\frac{\partial \mathcal{C}}{\partial z_2}\bigg)+\nonumber\\
   &\hspace{1cm}-\frac{1}{(z-z_4)^2}\bigg(2h_4\frac{\partial \mathcal{C}}{\partial z}+h_{(1,3)}\frac{\partial \mathcal{C}}{\partial z_4}\bigg)
    -2h_{(1,3)}\bigg[\frac{h_1}{(z-z_1)^3}+\frac{h_2}{(z-z_2)^3}+\frac{h_4}{(z-z_4)^3}\bigg]\mathcal{C}\,.\label{eq:dif0}
\end{align}
Furthermore, as we will see on two concrete examples below, one can generally use global conformal invariance to rewrite $\mathcal{C}$ in terms of one independent cross-ratio. \eqref{eq:dif0} then yields a third-order ODE so that one needs three pieces of data to fix a particular solution for a correlator. This procedure depends on the nature of the correlator in question, but is typically performed by identifying leading coefficients in various OPE channels which appear in the limits when $z$ collides with other insertion points. The integration constants are then fixed in terms of boundary and bulk-boundary OPE structure constants (which, for a fixed boundary condition $(\alpha,\beta)$, can be computed using \eqref{eq:Cboundary} and \eqref{eq:Bbulk_boundary}). 

For the applications we are interested in in this paper, it will be sufficient to work in the strict $m\to \infty$ limit, where $h_{(1,3)}=1$ and, recalling \eqref{eq:sq_int}, the dimensions of other boundary primaries are equal to squares of integers. That is, we can write 
$h_i = k_i^2$ for some $k_i\in \mathbb{Z}$ and $i=1,2,4$. See also \cite{Lauria:2023uca}, where some of the correlators considered below were computed at finite $m$.

\subsection{Boundary 4-point correlators}

Let us first discuss how \eqref{eq:dif0} can be used to obtain solutions for the 4-point correlators 
\begin{align}
    \mathcal{C}_{0,4}(x_1,x_2,x_3,x_4) = \big\langle \psi_1(x_1)\,\psi_2(x_2)\,\psi_{(1,3)}(x_3)\,\psi_4(x_4)\big\rangle
\end{align}
of boundary primary fields which include one insertion of $\psi_{(1,3)}$. Here the insertions are ordered to satisfy
\begin{align}
    x_1>x_2>x_3>x_4\label{eq:bndy_ord}
\end{align}
along the real line. 
One can first exploit the $SL(2;\mathbb{R})$ invariance to write
\begin{align}
    \mathcal{C}_{0,4}(x_1,x_2,x_3,x_4) = g\,\frac{x_{12}^{1-k_1^2-k_2^2+k_4^2}x_{14}^{1-k_1^2+k_2^2-k_4^2}x_{24}^{-1+k_1^2-k_2^2-k_4^2}}{x_{13}^2}\mathcal{F}(\xi(x_1,x_2,x_3,x_4))
\end{align}
where $g$ is the $g$-function and we define $x_{ij}=x_i-x_j$, as well as the cross-ratio
\begin{align}
    \xi(x_1,x_2,x_3,x_4) = \frac{x_{12}x_{34}}{x_{13}x_{24}}\,.
\end{align}
For the particular ordering \eqref{eq:bndy_ord}, we can show that $0<\xi<1$.
We can also recognize that, in fact,
\begin{align}
   g \mathcal{F}(\xi) = \big\langle \psi_1(\infty)\,\psi_2(1)\,\psi_{(1,3)}(\xi)\,\psi_4(0)\big\rangle\equiv \lim_{x_1\to\infty} x_1^{2k_1^2}\mathcal{C}_{0,4}(x_1,1,\xi,0)\,.
\end{align}
The null-state equation \eqref{eq:dif0} then implies the differential equation
\begin{align}
   0&=(1-\xi)^3 \xi^3 \,\frac{d^3\mathcal{F}}{d\xi^3} -4 (1-\xi)^2 \xi^2 (2 \xi-1)\,\frac{d^2\mathcal{F}}{d\xi^2}+\nonumber\\
   &\hspace{1cm}+2 (\xi-1) \xi \left[\xi \left(2 k_1^2 (\xi-1)+2 k_2^2-7 \xi+7\right)-2 k_4^2 (\xi-1)-1\right]\,\frac{d\mathcal{F}}{d\xi}+\nonumber\\[2mm]
   &\hspace{1cm}-2 \left[\xi \left(k_1^2 ((3-2 \xi) \xi-1)-3 \left(k_2^2+1\right) \xi+k_2^2+2 \xi^2+1\right)+k_4^2 (\xi-1) (3 \xi-2)\right]\,\mathcal{F}\label{eq:ODE_F}
\end{align}
satisfied by $\mathcal{F}(\xi)$ for $0<\xi<1$. For some fixed integers $k_1$, $k_2$ and $k_4$, this equation can generally be solved using the Frobenius method. Here we choose to present the general solution in the cases when 1.\ $\psi_2=\psi_{(1,3)}$, $\psi_1=\psi_4=\psi_{(i,j)}$ (Section \ref{sec:case1}) and 2.\ $\psi_1=\psi_{(1,3)}$, $\psi_2=\psi_4=\psi_{(i,j)}$ (Section \ref{sec:case2}) for some Kac labels $(i,j)\neq (1,1)$ (the case $(i,j)= (1,1)$ trivially reduces to a two-point function of $\psi_{(1,3)}$). This will eventually allow us to determine the ordered boundary correlators of the type 
\begin{align}
    \big\langle\mathsf{T}\{\psi_{(i,j)}(\infty)\,\psi_{(1,3)}(1)\,\psi_{(1,3)}(\xi)\,\psi_{(i,j)}(0)\} \big\rangle \,,\qquad -\infty <\xi <\infty\,,\label{eq:ordered1}
\end{align}
for some boundary primary field $\psi_{(i,j)}$ (Section \ref{sec:t1}). These correlators are the basic ingredients for the computation of the subleading anomalous dimension of $\psi_{(i,j)}$ under the $\psi_{(1,3)}$ boundary deformation, as well as for the subleading change in the $g$-function (when we set $(i,j)=(1,3)$) and the leading-order change of the structure constant $C_{(i,j)(i,j)(1,3)}$.

To give yet another example, below (Section \ref{sec:t2}) we will also present the result for the ordered correlator
\begin{align}
   \big\langle\mathsf{T}\{\psi_{(1,7)}(\infty)\,\psi_{(1,3)}(1)\,\psi_{(1,3)}(\xi)\,\psi_{(1,5)}(0)\} \big\rangle \,,\qquad -\infty <\xi <\infty\,,\label{eq:ordered2}
\end{align}
which enters the calculation of the change in the OPE structure constant $C_{(1,3)(1,5)(1,7)}$.

\subsubsection*{Case $\psi_2=\psi_{(1,3)}$, $\psi_1=\psi_4=\psi_{(i,j)}$}
\label{sec:case1}

In this kinematical configuration, one can find the general solution 
\begin{align}
\mathcal{F}(\xi)&=\frac{1}{ (1-\xi)^2 \xi }\Big[\big(2 k_{(i,j)}^2 (1-\xi)^2+\xi\big) C_1+\xi^{1-2 k_{(i,j)}}C_2+\xi^{1+2 k_{(i,j)}}C_3\Big]\,,\qquad 0<\xi<1\,,
\end{align}
of \eqref{eq:ODE_F} where $C_1$, $C_2$ and $C_3$ are some integration constants. Note that a separate treatment would be needed 
for the case $i=j$ (where $k_{(i,j)}$ vanishes). Since we will not need this case for the examples discussed in this paper, we will from now on assume $i\neq j$.

The constants $C_1$, $C_2$ and $C_3$ can be fixed in terms of OPE structure constants by looking at the limits $\xi\to 0$ and $\xi\to 1$. First, we note that 
\begin{align}
    \mathcal{F}(\xi)&\overset{\xi\to 1}{\sim} \frac{C_1+C_2+C_3}{(1-\xi)^2}+\frac{ 2k_{(i,j)}(C_2 - C_3)}{1-\xi}\,,
\end{align}
so that comparing with the OPE coefficients in the $\psi_{(1,3)}\psi_{(1,3)}$ and $\psi_{(i,j)}\psi_{(i,j)}$ collisions, we have to set
\begin{subequations}
    \begin{align}
        C_1+C_2+C_3&=1\,,\label{eq:e1}\\
        2k_{(i,j)}(C_2 - C_3)&=C_{(1,3)(1,3)(1,3)}C_{(i,j)(i,j)(1,3)}\,.\label{eq:e2}
    \end{align}
\end{subequations}
On the other hand, realizing that $k_{(i,j)}\geqslant 1$, we can write
\begin{align}
    \mathcal{F}(\xi)&\overset{\xi\to 0}{\sim} \frac{C_2}{\xi^{2 k_{(i,j)}}}+\ldots +2k_{(i,j)}\big(C_2+k_{(i,j)} C_1\big)\,\frac{1}{\xi}\,.\label{eq:xi1}
\end{align}
Comparing with the leading pole in the $\psi_{(1,3)}\psi_{(i,j)}$ collision, we therefore put
\begin{align}
    C_2&= \big(C_{(1,3)(i,j)(i,j-2)}\big)^2\,.\label{eq:e3}
\end{align}
Solving the three linear equations \eqref{eq:e1}, \eqref{eq:e2} and \eqref{eq:e3} for $C_1$, $C_2$ and $C_3$, we obtain
\begin{subequations}
    \begin{align}
        C_1&= 1-2\big(C_{(1,3)(i,j)(i,j-2)}\big)^2+\frac{C_{(1,3)(1,3)(1,3)}C_{(i,j)(i,j)(1,3)}}{2k_{(i,j)}}\,,\\
        C_2&=\big(C_{(1,3)(i,j)(i,j-2)}\big)^2\,,\\[1mm]
        C_3&=\big(C_{(1,3)(i,j)(i,j-2)}\big)^2-\frac{C_{(1,3)(1,3)(1,3)}C_{(i,j)(i,j)(1,3)}}{2k_{(i,j)}}\,,
    \end{align}
\end{subequations}
which, in turn, gives 
\begin{align}
    \mathcal{F}(\xi)&= \frac{k_{(i,j)} \big[C_{(1,3)(1,3)(1,3)} C_{(i,j)(i,j)(1,3)}+\big(2-4 C_{(1,3)(i,j)(i,j-2)}^2\big) k_{(i,j)}\big]}{\xi}+\nonumber\\
    &\hspace{-0.8cm}+\frac{2k_{(i,j)} C_{(1,3)(i,j)(i,j-2)}^2 \big(\xi^{ k_{(i,j)}}-\xi^{ -k_{(i,j)}}\big)^2 +{C_{(1,3)(1,3)(1,3)} C_{(i,j)(i,j)(1,3)} \big(1-\xi^{2 k_{(i,j)}}\big)}+2k_{(i,j)}}{2 (1-\xi)^2}\,.
\end{align}
At the same time, comparing the simple pole residue in \eqref{eq:xi1} with the propagation of $\psi_{(i,j)}$ in the $\psi_{(1,3)}\psi_{(i,j)}$ collision, we recover the crossing relation (valid in the limit $m\to \infty$)
\begin{align}
 k_{(i,j)} \Big[C_{(1,3)(1,3)(1,3)} C_{(i,j)(i,j)(1,3)}+C_{(1,3)(i,j)(i,j-2)}^2 \big(2-4 k_{(i,j)}\big)+2 k_{(i,j)}\Big]=C_{(i,j)(i,j)(1,3)}^2\label{eq:crossing}
\end{align}
for boundary OPE structure constants.
Solving \eqref{eq:crossing} for $C_{(1,3)(i,j)(i,j-2)}^2$ and substituting back into the solution \eqref{eq:xi1} for $\mathcal{F}(\xi)$, we finally obtain the result
\begin{align}
     &\frac{1}{g}\big\langle\psi_{(i,j)}(\infty)\,\psi_{(1,3)}(1)\,\psi_{(1,3)}(\xi)\,\psi_{(i,j)}(0) \big\rangle=\nonumber\\
     &\hspace{0cm}=\frac{1}{2 (1-\xi)^2}\bigg[2+\frac{C_{(1,3)(1,3)(1,3)} C_{(i,j)(i,j)(1,3)} \big(1-\xi^{2 k_{(i,j)}}\big)}{k_{(i,j)}}+\nonumber\\
    &\hspace{0.8cm}+\frac{\big(\xi^{2 k_{(i,j)}}+\xi^{-2 k_{(i,j)}}-2\big)  \big(C_{(1,3)(1,3)(1,3)} C_{(i,j)(i,j)(1,3)} k_{(i,j)}-C_{(i,j)(i,j)(1,3)}^2+2 k_{(i,j)}^2\big)}{k_{(i,j)} (2 k_{(i,j)}-1)}\bigg]+\nonumber\\
    &\hspace{5cm}+\frac{k_{(i,j)} \big(C_{(1,3)(1,3)(1,3)} C_{(i,j)(i,j)(1,3)}-2 C_{(i,j)(i,j)(1,3)}^2+2 k_{(i,j)}\big)}{(1-2 k_{(i,j)}) \xi}\label{eq:r1}
\end{align}
for $0<\xi<1$.

\subsubsection*{Case $\psi_1=\psi_{(1,3)}$, $\psi_2=\psi_4=\psi_{(i,j)}$}
\label{sec:case2}

The general solution of \eqref{eq:ODE_F} in the special case when $\psi_1=\psi_{(1,3)}$, $\psi_2=\psi_4=\psi_{(i,j)}$ can be written as
\begin{align}
    \mathcal{F}(\xi) &= \frac{2 k_{(i,j)}^2+\xi^2-\xi}{\xi(1-\xi) }\,C_1+
    \xi^{-2 k_{(i,j)}} (1-\xi)^{2 k_{(i,j)}}\,C_2+\xi^{2 k_{(i,j)}} (1-\xi)^{-2 k_{(i,j)}}\,C_3\,,\quad 0<\xi<1\,,
\end{align}
for some integration constants $C_1$, $C_2$ and $C_3$. In the limits $\xi\to 0$ and $\xi\to 1$, we can write 
\begin{align}
    \mathcal{F}(\xi)\overset{\xi\to 0}{\sim} \frac{C_2}{\xi^{2 k_{(i,j)}}}+\ldots+2k_{(i,j)}\big( k_{(i,j)} C_1- C_2\big)\,\frac{1}{\xi}
\end{align}
and
\begin{align}
    \mathcal{F}(\xi)\overset{\xi\to 1}{\sim} \frac{C_3}{(1-\xi)^{2 k_{(i,j)}}}+\ldots+2k_{(i,j)}\big( k_{(i,j)} C_1- C_3\big)\,\frac{1}{\xi}\,,
\end{align}
respectively. Comparing with the propagation channels of $\psi_{(i,j-2)}$ and $\psi_{(i,j)}$ in the $\psi_{(1,3)}\psi_{(i,j)}$ collision, we are obliged to set
\begin{subequations}
    \begin{align}
        C_2 &= C_3 = C_{(1,3)(i,j)(i,j-2)}^2\,,\\
       2k_{(i,j)}\big( k_{(i,j)} C_1- C_2\big)&=C_{(i,j)(i,j)(1,3)}^2 \,.
    \end{align}
\end{subequations}
Substituting the solution of the crossing relation \eqref{eq:crossing}, we therefore obtain
\begin{subequations}
\begin{align}
   C_1 &=\frac{C_{(1,3)(1,3)(1,3)} C_{(i,j)(i,j)(1,3)} k_{(i,j)}+2 C_{(i,j)(i,j)(1,3)}^2 (k_{(i,j)}-1)+2 k_{(i,j)}^2}{2 k_{(i,j)}^2 (2 k_{(i,j)}-1)}\,,\\
   C_2 &= \frac{C_{(1,3)(1,3)(1,3)} C_{(i,j)(i,j)(1,3)} k_{(i,j)}-C_{(i,j)(i,j)(1,3)}^2+2 k_{(i,j)}^2}{2 k_{(i,j)} (2 k_{(i,j)}-1)}=C_3\,,
\end{align}
\end{subequations}
which finally gives the result
\begin{align}
     &\frac{1}{g}\big\langle\psi_{(1,3)}(\infty)\,\psi_{(i,j)}(1)\,\psi_{(1,3)}(\xi)\,\psi_{(i,j)}(0) \big\rangle=\nonumber\\
     &\hspace{0.2cm}=\frac{1}{2 k_{(i,j)}^2 (2 k_{(i,j)}-1)}\Bigg[k_{(i,j)} \Big(\big(\tfrac{\xi}{1-\xi}\big)^{2 k_{(i,j)}}  +\big(\tfrac{\xi}{1-\xi}\big)^{-2 k_{(i,j)}} \Big)\times\nonumber\\
    &\hspace{4.9cm}\times\Big(C_{(1,3)(1,3)(1,3)} C_{(i,j)(i,j)(1,3)} k_{(i,j)}-C_{(i,j)(i,j)(1,3)}^2+2 k_{(i,j)}^2\Big)+\nonumber\\
    &\hspace{0.4cm}+{\bigg(\frac{2 k_{(i,j)}^2}{\xi(1-\xi)}- 1\bigg) \Big(C_{(1,3)(1,3)(1,3)} C_{(i,j)(i,j)(1,3)} k_{(i,j)}+2 C_{(i,j)(i,j)(1,3)}^2 (k_{(i,j)}-1)+2 k_{(i,j)}^2\Big)}\Bigg]\label{eq:r2}
\end{align}
for $0<\xi<1$.

\subsubsection*{Result for $ \langle\mathsf{T}\{\psi_{(i,j)}(\infty)\,\psi_{(1,3)}(1)\,\psi_{(1,3)}(\xi)\,\psi_{(i,j)}(0)\} \rangle $}
\label{sec:t1}

Let us now use the results \eqref{eq:r1} and \eqref{eq:r2} to write down an expression for the ordered correlator \eqref{eq:ordered1}. In the range $0<\xi<1$, this is simply given by \eqref{eq:r1}, that is
\begin{align}
     &\frac{1}{g}\big\langle\psi_{(i,j)}(\infty)\,\psi_{(1,3)}(1)\,\psi_{(1,3)}(\xi)\,\psi_{(i,j)}(0) \big\rangle=\nonumber\\
     &\hspace{0cm}=\frac{1}{2 (1-\xi)^2}\bigg[2+\frac{C_{(1,3)(1,3)(1,3)} C_{(i,j)(i,j)(1,3)} \big(1-\xi^{2 k_{(i,j)}}\big)}{k_{(i,j)}}+\nonumber\\
    &\hspace{0.8cm}+\frac{\big(\xi^{2 k_{(i,j)}}+\xi^{-2 k_{(i,j)}}-2\big)  \big(C_{(1,3)(1,3)(1,3)} C_{(i,j)(i,j)(1,3)} k_{(i,j)}-C_{(i,j)(i,j)(1,3)}^2+2 k_{(i,j)}^2\big)}{k_{(i,j)} (2 k_{(i,j)}-1)}\bigg]+\nonumber\\
    &\hspace{5cm}+\frac{k_{(i,j)} \big(C_{(1,3)(1,3)(1,3)} C_{(i,j)(i,j)(1,3)}-2 C_{(i,j)(i,j)(1,3)}^2+2 k_{(i,j)}\big)}{(1-2 k_{(i,j)}) \xi}
\end{align}
To compute \eqref{eq:ordered1} in the range $1<\xi<\infty$, we have to apply the $SL(2;\mathbb{R})$ map
\begin{align}
    z\longrightarrow w(z) = \xi z
\end{align}
on the correlator \eqref{eq:r1}, obtaining
\begin{align}
  \big\langle\psi_{(i,j)}(\infty)\,\psi_{(1,3)}(\xi)\,\psi_{(1,3)}(1)\,\psi_{(i,j)}(0) \big\rangle=  \tfrac{1}{\xi^{2}}\,\big\langle\psi_{(i,j)}(\infty)\,\psi_{(1,3)}(1)\, \psi_{(1,3)}(\tfrac{1}{\xi})\, \psi_{(i,j)}(0) \big\rangle
\end{align}
and therefore
\begin{align}
      &\frac{1}{g}\big\langle\psi_{(i,j)}(\infty)\,\psi_{(1,3)}(\xi)\,\psi_{(1,3)}(1)\,\psi_{(i,j)}(0) \big\rangle=\nonumber\\[2mm]
     &\hspace{0.5cm}=\frac{1}{2 (1-{\xi})^2}\bigg[2+\frac{C_{(1,3)(1,3)(1,3)} C_{(i,j)(i,j)(1,3)} \big(1-\xi^{-2 k_{(i,j)}}\big)}{k_{(i,j)}}+\nonumber\\
    &\hspace{1.2cm}+\frac{\big(\xi^{2 k_{(i,j)}}+\xi^{-2 k_{(i,j)}}-2\big) \big(C_{(1,3)(1,3)(1,3)} C_{(i,j)(i,j)(1,3)} k_{(i,j)}-C_{(i,j)(i,j)(1,3)}^2+2 k_{(i,j)}^2\big)}{k_{(i,j)} (2 k_{(i,j)}-1)}\bigg]+\nonumber\\
    &\hspace{5.0cm}+\frac{k_{(i,j)} \big(C_{(1,3)(1,3)(1,3)} C_{(i,j)(i,j)(1,3)}-2 C_{(i,j)(i,j)(1,3)}^2+2 k_{(i,j)}\big)}{(1-2 k_{(i,j)}) \xi}
\end{align}
for $1<\xi<\infty$. Finally, to obtain the ordered correlator \eqref{eq:ordered1} in the range $-\infty <\xi<0$, we have to transform the correlator \eqref{eq:r2} using suitable $SL(2;\mathbb{R})$ map.
Writing
   \begin{align}
         \langle\psi_{(i,j)}(\infty)\,\psi_{(1,3)}(1)\,\psi_{(i,j)}(0)\,\psi_{(1,3)}(\xi) \big\rangle=   \lim_{x_1\to\infty} z_1^{2k_{(i,j)}^2}\,\langle\psi_{(i,j)}(x_1)\,\psi_{(1,3)}(1)\,\psi_{(i,j)}(0)\,\psi_{(1,3)}(\xi) \big\rangle
    \end{align}
and defining the $SL(2;\mathbb{R})$ map
\begin{align}
    z\longrightarrow\tilde{w}(z)=\frac{\xi(1-x_1)x_1 z}{z(\xi-x_1^2)-x_1\xi+x_1^2}\,,
\end{align}
we can express
    \begin{align}
     &\lim_{x_1\to\infty}x_1^{2k_{(i,j)}^2}\,\big\langle\psi_{(i,j)}(x_1)\,\psi_{(1,3)}(1)\,\psi_{(i,j)}(0) \,\psi_{(1,3)}(\xi)\big\rangle=\nonumber\\[2mm]
     &\hspace{0.2cm}=  \lim_{x_1\to\infty}\Bigg[\Bigg(\frac{1-\tfrac{1}{x_1} }{1-\xi (1-\tfrac{1}{x_1}) -\tfrac{\xi}{x_1^2} }\Bigg)^2\,\times\nonumber\\
     &\hspace{3cm}\times x_1^2\,\Big\langle\tilde{w}\circ\psi_{(i,j)}(1)\,\tilde{w}\circ\psi_{(1,3)}\big(\tfrac{x_1 (\xi-x_1)}{\xi (x_1^2-x_1+1)-x_1^2}\big)\,\tilde{w}\circ\psi_{(i,j)}(0) \,\tilde{w}\circ\psi_{(1,3)}(x_1)\Big\rangle\Bigg]\,,
\end{align}
which gives
\begin{align}
   \big\langle\psi_{(i,j)}(\infty)\,\psi_{(1,3)}(1)\,\psi_{(i,j)}(0) \,\psi_{(1,3)}(\xi)\big\rangle=  \big(\tfrac{1}{1-\xi}\big)^2\big\langle\psi_{(1,3)}(\infty)\,\psi_{(i,j)}(1)\,\psi_{(1,3)}\big(\tfrac{1}{1-\xi}\big)\,\psi_{(i,j)}(0) \big\rangle\,.
\end{align}
Hence, we finally obtain
\begin{align}
  &\frac{1}{g}\big\langle\psi_{(i,j)}(\infty)\,\psi_{(1,3)}(1)\,\psi_{(i,j)}(0) \,\psi_{(1,3)}(\xi)\big\rangle=\nonumber\\[2mm]
     &\hspace{0.5cm}  =\frac{1}{2(1-\xi)^2}\Bigg[2+\frac{C_{(1,3)(1,3)(1,3)} C_{(i,j)(i,j)(1,3)} -\tfrac{2}{k_{(i,j)}} C_{(i,j)(i,j)(1,3)}^2}{k_{(i,j)}}+\nonumber\\
    &\hspace{1.2cm}+\frac{ \Big(\xi^{2 k_{(i,j)}}  +\xi^{-2 k_{(i,j)}}-2 \Big)\Big(C_{(1,3)(1,3)(1,3)} C_{(i,j)(i,j)(1,3)} k_{(i,j)}-C_{(i,j)(i,j)(1,3)}^2+2 k_{(i,j)}^2\Big)}{ k_{(i,j)} (2 k_{(i,j)}-1)}\Bigg]+\nonumber\\
    &\hspace{4.0cm}+\frac{ k_{(i,j)}\big(C_{(1,3)(1,3)(1,3)} C_{(i,j)(i,j)(1,3)} +2 C_{(i,j)(i,j)(1,3)}^2 (1-\tfrac{1}{k_{ij}})+2 k_{(i,j)}\big)}{ (1-2 k_{(i,j)})\xi}
    \end{align}
for $-\infty<\xi<0$. In the special case $(i,j)=(1,3)$ we can write
\begin{align}
 &\frac{1}{g}\big\langle\mathsf{T}\{\psi_{(1,3)}(\infty)\,\psi_{(1,3)}(1)\,\psi_{(1,3)}(\xi)\,\psi_{(1,3)}(0)\} \big\rangle=\nonumber\\[2mm]
 &\hspace{3cm}=\left\{\begin{array}{lc}
     -\frac{C_{(1,3)(1,3)(1,3)}^2-2}{(\xi-1) \xi}-\frac{C_{(1,3)(1,3)(1,3)}^2}{2}+\frac{1}{(\xi-1)^2 \xi^2}+1 & 0<\xi<1 \\[2mm]
    C_{(1,3)(1,3)(1,3)}^2 \big(\frac{1}{\xi-1}-\frac{1}{2 \xi^2}\big)+\frac{1}{\xi^2}+\frac{1}{(\xi-1)^2}+1  & 1<\xi<\infty\\[1mm]
   -\frac{C_{(1,3)(1,3)(1,3)}^2}{\xi}-\frac{C_{(1,3)(1,3)(1,3)}^2-2}{2 (\xi-1)^2}+\frac{1}{\xi^2}+1   &-\infty<\xi<0
 \end{array}\right.\,.
\end{align}
Given these results, we can now also evaluate the integrated correlator \eqref{eq:Amp4pt}. Performing the necessary subtractions of logarithmically-divergent terms and analytically continuing in the dimension of propagating tachyons as explained above, one eventually finds
\begin{subequations}
\begin{align}
&\tilde{\mathcal{A}}_{(i,j)(1,3)(i,j)(1,3)}=\nonumber\\
&\hspace{0.4cm}=\int\limits_{\substack{-\infty \\ \text{a.c.}}}^\infty d\xi\, \bigg[\big\langle\mathsf{T}\big\{\psi_{(i,j)}(\infty)\,\psi_{(1,3)}(1)\,\psi_{(i,j)}(0)\,V(\xi)\big\}\big\rangle-\Sigma_{(i,j)(1,3)(i,j)}(\xi)\bigg]\\
    &\hspace{0.4cm}=-\frac{C_{(1,3)(1,3)(1,3)}C_{(i,j)(i,j)(1,3)}}{2k_{(i,j)}}\int\limits_0^1d\xi\,\xi^{2k_{(i,j)}}\sum_{n=1}^{2(k_{(i,j)}-1)} n\xi^{-n-1}+\nonumber\\
    &\hspace{2cm}-\frac{C_{(1,3)(1,3)(1,3)}C_{(i,j)(i,j)(1,3)}}{2k_{(i,j)}}\int\limits_1^\infty d\xi\,\frac{1}{\xi^{2k_{(i,j)}}}\sum_{n=1}^{2(k_{(i,j)}-1)} n\xi^{n-1}+\nonumber\\
    &\hspace{2cm}+\frac{({k_{(i,j)}}C_{(1,3)(1,3)(1,3)}-2C_{(i,j)(i,j)(1,3)})C_{(i,j)(i,j)(1,3)}}{2k_{(i,j)}^2}\int\limits_{-\infty}^0 d\xi\,\frac{1}{(1-\xi)^2}\,,
\end{align}
\end{subequations}
that is
\begin{align}
   \tilde{\mathcal{A}}_{(i,j)(1,3)(i,j)(1,3)}&=-\frac{gC_{(1,3)(1,3)(1,3)}C_{(i,j)(i,j)(1,3)}}{|i-j|}\bigg[1+2|i-j|  \bigg(\frac{\Gamma^\prime(|i-j|)}{\Gamma(|i-j|)}+\gamma_\text{E} -1\bigg)\bigg]+\nonumber\\
    &\hspace{7cm}-\frac{4gC_{(i,j)(i,j)(1,3)}C_{(i,j)(i,j)(1,3)}}{(i-j)^2}\,,
    \label{integrated4ptCorrelator}
\end{align}
where $\gamma_\text{E}$ is the Euler-Mascheroni constant.

\subsubsection*{Result for $ \langle\mathsf{T}\{\psi_{(1,7)}(\infty)\,\psi_{(1,3)}(1)\,\psi_{(1,3)}(\xi)\,\psi_{(1,5)}(0)\} \rangle $}
\label{sec:t2}

Solving the differential equation \eqref{eq:ODE_F} and fixing integration constants by expanding in the OPE channels where $\xi$ approaches $0$ and $1$, we first obtain (for $0<\xi<1$)
\begin{subequations}
\begin{align}
  &\big\langle\psi_{(1,7)}(\infty)\,\psi_{(1,3)}(1)\,\psi_{(1,3)}(\xi)\,\psi_{(1,5)}(0) \big\rangle =\nonumber\\
  &\hspace{3cm}=-\frac{ 10 \xi^6-15 \xi^5+9 \xi-6}{2\xi (1-\xi) }\,gC_{(1,3)(1,3)(1,3)} C_{(1,3)(1,5)(1,7)}\,,\label{eq:312}\\  
  &\big\langle\psi_{(1,3)}(\infty)\,\psi_{(1,5)}(1)\,\psi_{(1,3)}(\xi)\,\psi_{(1,7)}(0) \big\rangle =\nonumber\\
  &\hspace{3cm}=-\frac{3  \left(2 \xi^6-6 \xi^5+10 \xi^3-15 \xi^2+9 \xi-2\right)}{2 \xi^6 (1-\xi) }\,gC_{(1,3)(1,3)(1,3)} C_{(1,3)(1,5)(1,7)}\,,\label{eq:123}\\
   &\big\langle\psi_{(1,5)}(\infty)\,\psi_{(1,7)}(1)\,\psi_{(1,3)}(\xi)\,\psi_{(1,3)}(0) \big\rangle =\nonumber\\
  &\hspace{3cm}=-\frac{ 10 \xi^6-45 \xi^5+75 \xi^4-50 \xi^3+6 \xi-2}{2\xi (1-\xi)^6 }\,gC_{(1,3)(1,3)(1,3)} C_{(1,3)(1,5)(1,7)}\,.\label{eq:231}
\end{align}
\end{subequations}
Along the way, we derive the crossing relations
\begin{subequations}
\begin{align}
    C_{(1,3)(1,5)(1,5)}&=3C_{(1,3)(1,3)(1,3)}\,,\\
    C_{(1,3)(1,7)(1,7)}&=6C_{(1,3)(1,3)(1,3)}\,.
\end{align}
\end{subequations}
which hold in the limit $m\to\infty$. To obtain the ordered correlator \eqref{eq:ordered2}, we have to apply suitable $SL(2;\mathbb{R})$ transformations on the results \eqref{eq:312}, \eqref{eq:123} and \eqref{eq:231}. On the interval $0<\xi<1$, the ordered correlator is simply given by \eqref{eq:312}. For $1<\xi<\infty$ , we can write
\begin{align}
 \big\langle\psi_{(1,7)}(\infty)\,\psi_{(1,3)}(\xi)\,\psi_{(1,3)}(1)\,\psi_{(1,5)}(0) \big\rangle  &=  \xi^{3}\, \big\langle\psi_{(1,7)}(z_1)\,\psi_{(1,3)}(1)\,\psi_{(1,3)}(\tfrac{1}{\xi})\,\psi_{(1,5)}(0) \big\rangle\,,
\end{align}
while, on the other hand, in the range $-\infty<\xi<0$, we have
\begin{align}
 \big\langle\psi_{(1,7)}(\infty)\,\psi_{(1,3)}(1)\,\psi_{(1,5)}(0)\,\psi_{(1,3)}(\xi) \big\rangle    =\tfrac{1}{(1-\xi)^2} \big\langle\psi_{(1,3)}(\infty)\,\psi_{(1,5)}(1)\,\psi_{(1,3)}\big(\tfrac{1}{1-\xi}\big)\,\psi_{(1,7)}(0) \big\rangle\,.
\end{align}
Altogether, upon substituting from \eqref{eq:312}, \eqref{eq:123} and \eqref{eq:231}, we therefore obtain
\begin{align}
    \frac{1}{g}\big\langle\mathsf{T}\big\{\psi_{(1,7)}(\infty)\,\psi_{(1,3)}(1)\,\psi_{(1,3)}(\xi)\,\psi_{(1,5)}(0)\big\} \big\rangle =\left\{
  \hspace{-2mm}
    \begin{array}{lc}
        -\frac{ 10 \xi^6-15 \xi^5+9 \xi-6}{2\xi (1-\xi) }\,C_{(1,3)(1,3)(1,3)} C_{(1,3)(1,5)(1,7)} &  \\
       -\frac{ 6 \xi^6-9 \xi^5+15 \xi-10}{2 \xi(1-\xi) }\,C_{(1,3)(1,3)(1,3)} C_{(1,3)(1,5)(1,7)}  & \\
     -  \frac{3 (2 \xi^6-3 \xi^5-3 \xi+2)}{2 \xi(1-\xi) }\, C_{(1,3)(1,3)(1,3)} C_{(1,3)(1,5)(1,7)} &
    \end{array}\right.\hspace{-6mm}\,.
\end{align}
One can also evaluate the boundary amplitude \eqref{eq:Amp4pt} as
\begin{align}
    \tilde{\mathcal{A}}_{(1,7)(1,3)(1,5)(1,3)}
    &=-\frac{101}{60}gC_{(1,3)(1,3)(1,3)}C_{(1,3)(1,5)(1,7)}\,.
    \label{17131513amp}
\end{align}

\subsection{Bulk-boundary 3-point correlators}

We would now like to compute the bulk-boundary correlators of the type
\begin{align}
    \big\langle\mathsf{T}\big\{\phi_{(a,b)}(i,-i)\,\psi_{(1,3)}(\xi)\,\psi_{(1,3)}(0)\big\}\big\rangle\label{eq:orderedBB}
\end{align}
for some fixed Cardy boundary condition labeled by $(\alpha,\beta)$. We will work in the strict $y=0$ limit. Again, our strategy will be to leverage the existence of a level-3 null state in the $(1,3)$ module which implies the equation \eqref{eq:dif0} satisfied by chiral 4-point correlation functions.

Let us first assume that $\xi>0$. We can use $SL(2;\mathbb{R})$ symmetry to parametrize
\begin{align}
    \big\langle \phi_{(a,b)}(i,-i)\,\psi_{(1,3)}(\xi)\,\psi_{(1,3)}(0)\big\rangle = g\xi^{-2}\,\mathcal{G}(\eta(\xi))\,,
\end{align}
where $\mathcal{G}(\eta)$ is some as-yet-undetermined function of the cross-ratio
\begin{align}
    \eta(\xi) = \frac{2i\xi}{1+i\xi}\,.
\end{align}
The null-state equation then turns into the 3rd-order ODE
\begin{align}
    0 = -\eta(1-\eta)^3\,\frac{d^3\mathcal{G}}{d\eta^3}+2(1+\eta)(1-\eta)^2\frac{d^2\mathcal{G}}{d\eta^2}-2(1-\eta)(1-2h_{(a,b)}\eta)\frac{d\mathcal{G}}{d\eta}-4h_{(a,b)}(2-\eta)\, \mathcal{G}\,,
\end{align}
which admits general solution
\begin{align}
    \mathcal{G}(\eta)=\bigg(2 h_{(a,b)}\frac{ \eta^2}{ \eta-1}-1\bigg)C_1+(\eta-1)^{-2 \sqrt{h_{(a,b)}}}\,C_2+ (\eta-1)^{2 \sqrt{h_{(a,b)}}}\,C_3\,.
\end{align}
This, in turn, gives
\begin{align}
    &\frac{1}{g}\big\langle \phi_{(a,b)}(i,-i)\,\psi_{(1,3)}(\xi)\,\psi_{(1,3)}(0)\big\rangle =\nonumber\\
    &\hspace{2cm}=\frac{1}{\xi^2}\Bigg[\bigg(8h_{(a,b)}\frac{\xi^2}{\xi^2+1}-1\bigg)\,C_1+\bigg(\frac{\xi+i}{\xi-i}\bigg)^{-2\sqrt{h_{(a,b)}}}C_2+\bigg(\frac{\xi+i}{\xi-i}\bigg)^{2\sqrt{h_{(a,b)}}}C_3\Bigg]\,.\label{eq:corr_gen}
\end{align}
Let us now attempt to fix the integration constants $C_1$, $C_2$ and $C_3$ by expanding \eqref{eq:corr_gen} in the limit $\xi\to 0$ where the two $\psi_{(1,3)}$ boundary insertions collide. In particular, we can write
\begin{align}
 &\frac{1}{g}\big\langle \phi_{(a,b)}(i,-i)\,\psi_{(1,3)}(\xi)\,\psi_{(1,3)}(0)\big\rangle =\nonumber\\
    &\hspace{1cm}=
   \bigg[C_2 (-1)^{-2 \sqrt{h_{(a,b)}}}+C_3 (-1)^{2 \sqrt{h_{(a,b)}}}-C_1\bigg]\,\frac{1}{\xi^2}+\nonumber\\
   &\hspace{1.4cm}+\bigg[4 i (-1)^{-2 \smash{\sqrt{h_{(a,b)}}}} \sqrt{h_{(a,b)}} \Big(C_2-C_3 (-1)^{4 \sqrt{h_{(a,b)}}}\Big)\bigg]\,\frac{1}{\xi}+\nonumber\\
   &\hspace{1.4cm}+8 h_{(a,b)} \bigg[C_1-(-1)^{-2 \sqrt{h_{(a,b)}}} \Big(C_3 (-1)^{4 \sqrt{h_{(a,b)}}}+C_2\Big)\bigg]\,\xi^0+\nonumber\\
   &\hspace{1.4cm}+\frac{1}{3} 4 i (-1)^{-2 \sqrt{h_{(a,b)}}} (8 h_{(a,b)}+1) \sqrt{h_{(a,b)}} \Big(C_3 (-1)^{4 \sqrt{h_{(a,b)}}}-C_2\Big) \,\xi+\nonumber\\
   &\hspace{1.4cm}+\frac{8}{3} h_{(a,b)}  \bigg[2 (-1)^{-2 \sqrt{h_{(a,b)}}} (2 h_{(a,b)}+1) \Big(C_3 (-1)^{4 \sqrt{h_{(a,b)}}}+C_2\Big)-3 C_1\bigg]\,\xi^2+\mathcal{O}(\xi^3)\,.
\end{align}
First, using the result \eqref{eq:Bbulk_boundary}, it is possible to show that
\begin{align}
    B_{(a,b)(1,3)}=\mathcal{O}(\tfrac{1}{m})\,,\qquad B_{(a,b)(1,5)}=\mathcal{O}(\tfrac{1}{m^2})\,,\label{eq:BBasympt}
\end{align}
for all bulk primaries $\phi_{(a,b)}$ and all boundary conditions $(\alpha,\beta)$. As we are computing the correlator \eqref{eq:orderedBB} in the strict $m\to\infty$ limit, this means that we have to ensure that neither $\psi_{(1,3)}$ nor $\psi_{(1,5)}$ propagate on the r.h.s.\ of \eqref{eq:corr_gen} as we take $\xi\to 0$. Only the identity channel will propagate. In particular, this implies
\begin{align}
    C_2-C_3 (-1)^{4 \sqrt{h_{(a,b)}}}=0\,.\label{eq:BBeq1}
\end{align}
Furthermore, the second-order pole comes from the propagation of boundary identity in the $\psi_{(1,3)}\psi_{(1,3)}$ collision, so that we should set
\begin{align}
    C_2 (-1)^{-2 \sqrt{h_{(a,b)}}}+C_3 (-1)^{2 \sqrt{h_{(a,b)}}}-C_1=4^{-h_{(a,b)}}B_{(a,b)(1,1)}\,.\label{eq:BBeq2}
\end{align}
Solving \eqref{eq:BBeq1} and \eqref{eq:BBeq2} for $C_2$ and $C_3$, we can re-expand the bulk-boundary correlator as
\begin{align}
    &\frac{1}{g}\big\langle \phi_{(a,b)}(i,-i)\,\psi_{(1,3)}(\xi)\,\psi_{(1,3)}(0)\big\rangle =\nonumber\\
    &\hspace{1cm}=\frac{4^{- h_{(a,b)}}B_{(a,b)(1,1)}}{\xi^2}- 2^{3-2 h_{(a,b)}} h_{(a,b)}B_{(a,b)(1,1)}+\nonumber\\
    &\hspace{2cm}+ \bigg[\tfrac{1}{3}\, 2^{4-2 h_{(a,b)}} h_{(a,b)} (2 h_{(a,b)}+1) \Big(B_{(a,b)(1,1)}+ 4^{h_{(a,b)}}C_1\Big)-8  h_{(a,b)}C_1\bigg]\,\xi^2+\mathcal{O}(\xi^3)\,.\label{eq:re_expanded}
\end{align}
We observe that in order to fix $C_1$, we have to expand the identity channel of the $\psi_{(1,3)}\psi_{(1,3)}$ boundary OPE and the $\phi_{(a,b)}$ bulk-boundary OPE up to level 4. In detail, we can write
\begin{align}
    \psi_{(1,3)}(\xi)\,\psi_{(1,3)}(0) &= \frac{1}{\xi^2}\bigg[\mathbf{1}(0)+2\xi^2 \Big(T(0)+\tfrac{1}{2}\xi\,\partial T(0)+\tfrac{3}{20}\xi^2 \partial^2 T(0)+\ldots\Big)+\nonumber\\
    &\hspace{9cm}+\tfrac{4}{9}\xi^4\Lambda^{(4)}(0)+\ldots\bigg]\,,\label{eq:OPE1}
\end{align}
(where $\ldots$ denote higher order terms, as well as non-identity channels), as well as
\begin{align}
    \phi_{(a,b)}(i,-i) &= 4^{- h_{(a,b)}}B_{(a,b)(1,1)}\bigg[\mathbf{1}(0)-8h_{(a,b)}\Big(T(0)-\tfrac{1}{10}\partial^2 T(0)+\ldots\Big)+\nonumber\\
    &\hspace{7cm}+\frac{32h_{(a,b)}(5h_{(a,b)}+1)}{27}\Lambda^{(4)}(0)+\ldots\bigg]\,,\label{eq:OPE2}
\end{align}
where $\Lambda^{(4)}$ denotes the level-4 quasi-primary state
\begin{align}
    |\Lambda^{(4)}\rangle =\Big(L_{-2}L_{-2}-\tfrac{3}{5}L_{-4}\Big)|0\rangle
\end{align}
in the boundary identity module. We can readily verify its norm
\begin{align}
    \||\Lambda^{(4)}\rangle\|^2 = \frac{c}{10}(5c+22)\longrightarrow \frac{27}{10}\quad\text{as\quad $m\to\infty$}\,.
\end{align}
Computing the bulk-boundary correlator on the l.h.s.\ of \eqref{eq:re_expanded} using the OPEs \eqref{eq:OPE1} and \eqref{eq:OPE2}, we obtain
\begin{align}
    &\frac{1}{g}\big\langle \phi_{(a,b)}(i,-i)\,\psi_{(1,3)}(\xi)\,\psi_{(1,3)}(0)\big\rangle =\nonumber\\
    &\hspace{1cm}=4^{- h_{(a,b)}}B_{(a,b)(1,1)} \bigg[
    \tfrac{1}{\xi^2}-8h_{(a,b)} +\tfrac{4}{5}\xi^2 h_{(a,b)}\Big(6+\tfrac{16}{9}(5h_{(a,b)}+1)\Big)+\mathcal{O}(\xi^4)
    \bigg]\,.
\end{align}
We can see that we have indeed reproduced the $\xi^0$ term on the r.h.s.\ of \eqref{eq:re_expanded}. In order to also match the coefficient of the $\xi^2$ term, we have to set
\begin{align}
  &\tfrac{1}{3}\, 2^{4-2 h_{(a,b)}} h_{(a,b)} (2 h_{(a,b)}+1) \Big(B_{(a,b)(1,1)}+ 4^{h_{(a,b)}}C_1\Big)-8  h_{(a,b)}C_1=\nonumber\\
  &\hspace{7cm}=\tfrac{4}{5}h_{(a,b)}\Big(6+\tfrac{16}{9}(5h_{(a,b)}+1)\Big)4^{- h_{(a,b)}}B_{(a,b)(1,1)}\,,
\end{align}
which finally gives
\begin{align}
     &\frac{1}{g}\big\langle \mathsf{T}\big\{\phi_{(a,b)}(i,-i)\,\psi_{(1,3)}(\xi)\,\psi_{(1,3)}(0)\big\}\big\rangle =\nonumber\\
    &\hspace{4cm}=\frac{4^{-h_{(a,b)}}B_{(a,b)(1,1)}}{3\xi^2}\bigg[1+2\cos\Big(4\sqrt{h_{(a,b)}}\arctan \xi\Big)-8h_{(a,b)}\frac{\xi^2}{1+\xi^2}\bigg]\,.\label{eq:BB_final}
\end{align}
We have noted that since only even powers of $\xi$ appear upon expanding the r.h.s.\ in small $\xi$, we could reintroduce the ordering operator $\mathsf{T}\{\ldots\}$ at no additional cost. One may also integrate \eqref{eq:BB_final} to obtain the corresponding bulk-boundary amplitude \eqref{eq:Amp3ptBB} as
\begin{subequations}
\begin{align}
    \tilde{\mathcal{A}}_{(a,b)(1,3)(1,3)} &= 2^{2h_{(a,b)}-2} \int\limits_{\substack{-\infty \\ \text{a.c.}}}^\infty d\xi\,\big\langle \mathsf{T}\big\{\phi_{(a,b)}(i,-i)\,\psi_{(1,3)}(\xi)\,\psi_{(1,3)}(0)\big\}\big\rangle\\
    &=\frac{gB_{(a,b)(1,1)}}{6}\int\limits_{-\infty}^\infty d\xi\, \frac{1}{\xi^2}\bigg[\cos\Big(4\sqrt{h_{(a,b)}}\arctan\xi\Big)-1-4h_{(a,b)}\frac{\xi^2}{1+\xi^2}\bigg]
\end{align}
\end{subequations}
that is
\begin{align}
     \tilde{\mathcal{A}}_{(a,b)(1,3)(1,3)} &=-\frac{\pi gB_{(a,b)(1,1)}}{3}|a-b|\bigg(1+\frac{1}{2} |a-b|\bigg)\,.\label{eq:BBIntegrated}
\end{align}
The boundary identity propagation in the $\psi_{(1,3)}\psi_{(1,3)}$ collision was taken care of by noting that 
\begin{align}
    \int\limits_{\substack{-\infty \\ \text{a.c.}}}^\infty d\xi\, \frac{1}{\xi^2}=0\,,
\end{align}
while the logarithmic divergence is absent (as a consequence of \eqref{eq:BBasympt}) so that the subtraction \eqref{eq:Sigmaai} vanishes. 

\printbibliography[]
\endgroup
\end{document}